\documentclass[a4paper,11pt]{article}
\pdfoutput=1 

\usepackage{jcappub} 
\usepackage{float}

\usepackage[T1]{fontenc} 
\usepackage{amsmath}
\usepackage{comment}
\usepackage{cleveref}
\usepackage{tikz}
\usetikzlibrary{calc, arrows.meta}
\usepackage[dvipsnames]{xcolor}
\usepackage{subcaption}

\newcommand{\hubble}{\mathcal{H}}

\newcommand{\conf}{^{\prime}}

\newcommand{\conv}{\mathbf{k}-\mathbf{p}}
\newcommand{\bareta}{\overline{\eta}}
\newcommand{\barx}{\overline{x}}

\newcommand{\tildey}{\tilde{y}}
\newcommand{\eps}{\epsilon}
\newcommand{\timeder}{\partial _{\eta}}
\newcommand{\timederbar}{\partial _{\overline{\eta}}}
\newcommand{\avant}{ \left ( \frac{3+3w}{5+3w} \right ) }
\allowdisplaybreaks

\title{\boldmath Suppression of the induced gravitational wave background due to third-order perturbations}

\author[a]{Rapha{\" e}l Picard,}
\emailAdd{r.h.j.picard@qmul.ac.uk}
\affiliation[a]{Queen Mary University of London,
London, E1 4NS, United Kingdom}

\author[a]{Luis E.~Padilla,}
\emailAdd{l.padilla@qmul.ac.uk}

\author[b]{Karim A.~Malik,}
\emailAdd{drkmalik@proton.me}
\affiliation[b]{Independent}
\author[a]{and David J. Mulryne}
\emailAdd{d.mulryne@qmul.ac.uk}

\abstract{In this work, we revisit and evaluate new source terms which contribute to the induced gravitational wave background. We study their respective contributions to the stochastic gravitational wave background by computing their spectral densities in a radiation-dominated universe. These terms appear at third order in cosmological perturbation theory, however, their correlations with primordial gravitational waves are non-trivial and appear at the same order as so-called scalar induced and scalar-tensor induced gravitational waves. We find that these gravitational wave sources suppress the spectral density at the scales we consider. Furthermore, similarly to scalar-tensor source terms at second order, we find that some terms are enhanced when the input primordial power spectrum of scalar fluctuations is not sufficiently peaked. Hence, where possible, we show that under certain limits the integrands of these terms diverge in the UV sector.}

\begin{document}
\maketitle
\flushbottom

\newpage

\section{Introduction}\label{sec:intro}
Detection of the stochastic gravitational wave background (SGWB) seems to be only a matter of time. Recently, we had strong hints from the pulsar timing array community ~\cite{NANOGrav:2023gor,EPTA:2023sfo,EPTA:2023fyk,EPTA:2023xxk,Xu:2023wog,Reardon:2023gzh,Zic:2023gta,NANOGrav:2023hvm} that such a background exists on large scales (waves which have much lower frequencies than LISA~\cite{Bartolo:2016ami} will be sensitive to). Computing and characterizing the various contributions to the SGWB becomes then both timely and essential. Beyond linear order, gravitational waves (GWs) are sourced by metric and matter fluctuations \cite{Tomita:1967wkp,Matarrese:1992rp,Matarrese:1993zf,Matarrese:1997ay}. So-called scalar-induced gravitational waves (SIGWs) are an example of this, GWs at second order in perturbation theory sourced by quadratic linear scalar fluctuations \cite{Ananda:2006af,Baumann:2007zm}. They have gained interest as the potential counterpart signal to primordial black holes \citep{Padilla:2024cbq,Inomata:2023zup,Clesse:2018ogk,PhysRevD.95.043511,Nakama:2020kdc,Vaskonen:2020lbd,DeLuca:2020agl} (which could represent a significant part of dark-matter, see Ref.~\cite{Yuan:2021qgz} and the references therein for a review) which form from the collapse of large density fluctuations in the early universe. For a general review on SIGWs see Ref.~\cite{Domenech:2021ztg}.

at first order, if tensor fluctuations are also enhanced~\cite{Barnaby:2011qe,Thorne:2017jft,Dimastrogiovanni:2016fuu}, they can leave an imprint on the spectral density of GWs through their coupling with scalar modes \cite{Chang:2022vlv,Bari:2023rcw, Picard:2023sbz,Yu:2023lmo}. Scalar-tensor induced gravitational waves (iGWs) become dominant compared to the amplitude of SIGWs at shorter scales for peaked-input power spectra. Furthermore, Ref.~\cite{Bari:2023rcw} showed that scalar-tensor iGWs suffer from an `unphysical enhancement' due to the divergent behaviour of the source terms in the UV limit. This was subsequently confirmed in Ref.~\cite{Picard:2023sbz}. This divergence is completely classical and therefore it is worth investigating if other source terms can cancel this divergence, particularly those at the same perturbative order.

Ref.~\cite{Chen:2022dah} identified `missing contributions' to the SGWB which contribute at the same perturbative order as scalar-tensor iGWs: correlations of third-order induced and primordial GWs. The authors looked at the behaviour of some source terms in the IR limit and found that on large scales they are scale-invariant and negative. However, since these new contributions are at the same order as SIGWs and scalar-tensor iGWs, it is worth investigating \textit{all} possible source terms at \textit{all scales} since we are also interested in their properties on small scales. In this paper, we will do exactly this. 

Our results show that taking into account correlations of third-order iGWs with primordial GWs heavily suppresses the spectral-density, when all source-terms are considered. Furthermore, some of these new terms also diverge at UV scales and their limiting behaviour does not cancel the divergence which was occurring for scalar-tensor iGWs. However, for some terms, it is not possible to come to definite conclusions, since they are made out of nested integrals of highly-oscillatory functions which are not possible to integrate analytically, and hence, come to definite conclusions about how they behave at the smallest scales.

Our paper is organised as follows. In Sec.~\ref{sec2}, we detail the source terms under consideration and explain the motivation for including them. In Sec.~\ref{sec3}, we present the solution to the third-order GW equation. Sec.~\ref{sec4} provides expressions for the power spectrum of the new contributions, valid for arbitrary equation-of-state parameters $w>-\tfrac{1}{3}$. Finally, in Sec.~\ref{sec5}, we focus on the radiation-dominated universe and examine the limiting behaviour of these terms in different regimes.

\textit{Conventions:} We work in conformal time, defined as $\mathrm{d}\eta \equiv \mathrm{d}t/a(t)$, where $t$ is coordinate time and $a$ the scale factor. A $'$ denotes a conformal time derivative. We use the `mostly plus' metric, Latin indices run over the spatial coordinates $1,2,3$, and we work in natural units, i.e.~$c=1$, and define $M_{Pl}^2=(8\pi G)^{-1}$.

\section{The need to go to third-order}\label{sec2}
In this section, we establish the formalism for third-order iGWs and provide an overview of the various contributions that require consideration for our computation.

Our work is carried out in the conformal Newtonian gauge
\begin{subequations}\label{metricdef}
    \begin{align}
        g_{00} &= -a^2(\eta)(1+2\Phi ^{(1)} + \Phi ^{(2)}) \, , \\
        g_{0i} &= a^2(\eta)\frac{1}{2}B_i^{(2)}=g_{i0}\, , \\
        g_{ij} &= a^2(\eta) \left ( (1-2\Psi ^{(1)} - \Psi ^{(2)})\delta _{ij}  +2h^{(1)}_{ij}+h^{(2)}_{ij}+\frac{1}{3}h^{(3)}_{ij} \right )\, ,
    \end{align}
\end{subequations}
where we followed the expansion in Ref.~\cite{Malik:2008im,Christopherson:2009fp}. $\Phi ^{(n)}$ and $\Psi ^{(n)}$ are the scalar degrees of freedom which are, respectively, the lapse and curvature perturbations. We have not included third-order scalar perturbations since they will not contribute to the GW equation at third order. $B_i^{(2)}$ is a second-order (transverse) vector perturbation and we have not included first-order vectors since they are diluted away during inflation. Similarly, the third-order vectors do not contribute to the equation of motion for the third-order iGWs{, and are therefore also neglected}. $h_{ij}^{(n)}$ are tensorial degrees of freedom which are transverse-traceless (TT). 

Throughout this work, we will be computing the correlation of tensor modes in Fourier space, known as the power spectrum, $P_{\lambda}^{\text{$(nm)$}} (\eta , k)$, defined by 
\begin{equation}\label{generalcorrelationoftensors}
     \langle h_{\mathbf{k},\lambda}^{(n)}h_{\mathbf{k \conf} ,\lambda \conf}^{(m)} \rangle= \delta ^{\lambda \lambda ^{\prime}} \delta (\mathbf{k}+ \mathbf{k^{\prime}}) P_{\lambda}^{\text{$(nm)$}} (\eta , k) = \delta ^{\lambda \lambda ^{\prime}} \delta (\mathbf{k}+ \mathbf{k^{\prime}}) \frac{2\pi ^2}{k^3}\mathcal{P}_{\lambda}^{\text{$(nm)$}} (\eta , k) \, .
\end{equation}
We refer the reader to Appendix~\ref{secondorderApp} for our Fourier conventions. In the above definition, $(n)$, $(m)$ refer to the order of perturbation and $\mathbf{k}$, $\lambda$ are, respectively, the wave-vector and polarisation associated with the GW. $\mathcal{P}_{\lambda}^{\text{$(nm)$}} (\eta , k)$ is the dimensionless power spectrum. To characterise the SGWB, we define \cite{Caprini:2018mtu}
\begin{equation}
    \Omega (\eta , k) = \frac{\rho _{GW}(\eta , k)}{\rho _{crit}(\eta)} \, ,
\end{equation}
where $\rho _{crit}=3\hubble ^2M_{Pl}^2/a^2$ is the critical density and the density of GWs $\rho _{GW}(\eta , k)$ is defined as 
\begin{equation}
    \rho _{GW}(\eta , k) = \frac{M_{Pl}^2}{4} \frac{k^2}{a^2(\eta)} \sum _{\lambda} \overline{\mathcal{P}^{(nm)}_{\lambda}(\eta , k)} \, .
\end{equation}
The overline denotes a time average. Focusing on the power series expansion of the tensor perturbation in the spatial part of the metric
\begin{equation}
    h_{ij}(\eta,\mathbf{x}) = h^{(1)}_{ij}(\eta,\mathbf{x})+\frac{1}{2}h^{(2)}_{ij}(\eta,\mathbf{x})+\frac{1}{6}h^{(3)}_{ij}(\eta,\mathbf{x}) \, ,
\end{equation}
and then considering the two point function in Fourier space of $\langle h_{\mathbf{k},\lambda}h_{\mathbf{k \conf} ,\lambda \conf} \rangle$, we can see what contributes to the GW background up to fourth order in the correlator
\begin{align}\label{tensortotalcorrelation}
    \begin{split}
        \langle h_{\mathbf{k},\lambda}h_{\mathbf{k \conf} ,\lambda \conf} \rangle &= \underbrace{ \langle h_{\mathbf{k},\lambda}^{(1)}h_{\mathbf{k \conf} ,\lambda \conf}^{(1)} \rangle }_{\text{Primordial}} + \frac{1}{4} \underbrace{ \langle h_{\mathbf{k},\lambda}^{(2)}h_{\mathbf{k \conf} ,\lambda \conf}^{(2)} \rangle }_{\text{Second-order GWs}} +   \underbrace{ \frac{1}{6}\langle h_{\mathbf{k},\lambda}^{(1)}h_{\mathbf{k \conf} ,\lambda \conf}^{(3)} \rangle   + \frac{1}{6}\langle h_{\mathbf{k},\lambda}^{(3)}h_{\mathbf{k \conf} ,\lambda \conf}^{(1)} \rangle }_{\text{Additional contributions}} \\
        &=  \langle h_{\mathbf{k},\lambda}^{(1)}h_{\mathbf{k \conf} ,\lambda \conf}^{(1)} \rangle +\frac{1}{4}  \langle h_{\mathbf{k},\lambda}^{(2)}h_{\mathbf{k \conf} ,\lambda \conf}^{(2)} \rangle +  \frac{1}{3}\langle h_{\mathbf{k},\lambda}^{(3)}h_{\mathbf{k \conf} ,\lambda \conf}^{(1)} \rangle \, .
    \end{split}
\end{align}
In the first line, we have assumed that tensor modes are drawn from some Gaussian distribution, so any odd-point primordial correlator vanishes. The second equality holds due to the fact that the first-order equation of motion are even under $\mathbf{k}\rightarrow -\mathbf{k}$ (see Eq.~\eqref{hevolutioneq}). Contributing to the iGW background up to fourth order in the correlator is therefore:
\begin{equation}\label{spectraldensitydef}
    \Omega (\eta , k) =\frac{k^2}{\mathcal{H}^2(\eta)}\sum _{\lambda} \left ( \frac{1}{12}\overline{\mathcal{P}^{(22)}_{\lambda}(\eta ,k )} + 2\times \frac{1}{18}\overline{\mathcal{P}^{(13)}_{\lambda}(\eta ,k )}\right ) \, .
\end{equation}
The second term in this equation is the focus of our work and was discussed in Ref.~\cite{Chen:2022dah}.

The content of our universe after inflation is modelled as a single, perfect (barotropic) fluid with adiabatic initial conditions and a constant equation of state. The stress-energy tensor, $T_{\mu \nu}$, then reads
\begin{equation}
\label{EMtensor}
    T_{\mu \nu} = (\rho + P ) u_{\mu} u_{\nu} + P g_{\mu \nu}\, ,
\end{equation}
where $\rho$ is the density, $P$ is the pressure, and $u^{\mu}$ is a comoving $4$-velocity. Furthermore, for simplicity, we will ignore extrinsic anisotropic stress at all orders. This means that, at first order, we have $\Psi ^{(1)} = \Phi ^{(1)}$. From this point onwards,  we take the convention of not showing the order of perturbation for first-order variables to reduce clutter. 

Having established the cosmological set-up and defined various quantities of interest, we now turn our attention to the terms contributing to the induced second-order GW background. 
\subsection{Thinking ahead: what source terms do we need to consider?}
at third order in perturbation theory, there are various products of first and second-order terms which source third-order GWs. To see this, we extract the TT part\footnote{This is done via the TT operator, $\Lambda ^{ij}_{ab}$, defined in Eq.~\eqref{tensorextractor}.} of the third-order spatial part of Einstein equation
\begin{equation}\label{thirdorderGWEQ}
    h_{ab}^{\prime \prime (3)} + 2\mathcal{H}h_{ab}^{\prime (3)} - \nabla ^2h_{ab}^{(3)}= 6\Lambda ^{ij}_{ab}S_{ij}^{(3)}\, ,
\end{equation}
and refer the reader to App.~\ref{appthirdorderEinstein} for the relevant part of the Einstein equations. $S_{ij}^{(3)}$ is the third-order source term, however, as we will show, not every term needs to be considered. 

For instance, the product of three first-order scalar perturbations,
\begin{equation}
      -8\Psi \partial _i \Psi \partial _j \Psi \subset S_{ij}^{(3)}  \, ,
\end{equation}
does source GWs (i.e. it has a non-vanishing TT contribution). Nonetheless, when we consider $\langle h_{\mathbf{k},\lambda}^{(3)}h_{\mathbf{k \conf} ,\lambda \conf}^{(1)} \rangle$ this term will not contribute since we are assuming first-order scalars and tensors do not correlate ($\langle h\psi \rangle=0$). Extending this logic to other cubic first-order terms, we see that we need to only consider terms that contain two scalars coupled to one tensor or cubic first-order tensor terms\footnote{From a perturbative point of view $hhh$ will contribute as $\mathcal{P}_h\mathcal{P}_h$, whilst $h\Psi \Psi$ as $\mathcal{P}_h\mathcal{P}_{\Psi}$. Since $\mathcal{P}_{\Psi} >\mathcal{P}_h$, the $hhh$ contributions will be subdominant. This was shown in Ref.~\cite{Picard:2023sbz} for induced second-order GWs for peaked input power spectra. We therefore ignore these contributions for the rest of our work.}. This line of thought extends to second-order perturbations. Schematically, we will therefore only consider terms that source $h^{(3)}$ as:
\begin{equation}
S^{(3)} \sim h  \Psi  \Psi  \text{ or } h \underbrace{X^{(2)}} _{\Psi   \Psi  } \text{ or } \Psi \underbrace{X^{(2)}}_{ h \Psi } \, ,
\end{equation}
where $X^{(2)}$ can be a second-order scalar, vector, or tensor. In the second to last term, we see that $X^{(2)}$ can be sourced by two first-order scalars, while in the last term it is sourced by one scalar and one tensor. To proceed, we require second-order equations of motion for second-order scalars, vectors, and tensors, sourced by first-order scalars and tensors. 

Using momentum conservation, a closer inspection of second-order perturbations sourced by two scalars coupled to a first-order tensor shows that the correlation $\langle h_{\mathbf{k}\conf , \lambda \conf}h^{(3)}_{\mathbf{k},\lambda}\rangle$ is proportional to
\begin{equation}
    \langle  h_{\mathbf{k}\conf , \lambda \conf} h_{\mathbf{k}-\mathbf{p}} \underbrace{X^{(2)}_{\mathbf{p}}} _{\Psi _{\mathbf{q}} \Psi _{\mathbf{p}-\mathbf{q}}} \rangle \propto \langle  h_{\mathbf{k}\conf , \lambda \conf} h_{\mathbf{k}-\mathbf{p}} \Psi _{\mathbf{q}} \Psi _{\mathbf{p}-\mathbf{q}} \rangle \propto \delta (\mathbf{k}\conf  + \mathbf{k}-\mathbf{p}) \delta (\mathbf{q}+\mathbf{p}-\mathbf{q}) \, ,
\end{equation}
which implies $\mathbf{p}=0$. Therefore, we see that we do not need to consider second-order scalars, vectors and tensors sourced by terms quadratic in first-order scalar fluctuations, when said second-order perturbations are sourcing third-order iGWs.

Bringing all of this together, we see that the source terms in Eq.~(\ref{thirdorderGWEQ}) we need to consider are
\begin{equation}
    S^{(3)}_{ij} =  S^{h\Psi \Psi}_{ij}  + S^{h^{(2)}\Psi}_{ij} + S^{B^{(2)}\Psi}_{ij} + S^{\Psi^{(2)}\Psi}_{ij} \, ,
\end{equation}
where for clarity: $S^{h\Psi \Psi}_{ij}$ is made of terms with one tensor perturbation and quadratic scalar perturbations, $S^{h^{(2)}\Psi}_{ij}$ a second-order tensor with a first-order scalar, $S^{B^{(2)}\Psi}_{ij}$ a second-order vector with a first-order scalar, and $S^{\Psi^{(2)}\Psi}_{ij}$ a second-order scalar with a first-order scalar. We now proceed to give the exact form of these source terms. 
\subsection{Relevant source terms}
In this subsection, we present in real space all source term that we need to consider throughout our work. These source terms have been simplified as far as possible and we have used $c_s^2=w$. For source terms which couple second-order perturbations to first-order ones, we also give equations of motion for induced second-order quantities. 
\subsubsection{Source term made of first-order fluctuations}
There is one source term made purely of first-order perturbations. The scalar-scalar-tensor source is
\begin{align}\label{sourcetermssh}
    \begin{split}
        S^{h\Psi \Psi}_{ij} &= -4h_{ij}\conf \Psi  \Psi \conf -\frac{1}{3(1+w)}h_{ij} \bigg ( 3(1+4w+3w^2)\Psi \conf \Psi \conf + 24(w^2-1)\Psi \nabla ^2 \Psi \\
        &+(-17-16w+9w^2)\partial _m \Psi \partial ^m \Psi \bigg ) - 16 h_j^m\Psi \partial _m \partial _i \Psi + 12 \Psi \partial _m h_{ij} \partial ^m \Psi \\
        &-8\Psi \partial ^m \Psi \partial _i h_{jm} - 16 h_{jm}\partial ^m \Psi \partial _i \Psi - 8\hubble h_{ij} \Psi \Psi \conf + \frac{8}{3(1+w)\hubble} \bigg ((w-1)h_{ij} \partial _m \Psi \partial ^m \Psi \conf \\
        &-h_{jm}\conf \partial ^m \Psi \partial _i \Psi +  h_{jm} \partial ^m \Psi \conf \partial _i \Psi   \bigg ) + \frac{4}{3(1+w)\hubble ^2} \bigg ((w-1)h_{ij}\partial _m \Psi \conf \partial ^m \Psi \conf \\
        &-2h_{jm}  \conf \partial ^m \Psi \partial _ i \Psi \conf + 2h_{jm} \partial ^m \Psi \conf \partial _i \Psi \conf  \bigg ) \, . 
    \end{split}
\end{align}
Ref.~\cite{Chen:2022dah} has considered the $S^{\Psi \Psi h}_{ij}$ source term; however, the authors use a different metric expansion than we do. 
\subsubsection{Source terms consisting of second-order scalar perturbations coupled to first-order perturbations}\label{inducedscalar}
The source term which couples second-order scalar perturbations to first-order scalars reads
\begin{align}
    \begin{split}\label{realspacesourcepsi2psi}
        S^{\Psi^{(2)}\Psi}_{ij} &= \frac{4}{3(1+w)\mathcal{H}^2} \partial _i \Psi ^{\prime} \partial _j \Psi ^{(2)\prime} + \frac{4}{3(1+w)\mathcal{H}} \left ( \partial _i \Psi ^{\prime} \partial _j \Phi ^{(2)} + \partial _i \Psi \partial _j \Psi ^{(2)\prime} \right )  \\
        &+\frac{2}{3}\frac{(5+3w)}{(1+w)}\partial _i \Psi \partial _j \Phi ^{(2)} \, .
    \end{split}
\end{align}
This term was considered in Ref.~\cite{Zhou:2021vcw} and our expression above is identical. To proceed, we need equations of motion for $\Psi ^{(2)}$ and $\Phi ^{(2)}$, which can be obtained by extracting two different parts of the spatial Einstein equations (see App.~\ref{appsecondorder}). The trace of the spatial part of the Einstein equations gives us an evolution equation for $\Psi^{(2)}$ in terms of $\Phi^{(2)}$ and first-order fluctuations
\begin{align}\label{eqtracerealspace}
    \begin{split}
        &3\Psi ^{\prime \prime (2)} + 3(2+3w)\hubble \Psi ^{\prime (2)} - (3w+1)\nabla ^2 \Psi ^{(2)} + 3\hubble \Phi ^{\prime (2)} + \nabla ^2 \Phi ^{(2)} = S^{\Psi ^{(2)}_{st}}  \, ,
    \end{split}
\end{align}
where 
\begin{align}
    \begin{split}
        S^{\Psi^{(2)}_{st}} & =2(1-3w)h^{ij}\partial _i \partial _j \Psi \text{ .}
    \end{split}
\end{align}
The other equation is obtained by extracting the symmetric trace-free part, obtained via the operator $D_{ij}$ defined in Eq.~\eqref{scalarextractor}, and gives us a relation between the two second-order scalar variables
\begin{equation}\label{eqanisrealspace}
    \nabla ^2 \nabla ^2 \Phi ^{(2)} - \nabla ^2 \nabla ^2 \Psi ^{(2)} = D^{ij}  \Pi _{ij}^{(2),st} \, ,
\end{equation}
with 
\begin{align}
    \begin{split}
        \Pi^{(2),st}_{ij} & = 12h_i^m \partial _j \partial _m \Psi -6(1+3w)\hubble h_{ij}\Psi \conf +6(w-1)h_{ij} \nabla ^2 \Psi -12\partial ^m \Psi \partial _m h_{ij} \\
    &+12 \partial ^m \Psi \partial _j h_{im}- 12 \Psi \nabla ^2 h_{ij} \, .
    \end{split}
\end{align}
As previously stated, in the absence of anisotropic stress the two scalar degrees of freedom in the metric are equal at first order. However, this is no longer the case at higher orders due to mode mixing. We point out that in Ref.~\cite{Chen:2022dah}, the authors set $\Phi ^{(2)}$ and $\Psi ^{(2)}$ to be the same, since they focus on the IR limit of the iGW background. 
\subsubsection{Source terms consisting of second-order vector perturbations coupled to first-order perturbations}\label{inducedvector}
Another source term which we need to consider contains second-order vector perturbations, $B^{(2)}_{i}$, coupled to first-order scalars  
\begin{align}\label{realspacesourcevectorpsi}
    \begin{split}
        S^{B^{(2)} \Psi}_{ij}\ &= -\Psi ^{\prime} \partial _i B_j^{(2)} - \Psi \partial _i B^{(2)\prime}_j + B^{(2)\prime}_j\partial _i \Psi - 2\mathcal{H}\Psi \partial _i B_j^{(2)} + 2\mathcal{H} B_j^{(2)} \partial _i \Psi + B_j^{(2)} \partial _i \Psi ^{\prime} \\
        &-\frac{1}{3(1+w)\mathcal{H}^2}\nabla ^2 B^{(2)}_i \left (\mathcal{H} \partial _j \Psi + \partial _j \Psi ^{\prime} \right ) \, ,
    \end{split}
\end{align}
also considered in Ref.~\cite{Zhou:2021vcw} and identical to ours. The second-order equation of motion for induced vectors is extracted by applying a transverse operator to the spatial part of the second-order Einstein equations. The transverse operator, $V ^{ij}_{a}$, is defined in Eq.~\eqref{vectorextractor}. The evolution equation reads
\begin{equation}\label{eqinducedvectorsrealspace}
    B^{(2)\prime}_a + 2\mathcal{H}B^{(2)}_a= -4 V ^{ij}_{a} S_{ij}^{B^{(2)}_{st}} \, ,
\end{equation}
and its second-order source term is given by
\begin{align}\label{sourcetermsV}
    \begin{split}
        S_{ij}^{B^{(2)}_{st}} & = -4h_j^m\partial _m \partial _i \Psi + 4 \partial _m h_{ij} \partial ^m \Psi + 4 \Psi \nabla ^2 h_{ij} -4\partial ^m \Psi \partial _j h_{im} \\
    &+2h_{ij} \left ((1+3w)\mathcal{H}\Psi ^{\prime} + (1-w) \nabla ^2 \Psi  \right ) \, .
    \end{split}
\end{align}
\subsubsection{Source terms consisting of second-order tensor perturbations coupled to first-order perturbations}\label{inducedtensor}
Finally, the last source term couples second-order tensor perturbations, $h^{(2)}_{ij}$, to first-order scalars
\begin{align}
    \begin{split}\label{realspaceh2psi}
        S^{h^{(2)} \Psi}_{ij} &= \left ( h_{ij}^{(2)\prime \prime} + 2\mathcal{H}h_{ij}^{(2)\prime} - \nabla ^2 h_{ij}^{(2)} \right )\Psi + 2\Psi \nabla ^2 h^{(2)}_{ij}+ (3w+1)\mathcal{H}h^{(2)}_{ij}\Psi ^{\prime} \\
        & - (w-1)h^{(2)}_{ij}\nabla ^2 \Psi  + 2 \partial ^m \Psi \partial _m h^{(2)}_{ij}  \, ,
    \end{split}
\end{align}
In a RD universe, our source term agrees with Ref.~\cite{Zhou:2021vcw}. Second-order tensors are sourced by first-order scalars and tensors, via
\begin{equation}
\label{GW2eq}
    h_{ab}^{\prime \prime (2)} + 2\mathcal{H}h_{ab}^{\prime (2)} - \nabla ^2h_{ab}^{(2)}=4 \Lambda ^{ij}_{ab} S_{ij}^{h^{(2)}_{st}}\, ,
\end{equation}
where we defined
\begin{equation}\label{sourcetermsGW}
    S_{ij}^{h^{(2)}_{st}}   = 2\Psi\nabla^2h_{ij} + 2\partial _mh_{ij}\partial ^m \Psi + h_{ij} (   \mathcal{H}(1+w) \Psi^{\prime} +(1-w)\nabla^2\Psi ) \, .
\end{equation}

Having listed all the contributions at third order we have to consider and also showed how the second-order variables are sourced from first-order perturbations, we now move on to solving the third-order GW equation for each source. 
\section{Solution to the third-order induced gravitational wave equation}\label{sec3}
We now proceed to solving the third-order iGW equation, Eq.~\eqref{thirdorderGWEQ}, in Fourier space  
\begin{equation}\label{GW3eqfourier}
     h_{\mathbf{k}, \lambda}^{\prime \prime (3)} + 2\mathcal{H}h_{\mathbf{k}, \lambda}^{\prime (3)} + k ^2h_{\mathbf{k}, \lambda}^{(3)}=6 \epsilon ^{ij}_{\lambda}(\mathbf{k})^{*}\mathcal{S}_{ij}^{(3)}(\eta , \mathbf{k})\equiv 6\mathcal{S}_{\mathbf{k},\lambda}^{(3)} (\eta)\, .
\end{equation}
We can obtain the solution for the third-order tensor mode using Green's method:
\begin{equation}\label{GWsol3rdorder}
    h^{(3)}_{\mathbf{k}, \lambda} (\eta ) = 6 \int _{\eta _i}^{\infty} {\rm d}\overline{\eta} \, \frac{a(\overline{\eta})}{a(\eta)}  G^h_k (\eta , \overline{\eta})\mathcal{S}_{\mathbf{k},\lambda}^{(3)} (\overline{\eta}) \, .
\end{equation}
Here $\eta _i$ corresponds to some initial conformal time. The solution is constructed in such a way that $h^{(3)}_{\mathbf{k}, \lambda} (\eta _i )=0$ and $\partial _{\eta}h^{(3)}_{\mathbf{k}, \lambda} (\eta _i )=0$. The Green’s function $G^h_k (\eta , \overline{\eta})$ is defined in Eq.\eqref{Greenstensor}; its explicit form depends on the background equation of state, which we leave unspecified for now.

In the following subsections, we present the solutions for each contribution in Fourier space. Since the solution to Eq.~\eqref{GW3eqfourier} depends on the equations of motion and solutions of both first- and second-order perturbations, we first provide a brief overview of how the corresponding third-order solutions are constructed.

For adiabatic initial conditions, first-order perturbations can be split up as:
\begin{equation}\label{firstordersplit}
    \Psi (\eta , \mathbf{k}) = \left ( \frac{3+3w}{5+3w} \right )T_{\Psi}(\sqrt{w}x) \mathcal{R} _{\mathbf{k}} \quad \text{and} \quad  h^{\lambda}(\eta , \mathbf{k}) = T_h(x)h_{\mathbf{k}}^{\lambda} \, ,
\end{equation}
where $x=k\eta$. $T_{\Psi / h}$ denote the transfer functions for scalar and tensor modes, which encode the linear evolution of the perturbations at horizon re-entry\footnote{For their explicit expressions, we refer the reader to Eq.~\eqref{psievolutioneq} and Eq.~\eqref{hevolutioneq}.}. $\mathcal{R} _{\mathbf{k}}$, the comoving curvature perturbation, and $h_{\mathbf{k}}^{\lambda}$ are the primordial (constant) values of scalar and tensor perturbations, respectively, set during inflation. The scalar transfer function depends on the equation of state of the universe; however, from here-on we drop the $w$ dependence in the transfer functions to reduce clutter.

A more complicated decomposition applies for second-order perturbations. While, it is still possible to isolate the primordial value of the perturbation, one must also account for the non-linear evolution of the second-order variables. The time evolution of second-order perturbations is encoded in a kernel, as opposed to a transfer function at linear order, and so this will give rise to nested kernels. For this reason, in what follows, we distinguish between source terms made of cubic first-order fluctuations and ones containing second-order perturbations. 

\subsection{Solution to the gravitational-wave equation for the pure first-order contribution}
There is one source term composed exclusively of first-order perturbations, corresponding to the coupling of quadratic scalar fluctuations to tensor modes, given in Eq.~\eqref{sourcetermssh}. This leads to the following solution in Fourier space:
\begin{align}\label{solhpsipsi}
    \begin{split}
        h^{\lambda \, (3)}_{\Psi \Psi h} (\eta , \mathbf{k}) &= \frac{6}{(2\pi)^3} \int {\rm d} ^3 \mathbf{q} \int {\rm d} ^3 \mathbf{p} \, \, \epsilon^{ij}_{\lambda}(\mathbf{k}) ^* \sum _{\lambda _1} \bigg ( \epsilon_{ij}^{\lambda _1}(\mathbf{k}-\mathbf{p}-\mathbf{q}) I_1^{ssh}(\eta , k,p,q) \\
        &+ 16 \epsilon_{jm}^{\lambda _1}(\mathbf{k}-\mathbf{p}-\mathbf{q})p^mp_i I_2^{ssh}(\eta , k,p,q) - 8\epsilon_{jm}^{\lambda _1}(\mathbf{k}-\mathbf{p}-\mathbf{q})p^m(q_i+p_i) I_3^{ssh}(\eta , k,p,q) \\
        &+8\epsilon_{jm}^{\lambda _1}(\mathbf{k}-\mathbf{p}-\mathbf{q})q^mp_i I_4^{ssh}(\eta , k,p,q)\bigg ) h^{\lambda _1}_{\mathbf{k}-\mathbf{p}-\mathbf{q}}\mathcal{R} _{\mathbf{q}} \mathcal{R} _{\mathbf{p}} \, .
    \end{split}
\end{align}
Above, we have decomposed the first-order perturbations into their primordial values and transfer functions (and polarisation functions for tensors) according to Eq.~\eqref{firstordersplit}. The factor of $\epsilon^{ij}_{\lambda}(\mathbf{k}) ^*$ comes from the TT projection operator, and we note that there is a sum over the two polarisation states of the first-order GW, denoted $\lambda _1$. Furthermore, we have collected all the time dependence into the kernels $I_{\textit{i}}^{ssh}(\eta , k,p,q)$ ($\textit{i}=1,2,3,4$), defined as
\begin{equation}
    I_{\textit{i}}^{\Psi \Psi h}(\eta , k,p,q) =  \int _{\eta _i}^{\infty} {\rm d} \overline{\eta} \, \frac{a(\overline{\eta})}{a(\eta)} G^h_{k}(\eta , \overline{\eta}) f^{\Psi \Psi h}_{\textit{i}}(\overline{\eta},k,p,q) \, ,
\end{equation}
with
\begin{subequations} \label{sshfunctions}
    \begin{align}
        \begin{split}
            f_1^{\Psi \Psi h}(\eta ,k,p,q) &= \left ( \frac{3+3w}{5+3w} \right )^2 \bigg ( -4T\conf _h(\eta|\mathbf{k}-\mathbf{p}-\mathbf{q}|)T_{\Psi}(\eta q) T\conf _{\Psi}(\eta p) - \frac{1}{3(1+w)}\\
            &\times T_h(\eta|\mathbf{k}-\mathbf{p}-\mathbf{q}|)  \bigg ( 3(1+4w+3w^2) T\conf _{\Psi}(\eta q) T \conf _{\Psi}(\eta p) -24(w^2-1)T_{\Psi}(\eta q)T_{\Psi}(\eta p) \\
            &- (-17-16w+9w^2)\, \mathbf{p}\cdot \mathbf{q}\, T_{\Psi}(\eta q)T_{\Psi}(\eta p) \bigg ) - 12( \mathbf{k}\cdot \mathbf{p} -  \mathbf{q}\cdot \mathbf{p} -p^2) \\
            &\times T_h(\eta|\mathbf{k}-\mathbf{p}-\mathbf{q}|)T_{\Psi}(\eta q) T_{\Psi}(\eta p) - 8 \hubble T_h(\eta|\mathbf{k}-\mathbf{p}-\mathbf{q}|)T_{\Psi}(\eta q) T\conf _{\Psi}(\eta p) \\
            &-\frac{8(w-1)}{3(1+w)\hubble} \, \mathbf{q} \cdot \mathbf{p} \, T_h(\eta|\mathbf{k}-\mathbf{p}-\mathbf{q}|)T_{\Psi}(\eta q) T\conf _{\Psi}(\eta p) \\
            &-\frac{4(w-1)}{3(1+w)\hubble ^2} \, \mathbf{q}\cdot \mathbf{p} \, T_h(\eta|\mathbf{k}-\mathbf{p}-\mathbf{q}|)T \conf _{\Psi}(\eta q) T\conf _{\Psi}(\eta p) \bigg ) \, , 
        \end{split}
        \\
        \begin{split}
             f_2^{\Psi \Psi h}(\eta ,k,p,q) &=  f_3^{\Psi \Psi h}(\eta ,k,p,q)  = \left ( \frac{3+3w}{5+3w} \right )^2 \bigg ( T_h(\eta|\mathbf{k}-\mathbf{p}-\mathbf{q}|)T_{\Psi}(\eta q) T_{\Psi}(\eta p) \bigg ) \, ,
        \end{split}
        \\
        \begin{split}
             f_4^{\Psi \Psi h}(\eta ,k,p,q) &=\left ( \frac{3+3w}{5+3w} \right )^2 \bigg ( 2 T_h(\eta|\mathbf{k}-\mathbf{p}-\mathbf{q}|)T_{\Psi}(\eta q) T_{\Psi}(\eta p) + \frac{1}{3(1+w)\hubble }T \conf_h(\eta|\mathbf{k}-\mathbf{p}-\mathbf{q}|) \\
             &\times T_{\Psi}(\eta q) T_{\Psi}(\eta p) -  \frac{1}{3(1+w)\hubble }T_h(\eta|\mathbf{k}-\mathbf{p}-\mathbf{q}|)T \conf _{\Psi}(\eta q) T_{\Psi}(\eta p) +  \frac{1}{3(1+w)\hubble ^2} \\
             &\times T \conf_h(\eta|\mathbf{k}-\mathbf{p}-\mathbf{q}|)T_{\Psi}(\eta q)T \conf _{\Psi}(\eta q) - \frac{1}{3(1+w)\hubble ^2} T_h(\eta|\mathbf{k}-\mathbf{p}-\mathbf{q}|)T \conf _{\Psi}(\eta q) T\conf _{\Psi}(\eta p) \bigg ) \, .
        \end{split}
    \end{align}
\end{subequations}
We observe that the structure of these solutions remains largely unchanged compared to the case of second-order iGWs, except that there are now two unconstrained momenta, $\mathbf{p}$ and $\mathbf{q}$. 
\subsection{Solution to the gravitational-wave equation for source terms with second-order contributions}
We now present the solutions for the source terms that involve second-order perturbations coupled to first-order ones. We begin with sources containing second-order scalars, then consider those coupling second-order vectors to first-order perturbations, and finally address the source terms involving second-order tensors.
\subsubsection{Source term containing second-order scalars}
The source term that couples second-order scalars to first-order scalars, Eq.~\eqref{realspacesourcepsi2psi}, in Fourier space reads
\begin{align}
    \begin{split}
        \mathcal{S}_{\lambda }^{\Psi^{(2)}\Psi}(\eta , \mathbf{k}) &= \frac{1}{(2\pi)^{\frac{3}{2}}} \int {\rm d} ^3 \mathbf{p}  \, \eps ^{ij}_{\lambda}(\mathbf{k})^* p_ip_j \bigg ( \frac{4}{3(1+w)\hubble ^2} \timeder \Psi (\eta,\mathbf{k}-\mathbf{p}) \timeder \Psi ^{(2)}_{st}(\eta,\mathbf{p}) \\
        &+ \frac{4}{3(1+w)\hubble}\left (\timeder \Psi (\eta,\mathbf{k}-\mathbf{p}) \Phi ^{(2)}_{st}(\eta,\mathbf{p}) +  \Psi (\eta,\mathbf{k}-\mathbf{p}) \timeder \Psi ^{(2)}_{st}(\eta,\mathbf{p})  \right ) \\
        &+ \frac{2(5+3w)}{3(1+w)}  \Psi (\eta,\mathbf{k}-\mathbf{p})  \Phi ^{(2)}_{st}(\eta,\mathbf{p})  \bigg ) \, .
    \end{split}
\end{align}
We refer the reader to Appendix~\ref{AppSOscalars} where we have derived all the necessary equations needed for our computation, i.e. induced second-order scalars equations of motion in Fourier space. We recall that second order perturbations coupled to first-order scalar fluctuations will only contribute to the GW background if these second-order scalar fluctuations are induced by first-order tensors coupled to first-order scalars. We can write the above source term exclusively in terms of $\Psi^{(2)}$, via Eq.~(\ref{anis}), which leads to 
\begin{align}
    \begin{split}
        \mathcal{S}_{\lambda }^{\Psi^{(2)}\Psi}(\eta , \mathbf{k}) &= \frac{1}{(2\pi)^{\frac{3}{2}}}  \int {\rm d} ^3 \mathbf{p}  \, \eps ^{ij}_{\lambda}(\mathbf{k})^* p_ip_j \bigg ( \frac{4}{3(1+w)\hubble ^2} \timeder \Psi (\eta,\mathbf{k}-\mathbf{p}) \timeder \Psi ^{(2)}_{st}(\eta,\mathbf{p}) + \frac{4}{3(1+w)\hubble }\\
        &\times \left (\timeder \Psi (\eta,\mathbf{k}-\mathbf{p}) \Psi ^{(2)}_{st}(\eta,\mathbf{p}) + \frac{1}{p^4}\timeder \Psi (\eta,\mathbf{k}-\mathbf{p}) \Pi ^{(2)}_{st}(\eta , \mathbf{p}) +  \Psi (\eta,\mathbf{k}-\mathbf{p}) \timeder \Psi ^{(2)}_{st}(\eta,\mathbf{p})  \right ) \\
        &+ \frac{2(5+3w)}{3(1+w)} \bigg (  \Psi (\eta,\mathbf{k}-\mathbf{p})  \Psi ^{(2)}_{st}(\eta,\mathbf{p})+\frac{1}{p^4}\Psi (\eta,\mathbf{k}-\mathbf{p})  \Pi ^{(2)}_{st}(\eta , \mathbf{p})\bigg)  \bigg ) \, .
    \end{split}
\end{align}
The above source term is then substituted into the solution of the third-order iGW equation, Eq.~\eqref{GWsol3rdorder}, and we arrive at 
\begin{align}
    \begin{split} \label{GWeqsolpsi2psi1}
        h_{\lambda }^{\Psi^{(2)}\Psi}(\eta , \mathbf{k}) &= \frac{6}{(2\pi)^3} \int {\rm d} ^3 \mathbf{q} \int {\rm d} ^3 \mathbf{p} \sum _{ \lambda _1} \epsilon^{ij}_{\lambda}(\mathbf{k})^* p_ip_jQ_{\lambda _1}^{\Psi ^{(2)} _{st}}(\mathbf{p},\mathbf{q}) I^{\Psi ^{(2)}\Psi} (\eta , k,p,q) \\
        &\times  \mathcal{R} _{\mathbf{k}-\mathbf{p}}h_{\mathbf{q}}^{\lambda _1}  \mathcal{R} _{\mathbf{p}-\mathbf{q}} \, .
    \end{split}
\end{align}
To arrive at the expression above, we have substituted in the equation of motion for $ \Psi ^{(2)}_{st}(\eta,\mathbf{p})$, given in Eq.~\eqref{psiinducedst}, the expression for $\Pi ^{(2)}_{st}(\eta , \mathbf{p})$ in Eq.~\eqref{anisinducedst} and split up the first-order perturbations according to Eq.~\eqref{firstordersplit}. $ Q_{\lambda _1}^{\Psi ^{(2)} _{st}}(\mathbf{p},\mathbf{q})$ is defined in Eq.~\eqref{polinducedscalars} and there is a sum over the polarisation of the first-order GW since they are inducing the second-order scalars. The time dependence of the iGW is in the kernel, $I^{\Psi ^{(2)}\Psi} (\eta , k,p,q)$, defined as
\begin{equation} \label{kernelpsi2psi1}
    I^{\Psi ^{(2)}\Psi}(\eta , k,p,q) =  \int _{\eta _i}^{\infty} {\rm d} \overline{\eta} \, \frac{a(\overline{\eta})}{a(\eta)} G^h_k(\eta , \overline{\eta}) f^{\Psi^{(2)}\Psi}(\overline{\eta},k,p,q) \, ,
\end{equation}
where
\begin{align}\label{psi2psifunction}
    \begin{split}
        &f^{\Psi^{(2)}\Psi}(\bareta ,k,p,q) = \avant ^2 \bigg ( \frac{4}{3(1+w)\hubble ^2} T_{\Psi}\conf (\bareta |\conv | ) \timederbar I^{\Psi ^{(2)}}_{st}(\bareta , p ,q ) +  \frac{4}{3(1+w)\hubble } \\
        &\times \left ( T_{\Psi}\conf (\bareta |\conv | )  I^{\Psi ^{(2)}}_{st}(\bareta , p ,q ) + \frac{1}{p^4} T_{\Psi}\conf (\bareta |\conv | ) f^{\Pi ^{(2)}}_{st}(\bareta , p , q) + T_{\Psi}(\bareta | \conv |) \timederbar I^{\Psi ^{(2)}}_{st}(\bareta , p ,q )\right  )\\
        &  + \frac{2(5+3w)}{3(1+w)} \left (T_{\Psi}(\bareta | \conv |) I^{\Psi ^{(2)}}_{st}(\bareta , p ,q ) + \frac{1}{p^4}T_{\Psi}(\bareta | \conv |)f^{\Pi ^{(2)}}_{st}(\bareta , p , q)   \right ) \bigg ) \, .
    \end{split}
\end{align}
We now see explicitly that the third-order kernel, $I^{\Psi ^{(2)}\Psi} (\eta , k,p,q)$ depends on the second-order scalar kernel $I^{\Psi ^{(2)}}_{st}(\bareta , p ,q )$, defined in Eq.~\eqref{kernelinducedstscalars}, giving rise to nested kernels. The function $f^{\Pi ^{(2)}}_{st}(\bareta , p , q)$ is defined in Eq.~\eqref{fpistscalar}.

\subsubsection{Source term containing second-order vectors}
In this subsection, we present the solution to the third-order gravitational wave equations, where the source consists of second-order vectors coupled to first-order perturbations. Equations of motion and solutions for induced second-order vectors can be found in Appendix~\ref{AppSOvectors}.

The source term, given in real space by Eq.~\eqref{realspacesourcevectorpsi}, in Fourier space reads
\begin{align}
    \begin{split}
        \mathcal{S}_{\lambda }^{B^{(2)}\Psi}(\eta , \mathbf{k}) &= -i\sum _ r\frac{1}{(2\pi)^{\frac{3}{2}}} \int {\rm d} ^3 \mathbf{p}  \, \eps ^{ij}_{\lambda}(\mathbf{k})^* p_ie^r_j(\mathbf{p}) \bigg ( 2B^{(2)st}_r(\eta, \mathbf{p}) \partial _{\eta}\Psi (\eta,\mathbf{k}-\mathbf{p}) 
        \\
        &+ 2\partial _{\eta}B^{(2)st}_r(\eta, \mathbf{p})  \Psi (\eta,\mathbf{k}-\mathbf{p}) +4\hubble B^{(2)st}_r(\eta, \mathbf{p})\Psi (\eta,\mathbf{k}-\mathbf{p}) + \frac{1}{3(1+w)\hubble ^2}p^2 
        \\
        &\times B^{(2)st}_r(\eta, \mathbf{p}) \left ( \hubble \Psi (\eta,\mathbf{k}-\mathbf{p}) + \partial _{\eta} \Psi (\eta,\mathbf{k}-\mathbf{p})  \right )  \bigg ) \, ,
    \end{split}
\end{align}
where we sum over $r$, the parity of the second-order vectors. After substituting the equation of motion for second-order vectors sourced by first-order scalars coupled to tensors, i.e. Eq.~\eqref{2ndordervecst}, and splitting up first-order perturbations into transfer functions and their corresponding primordial values, we can write down the third-order solution as
\begin{align}\label{GWeqsolb2psi1}
    \begin{split} 
        h_{\lambda }^{B^{(2)}\Psi}(\eta , \mathbf{k}) &= \frac{6}{(2\pi)^3} \sum _{\lambda _1} \sum _r  \int {\rm d} ^3 \mathbf{p}  \int {\rm d} ^3 \mathbf{q}  \, \bigg ( Q^{B^{(2)}\Psi}_{\lambda ,r}(\mathbf{k},\mathbf{p})Q^{B^{(2)}_{st1}}_{r,\lambda _1}(\mathbf{p},\mathbf{q}) I^{B^{(2)}\Psi}_1(\eta , k,p,q) \\
        &+Q^{B^{(2)}\Psi}_{\lambda ,r}(\mathbf{k},\mathbf{p})Q^{B^{(2)}_{st2}}_{r,\lambda _1}(\mathbf{p},\mathbf{q}) I^{B^{(2)}\Psi}_2(\eta , k,p,q) \bigg ) \mathcal{R} _{\mathbf{k}-\mathbf{p}}h^{\lambda _1}_{\mathbf{q}}\mathcal{R} _{\mathbf{p}-\mathbf{q}} \, .
    \end{split}
\end{align}
We defined a polarisation function 
\begin{equation}
    Q^{B^{(2)}\Psi}_{\lambda ,r}(\mathbf{k},\mathbf{p}) =  \eps ^{ij}_{\lambda}(\mathbf{k})^* p_ie^r_j(\mathbf{p}) \, ,
\end{equation}
whilst $Q^{B^{(2)}_{st1}}_{r,\lambda _1}(\mathbf{p},\mathbf{q})$ and  $Q^{B^{(2)}_{st2}}_{r,\lambda _1}(\mathbf{p},\mathbf{q})$ are defined respectively in Eq.~\eqref{pol2ndvecst1} and Eq.~\eqref{pol2ndvecst2}. The kernels are defined as follows, ($i=1,2$),
\begin{equation}\label{kernelb2psi1}
    I^{B^{(2)}\Psi}_{\textit{i}}(\eta , k,p,q) =  \int _{\eta _i}^{\infty} {\rm d} \overline{\eta} \, \frac{a(\overline{\eta})}{a(\eta)} G^h_{k}(\eta , \overline{\eta}) f^{B^{(2)\Psi}}_{\textit{i}}(\overline{\eta},k,p,q) \, ,
\end{equation}
where
\begin{subequations}
    \begin{align}
        \begin{split}
            &f^{B^{(2)\Psi}}_1(\overline{\eta},k,p,q) = \avant ^2 \bigg ( \frac{2}{p^2} I^{B^{(2)}}_{st1}(\overline{\eta} ,k,p,q) T^{\prime}_{\Psi}(\overline{\eta} |\mathbf{k}-\mathbf{p}|) + \frac{2}{p^2}\frac{\partial}{\partial _{\overline{\eta}}} I^{B^{(2)}}_{st1}(\overline{\eta} ,k,p,q) \\
            &\times T_{\Psi}(\overline{\eta} |\mathbf{k}-\mathbf{p}|) + \frac{4\hubble}{p^2} I^{B^{(2)}}_{st1}(\overline{\eta} ,k,p,q) T_{\Psi}(\overline{\eta} |\mathbf{k}-\mathbf{p}|) + \frac{1}{3(1+w)\hubble} I^{B^{(2)}}_{st1}(\overline{\eta} ,k,p,q) T_{\Psi}(\overline{\eta} |\mathbf{k}-\mathbf{p}|) \\
            &+  \frac{1}{3(1+w)\hubble ^2} I^{B^{(2)}}_{st1}(\overline{\eta} ,k,p,q) T^{\prime}_{\Psi}(\overline{\eta} |\mathbf{k}-\mathbf{p}|)   \bigg ) \, ,
        \end{split}
        \\
        \begin{split}
            &f^{B^{(2)\Psi}}_2(\overline{\eta},k,p,q) =  \avant ^2 \bigg ( \frac{2}{p^2} I^{B^{(2)}}_{st2}(\overline{\eta} ,k,p,q) T^{\prime}_{\Psi}(\overline{\eta} |\mathbf{k}-\mathbf{p}|) + \frac{2}{p^2}\frac{\partial}{\partial _{\overline{\eta}}} I^{B^{(2)}}_{st2}(\overline{\eta} ,k,p,q) \\
            &\times T_{\Psi}(\overline{\eta} |\mathbf{k}-\mathbf{p}|) + \frac{4\hubble}{p^2} I^{B^{(2)}}_{st2}(\overline{\eta} ,k,p,q) T_{\Psi}(\overline{\eta} |\mathbf{k}-\mathbf{p}|) + \frac{1}{3(1+w)\hubble} I^{B^{(2)}}_{st2}(\overline{\eta} ,k,p,q) T_{\Psi}(\overline{\eta} |\mathbf{k}-\mathbf{p}|) \\
            &+  \frac{1}{3(1+w)\hubble ^2} I^{B^{(2)}}_{st2}(\overline{\eta} ,k,p,q) T^{\prime}_{\Psi}(\overline{\eta} |\mathbf{k}-\mathbf{p}|)   \bigg ) \, .
        \end{split}
    \end{align}
\end{subequations}
The nested kernels $ I^{B^{(2)}}_{st1}(\overline{\eta} ,k,p,q)$ and $ I^{B^{(2)}}_{st2}(\overline{\eta} ,p,q)$ are defined in Eqs.~\eqref{kernelvectorst1} and \eqref{kernelvectorst2}, respectively.
\subsubsection{Source term containing second-order tensors}
Finally, the last type of source terms we need to consider contains second-order tensor perturbations, whose equations of motion in Fourier space can be found in Appendix~\ref{AppSOtensors}.

The source term in Eq.~\eqref{realspaceh2psi} is expressed as a Fourier integral such that
\begin{align}
    \begin{split}
        \mathcal{S}_{\lambda }^{h^{(2)}\Psi}(\eta , \mathbf{k}) &= \frac{1}{(2\pi)^{\frac{3}{2}}}\sum _{\sigma }  \int {\rm d} ^3 \mathbf{p}  \, \eps ^{ij}_{\lambda}(\mathbf{k})^* \eps _{ij}^{\sigma}(\mathbf{p}) \bigg ( \big ( \timeder ^2 h^{(2)st}_{\sigma}(\eta , \mathbf{p}) + 2 \hubble \timeder h^{(2)st}_{\sigma}(\eta , \mathbf{p}) + p^2 h^{(2)st}_{\sigma}(\eta , \mathbf{p}) \big ) \\
        &\times \Psi (\eta,\mathbf{k}-\mathbf{p}) -2p^2h^{(2)st}_{\sigma}(\eta , \mathbf{p})\Psi (\eta,\mathbf{k}-\mathbf{p}) + (3w+1)\hubble h^{(2)st}_{\sigma}(\eta , \mathbf{p}) \timeder \Psi (\eta,\mathbf{k}-\mathbf{p})  \\
        & + (w-1)|\mathbf{k}-\mathbf{p}|^2 h^{(2)st}_{\sigma}(\eta , \mathbf{p})\Psi (\eta,\mathbf{k}-\mathbf{p})-2(k_m-p_m)p^m h^{(2)st}_{\sigma}(\eta , \mathbf{p})\Psi (\eta,\mathbf{k}-\mathbf{p})  \bigg )\, ,
    \end{split}
\end{align}
where we have summed over the polarisation states ($\sigma$) of the second-order tensor. In the first line above, we recognise the second-order equation of motion for tensors and substitute in the scalar-tensor source term (see Eq.~\eqref{GW2eqfourier}) and also substitute in the solution for scalar-tensor induced waves, Eq.~\eqref{solh2st}, in place of $h^{(2)st}_{\sigma}(\eta , \mathbf{p})$. We arrive at the third-order solution
\begin{align}\label{GWeqsolh2psi1}
    \begin{split}
        h_{\lambda }^{h^{(2)}\Psi}(\eta , \mathbf{k}) &= \frac{6}{(2\pi)^3} \int {\rm d} ^3 \mathbf{q} \int {\rm d} ^3 \mathbf{p} \sum _{\sigma , \lambda _1} \epsilon^{ij}_{\lambda}(\mathbf{k})^* \epsilon_{ij}^{\sigma}(\mathbf{p})\epsilon^{mn}_{\sigma}(\mathbf{p})^*\epsilon_{mn}^{\lambda _1}(\mathbf{q})I^{h^{(2)}\Psi}(\eta , k,p,q) \\
        &\times \mathcal{R} _{\mathbf{k}-\mathbf{p}} h_{\mathbf{q}}^{\lambda _1} \mathcal{R} _{\mathbf{p}-\mathbf{q}} \, ,
    \end{split}
\end{align}
where the kernels, $I^{h^{(2)}\Psi}(\eta  , k,p,q)$, are defined as  
\begin{equation} \label{kernelh2psi1}
    I^{h^{(2)}\Psi}(\eta  , k,p,q) =  \int _{\eta _i}^{\infty} {\rm d} \overline{\eta} \, \frac{a(\overline{\eta})}{a(\eta)} G^h_{k}(\eta , \overline{\eta}) f^{h^{(2)\Psi}}(\overline{\eta},k,p,q) \, ,
\end{equation}
and
\begin{align}\label{ffuntionh2psi1}
    \begin{split}
        &f^{h^{(2)\Psi}}(\overline{\eta},k,p,q) = \left ( \frac{3+3w}{5+3w} \right )^2 \bigg ( -8q^2 T_{\Psi} (\overline{\eta} |\mathbf{p}-\mathbf{q}|) T_h(\overline{\eta} q) T_{\Psi} (\overline{\eta} |\mathbf{k}-\mathbf{p}|)  - 8 (p_i - q_i)q^i \\
        &\times T_{\Psi} (\overline{\eta} |\mathbf{p}-\mathbf{q}|) T_h(\overline{\eta} q) T_{\Psi} (\overline{\eta} |\mathbf{k}-\mathbf{p}|) + 4\hubble (1+3w) \partial _{\bareta} T_{\Psi} (\overline{\eta} |\mathbf{p}-\mathbf{q}|) T_h(\overline{\eta} q) T_{\Psi} (\overline{\eta} |\mathbf{k}-\mathbf{p}|) \\
        &-4(1-w)|\mathbf{p}-\mathbf{q}|^2 T_{\Psi} (\overline{\eta} |\mathbf{p}-\mathbf{q}|) T_h(\overline{\eta} q) T_{\Psi} (\overline{\eta} |\mathbf{k}-\mathbf{p}|) -8p^2I^{h^{(2)}}_{st}(\overline{\eta} , p ,q)T_{\Psi} (\overline{\eta} |\mathbf{k}-\mathbf{p}|) \\
        &+4(3w+1)\hubble I^{h^{(2)}}_{st}(\overline{\eta} , p ,q)\partial _{\bareta} T_{\Psi} (\overline{\eta} |\mathbf{k}-\mathbf{p}|) + 4(w-1)|\mathbf{k}-\mathbf{p}|^2 I^{h^{(2)}}_{st}(\overline{\eta} , p ,q)T_{\Psi} (\overline{\eta} |\mathbf{k}-\mathbf{p}|) \\
        &-8(k_i-p_i)p^iI^{h^{(2)}}_{st}(\overline{\eta} , p ,q)T_{\Psi} (\overline{\eta} |\mathbf{k}-\mathbf{p}|)  \bigg ) \, .
    \end{split}
\end{align}
The nested kernel, $I^{h^{(2)}}_{st}(\overline{\eta} , p ,q)$, is defined in Eq.~\eqref{h2stkernel}. 

This concludes our discussion of all terms and their solutions in Fourier space; we now turn to their contributions to the power spectrum. 

\section{Power spectra of the different contributions}\label{sec4}
In this section, we compute the power spectrum for each contribution we have considered. In order to do this, we substitute in the third-order solution to Eq.~\eqref{generalcorrelationoftensors} and correlate it with a first-order GW generated in inflation. Explicitly, we are after $\mathcal{P}^{\text{$(13)$}} (\eta , k)$, which can be found by inserting our third-order iGW equation solutions into (omitting time-dependence, these are equal-time correlators)
\begin{equation}\label{correlation31}
     \langle h_{\lambda}(\mathbf{k})h_{\lambda \conf}^{(3)}(\mathbf{k \conf}) \rangle = \delta ^{\lambda \lambda ^{\prime}} \delta (\mathbf{k}+ \mathbf{k^{\prime}}) \frac{2\pi ^2}{k^3}\mathcal{P}_{\lambda}^{\text{$(13)$}} (\eta , k) \, .
\end{equation}
There are four different contributions, specifically:
\begin{align}
    \begin{split}
        \mathcal{P}^{(13)}(\eta,k) &=  \mathcal{P}_{\Psi \Psi h}^{(13)}(\eta,k) + \mathcal{P}_{\Psi ^{(2)}\Psi}^{(13)}(\eta,k) + \mathcal{P}_{B ^{(2)}\Psi}^{(13)}(\eta,k)  + \mathcal{P}_{h^{(2)}\Psi}^{(13)}(\eta,k)  \, .
    \end{split}
\end{align}
In what follows, we have dedicated a subsection to each of these terms. 

Beforehand, we define the power spectrum of the comoving curvature perturbation
\begin{equation}\label{scalarcorrelation}
    \langle \mathcal{R} _{\mathbf{k}} \mathcal{R} _{\mathbf{k \conf}} \rangle = \delta (\mathbf{k} + \mathbf{k \conf}) P_{\mathcal{R}} (\eta , k) \, .
\end{equation}
Additionally, since we are carrying-out our calculation in an flat FLRW universe, we parametrise vectors in spherical coordinates (see Appendix~\ref{secondorderApp}). Isotropy also implies we are free to align the wave-vector $\mathbf{k}$ (i.e. the wave-vector of the third-order iGW) with the $z$-axis, which means from now on we set $\theta _k$ and $\phi _k$ to zero in Eq.~\eqref{kwavevector}. Moreover, we will see that the correlation constrains the ingoing wave-vectors to be $\mathbf{p}$ and $\mathbf{k}-\mathbf{p}$ (through conservation of momentum), and so it is useful to introduce the scalar rescaled momentum coordinates
\begin{equation}\label{vandudef}
    v=\frac{p}{k}\quad \text{and} \quad  u=\frac{|\mathbf{k}-\mathbf{p}|}{k} \, .
\end{equation}
In the following, we use the following relations extensively
\begin{equation}
    \sin^2{\theta _p} = 1-\frac{(1+v^2-u^2)^2}{4v^2} = 1- \cos^2{\theta _p}\, ,
\end{equation}
and for the nested kernels it is useful to introduce $\tilde{y}=p\tilde{\eta}$. Also, the subsequent relations are helpful
\begin{equation}\label{yrelations}
    \quad p\overline{\eta}=v\overline{x}, \quad |\mathbf{k}-\mathbf{p}|\tilde{\eta}=\frac{\tilde{y}u}{v}\quad \text{and} \quad k\tilde{\eta} = \frac{\tilde{y}}{v}\, .
\end{equation}
Finally, a useful parameter is 
\begin{equation} \label{defb}
    b=\frac{1-3w}{1+3w} \, .
\end{equation}

We now proceed to compute each contribution. The results in this section hold for a general equation of state parameter $w \neq  -1/3 $. 

\subsection{The power spectrum of the scalar-scalar-tensor first-order contribution: 
\texorpdfstring{$\mathcal{P}^{(13)}_{\Psi \Psi h}$}{P(13)PsiPsiH}}
\begin{figure}[h!]
  \centering
  \begin{tikzpicture}

    \coordinate (A) at (0,0);
    \coordinate (B) at (1,2);
    \coordinate (C) at (3,2.5);
    \coordinate (D) at (4,0);


    \draw[thick, -{Stealth}, blue] (A) -- (D);
    \draw[ -{Stealth}, black] (C) -- (D);
    \draw[dashed, -{Stealth}, black] (A) -- (B);
    \draw[dashed, -{Stealth}, black] (B) -- (C);

    \node[above, text=blue] at ($(A)!0.5!(D)$) {$\mathbf{k}$};
    \node[left] at ($(A)!0.5!(B)$) {$\mathbf{p}$};
    \node[above] at ($(B)!0.5!(C)$) {$\mathbf{q}$};
    \node[right] at ($(C)!0.5!(D)$) {$\mathbf{k}-\mathbf{p}-\mathbf{q}$};

    \coordinate (E) at (6.5,0);
    \coordinate (F) at (6.55,0);
    \coordinate (G) at (6.5,2.5);
    \coordinate (H) at (6.55,2.5);
    \coordinate (I) at (10,0);
    \coordinate (J) at (6.5,-0.1);
    \coordinate (K) at (10,-0.1);
    \coordinate (L) at (6.5,-0.2);
    \coordinate (M) at (10,-0.2);

    \draw[dashed, -{Stealth}, black] (E) -- (G);
    \draw[dashed, -{Stealth}, black] (H) -- (F);
    \draw[ -{Stealth}, black] (E) -- (I);
    \draw[ -{Stealth}, blue] (J) -- (K);
    \draw[ -{Stealth}, blue] (M) -- (L);


    \node[text=blue] at (10.1,-0.05) {$\mathbf{k}$};
    \node[below, text=blue] at ($(M)!0.5!(L)$) {$\mathbf{k^{\prime}}$};
    \node[left] at ($(E)!0.5!(G)$) {$\mathbf{p}$};
    \node[right] at ($(F)!0.5!(H)$) {$\mathbf{q}$};
    \node[above] at ($(E)!0.5!(I)$) {$\mathbf{k}-\mathbf{p}-\mathbf{q}$};
  \end{tikzpicture}
  
  \caption{\footnotesize{\textit{Left}. Momentum configuration of two scalar modes (dashed black) and a tensor mode (solid black) inducing a third-order tensor mode (solid blue). \textit{Right}. Geometrical configuration of the scalar-scalar-tensor two-point function when contracted with a first-order tensor mode. The two scalar modes must have opposite momentum and the first-order tensor mode has equal momentum to the out-going (third-order) one.}}
  \label{fig:ssh2point}
\end{figure}
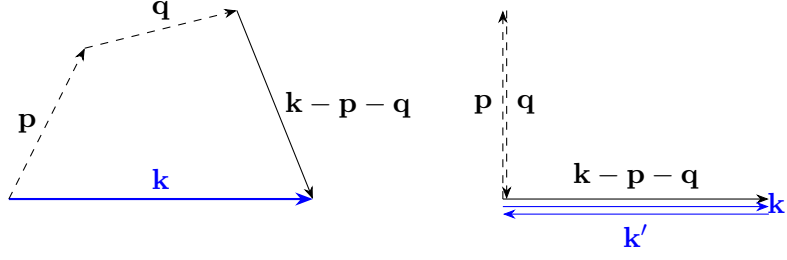

In this subsection, we compute the correlation of first and third-order tensor modes sourced by scalar-scalar-tensor interactions. We substitute Eq.~\eqref{solhpsipsi} into Eq.~\eqref{correlation31}
\begin{align}\label{h1h3psipsih}
    \begin{split}
            \langle h_{\lambda ^{\prime}}(\mathbf{k^{\prime}})h^{\Psi \Psi h}_{\lambda }(\mathbf{k}) \rangle &= \frac{6}{(2\pi)^3} \int {\rm d} ^3 \mathbf{q} \int {\rm d} ^3 \mathbf{p} \, T_h(\eta k \conf ) \epsilon^{ij}_{\lambda}(\mathbf{k})^*  \sum _{\lambda _1} \bigg ( \epsilon_{ij}^{\lambda _1}(\mathbf{k}-\mathbf{p}-\mathbf{q}) I_1^{ssh}(\eta , k,p,q) \\
            &+ 16 \epsilon_{jm}^{\lambda _1}(\mathbf{k}-\mathbf{p}-\mathbf{q})p^mp_i I_2^{ssh}(\eta , k,p,q) - 8\epsilon_{jm}^{\lambda _1}(\mathbf{k}-\mathbf{p}-\mathbf{q})p^m(q_i+p_i)  \\
            &\times I_3^{ssh}(\eta , k,p,q) +8\epsilon_{jm}^{\lambda _1}(\mathbf{k}-\mathbf{p}-\mathbf{q})q^mp_i I_4^{ssh}(\eta , k,p,q)\bigg )  \langle h^{\lambda \conf} _{\mathbf{k \conf}}h^{\lambda _1}_{\mathbf{k}-\mathbf{p}-\mathbf{q}}\mathcal{R} _{\mathbf{q}} \mathcal{R} _{\mathbf{p}} \rangle .
    \end{split}
\end{align}
The four-point function is broken down as
\begin{equation}
     \langle h^{\lambda \conf} _{\mathbf{k \conf}}h^{\lambda _1}_{\mathbf{k}-\mathbf{p}-\mathbf{q}}\mathcal{R} _{\mathbf{q}} \mathcal{R} _{\mathbf{p}} \rangle = \delta ^{\lambda \conf \lambda _1} \delta (\mathbf{k \conf} + \mathbf{k} - \mathbf{p} - \mathbf{q}) \delta (\mathbf{p}+ \mathbf{q}) P^{\lambda \conf}_h(\eta k\conf) P_{\mathcal{R}} (\eta q) \, ,
\end{equation}
where we recall that we are disregarding any connected contributions that could arise from inflation. The above suggests that, physically, out of the three modes responsible for inducing the third-order gravitational wave, one must share the same momentum as the outgoing third-order wave (or the first-order gravitational wave we are correlating with), while the other two modes must have opposite momenta, see Fig.~\ref{fig:ssh2point}. This physical picture contrasts with the usual second-order production mechanism encountered in SIGWs; for more details, see Ref.~\cite{Espinosa:2018eve}. We now substitute the expression for the four-point function into Eq.~\eqref{h1h3psipsih} and integrate over $\mathbf{p}$ 
\begin{align}
    \begin{split}
            \langle h_{\mathbf{k^{\prime}},\lambda ^{\prime}}h^{\Psi \Psi h}_{\mathbf{k},\lambda } \rangle &= \frac{6}{(2\pi)^3} \int {\rm d} ^3 \mathbf{q}\sum _{\lambda _1} \bigg ( \epsilon^{ij}_{\lambda}(\mathbf{k})^* \epsilon_{ij}^{\lambda _1}(\mathbf{k})  T_h(\eta k \conf )I_1^{ssh}(\eta , k,q) \\
            &+ 16\epsilon^{ij}_{\lambda}(\mathbf{k}) ^*\epsilon_{jm}^{\lambda _1}(\mathbf{k})q^mq_i  T_h(\eta k \conf )I_2^{ssh}(\eta , k,q)   \\
            & -8\epsilon^{ij}_{\lambda}(\mathbf{k}) ^*\epsilon_{jm}^{\lambda _1}(\mathbf{k})q^mq_i  T_h(\eta k \conf )I_4^{ssh}(\eta , k,q)\bigg )  \delta ^{\lambda \conf \lambda _1} \delta (\mathbf{k} + \mathbf{k \conf} ) P^{\lambda \conf}_h(\eta k\conf) P_{\mathcal{R}} (\eta q) \, .
    \end{split}
\end{align}
Using the normalisation condition of the polarisation tensors and the result
\begin{equation}
    \epsilon^{ij}_{\lambda}(\mathbf{k}) ^*\epsilon_{jm}^{\lambda _1}(\mathbf{k})q^mq_i  = \frac{1}{2}q^2 \sin ^2\theta _q \delta ^{\lambda  \lambda _1} \, ,
\end{equation}
the two-point function finally reduces to 
\begin{align}
    \begin{split}
            \langle h_{\mathbf{k^{\prime}},\lambda ^{\prime}}h^{(3)}_{\mathbf{k},\lambda } \rangle &= \frac{6}{(2\pi)^3} \int _0^{\infty} {\rm d} q \int _0^{\pi} {\rm d} \theta _q \int _0^{2\pi} {\rm d} \phi _q  \, q^2 \sin \theta _q \bigg (  T_h(\eta k \conf )I_1^{ssh}(\eta , k,q) + 8q^2 \sin ^2\theta _q  T_h(\eta k \conf ) \\
            &\times I_2^{ssh}(\eta , k,q) -4q^2 \sin ^2\theta _q  T_h(\eta k \conf )I_4^{ssh}(\eta , k,q)\bigg )  \delta ^{\lambda \conf \lambda _1}\delta ^{\lambda  \lambda _1} \delta (\mathbf{k} + \mathbf{k \conf} ) P^{\lambda \conf}_h(k\conf) P_{\mathcal{R}} ( q) \, .
    \end{split}
\end{align}
An important point to note, specific to the source terms involving pure linear perturbations, is that once we have integrated over one of the internal momenta, there is only a trivial angular dependence remaining. This means that after expressing the remaining internal momenta in spherical coordinates, we can integrate over both the polar and azimuthal angles, as the only angular dependence will reside in the polarization functions, see Fig.~\ref{fig:ssh2point} for a geometrical interpretation.  

Furthermore, unlike second-order iGWs, the primordial tensor power spectrum appears outside the integral and does not depend on the modes inducing the third-order perturbations. Therefore, these contributions act as a modulation to the primordial gravitational wave power spectrum. By integrating over the angular directions in spherical coordinates, relabelling $\mathbf{q}$ to $\mathbf{p}$ we can extract the power spectrum as follows
\begin{align}
    \begin{split}
        P^{(13)}_{R/L}(\eta , k) &= k^3\frac{6}{(2\pi )^2} P^{(11)}_{R/L}( k) \int _0 ^{\infty} {\rm d}v \, v^2 \bigg (\mathcal{I}^{ssh}_1(x,k,v) + \frac{32}{3}v^2k^2\mathcal{I}^{ssh}_2(x,k,v) \\
        &- \frac{16}{3}v^2k^2 \mathcal{I}^{ssh}_4(x,k,v) \bigg ) P_{\mathcal{R}}( vk) \, ,
    \end{split}
\end{align}
where we have further defined 
\begin{equation}
    \mathcal{I}^{ssh}_{\textit{i}}(x,k,v) = T_h(x )  I_{\textit{i}}^{ssh}(x,k,v) \, .
\end{equation}
Finally, the dimensionless power spectrum is given by:
\begin{align}\label{powerspectrumsshfinal}
    \begin{split}
        \mathcal{P}^{(13)}_{\Psi \Psi h}(\eta , k) &= 3 \left ( \mathcal{P}^R_h(k) + \mathcal{P}^L_h(k) \right ) \int _0 ^{\infty} {\rm d}v \, \frac{1}{v} \bigg (\mathcal{I}^{ssh}_1(x , k , v) + \frac{32}{3}v^2k^2\mathcal{I}^{ssh}_2(x , k , v) \\
        &- \frac{16}{3}v^2k^2 \mathcal{I}^{ssh}_4(x , k , v) \bigg ) \mathcal{P}_{\mathcal{R}}( vk) \, ,
    \end{split}
\end{align}
and the functions defined in Eq.~\eqref{sshfunctions} are 
\begin{subequations} \label{sshfunctionsfinal} 
    \begin{align}
        \begin{split}
            f_1^{\Psi \Psi h}(\barx , k ,v) &= \left ( \frac{3+3w}{5+3w} \right )^2 \bigg ( -4k^2 \frac{\partial}{\partial \barx}T _h(\barx)T_{\Psi}(\barx v) \frac{\partial}{\partial \barx}T_{\Psi}(\barx v) - \frac{1}{3(1+w)}T_h(\barx )\\
            &\times   \bigg ( 3k^2(1+4w+3w^2) \frac{\partial}{\partial \barx}T_{\Psi}(\barx v)\frac{\partial}{\partial \barx}T _{\Psi}(c_s\barx v) -24(w^2-1)T_{\Psi}(\barx v)T_{\Psi}(\barx v) \\
            &+ (-17-16w+9w^2)\, v^2k^2\, T_{\Psi}(\barx v)T_{\Psi}(\barx v) \bigg ) - 8 \frac{k^2}{\barx}T_h(\barx)T_{\Psi}(\barx v)\frac{\partial}{\partial \barx}T_{\Psi}(\barx v)\\
            &+\frac{8(w-1)}{3(1+w)} \barx v^2k^2 \, T_h(\barx)T_{\Psi}(\barx v) \frac{\partial}{\partial \barx}T_{\Psi}(\barx v)+\frac{4(w-1)}{3(1+w)} \, \barx ^2 v^2 k^2 \, \frac{\partial}{\partial \barx}T_h(\barx) \\
            &\times T_{\Psi}(\barx v)\frac{\partial}{\partial \barx}  T_{\Psi}(\barx v)\bigg ) \, , 
        \end{split}
        \\
        \begin{split}
             f_2^{\Psi \Psi h}(\barx , k ,v) &=  \left ( \frac{3+3w}{5+3w} \right )^2  T_h(\barx)T_{\Psi}(\barx v) T_{\Psi}(\barx v) \, ,
        \end{split}
        \\
        \begin{split}
             f_4^{\Psi \Psi h}(\barx , k ,v) &=\left ( \frac{3+3w}{5+3w} \right )^2 \bigg ( 2 T_h(\barx)T_{\Psi}(\barx v) T_{\Psi}(\barx v) + \frac{1}{3(1+w) }\barx \frac{\partial}{\partial \barx}T_h(\barx)T_{\Psi}(\barx v) T_{\Psi}(\barx v) \\
             &  -  \frac{1}{3(1+w) }\barx T_h(\barx) \frac{\partial}{\partial \barx}T_{\Psi}(\barx v) T_{\Psi}(\barx v) +  \frac{1}{3(1+w)} \barx ^2 \frac{\partial}{\partial \barx}T_h(\barx) T_{\Psi}(\barx v) \frac{\partial}{\partial \barx}T_{\Psi}(\barx v) \\
             & - \frac{1}{3(1+w)\hubble ^2}\barx ^2 T_h(\barx)\frac{\partial}{\partial \barx} T_{\Psi}(\barx v) \frac{\partial}{\partial \barx}T_{\Psi}(\barx v) \bigg ) \, .
        \end{split}
    \end{align}
\end{subequations}
\subsection{The power spectrum of the second-order scalars coupled to first-order scalars: 
\texorpdfstring{$\mathcal{P}^{(13)}_{\Psi ^{(2)} \Psi}$}{P(13)Psi2Psi}}
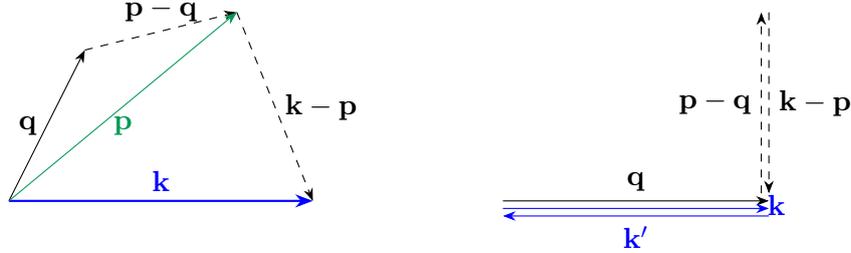
\begin{figure}[h!]
  \centering
  \begin{tikzpicture}

    \coordinate (A) at (0,0);
    \coordinate (B) at (1,2);
    \coordinate (C) at (3,2.5);
    \coordinate (D) at (4,0);


    \draw[thick, -{Stealth}, blue] (A) -- (D);
    \draw[dashed, -{Stealth}, black] (C) -- (D);
    \draw[ -{Stealth}, black] (A) -- (B);
    \draw[dashed, -{Stealth}, black] (B) -- (C);
    \draw[ -{Stealth}, ForestGreen] (A) -- (C);

    \node[above, text=blue] at ($(A)!0.5!(D)$) {$\mathbf{k}$};
    \node[left] at ($(A)!0.5!(B)$) {$\mathbf{q}$};
    \node[above] at ($(B)!0.5!(C)$) {$\mathbf{p}-\mathbf{q}$};
    \node[right] at ($(C)!0.5!(D)$) {$\mathbf{k}-\mathbf{p}$};
    \node[below, text=ForestGreen] at ($(A)!0.5!(C)$) {$\mathbf{p}$};

    \coordinate (E) at (6.5,0);
    \coordinate (F) at (6.55,0);
    \coordinate (G) at (6.5,2.5);
    \coordinate (H) at (6.55,2.5);
    \coordinate (I) at (10,0);
    \coordinate (J) at (6.5,-0.1);
    \coordinate (K) at (10,-0.1);
    \coordinate (L) at (6.5,-0.2);
    \coordinate (M) at (10,-0.2);
    \coordinate (O) at (9.9,0.1);
    \coordinate (P) at (9.9,2.5);
    \coordinate (Q) at (10,0.1);
    \coordinate (R) at (10,2.5);

    \draw[dashed, -{Stealth}, black] (O) -- (P);
    \draw[dashed, -{Stealth}, black] (R) -- (Q);
    \draw[ -{Stealth}, black] (E) -- (I);
    \draw[ -{Stealth}, blue] (J) -- (K);
    \draw[ -{Stealth}, blue] (M) -- (L);


    \node[text=blue] at (10.1,-0.05) {$\mathbf{k}$};
    \node[below, text=blue] at ($(M)!0.5!(L)$) {$\mathbf{k^{\prime}}$};
    \node[left] at ($(O)!0.5!(P)$) {$\mathbf{p}-\mathbf{q}$};
    \node[right] at ($(Q)!0.5!(R)$) {$\mathbf{k}-\mathbf{p}$};
    \node[above] at ($(E)!0.5!(I)$) {$\mathbf{q}$};
  \end{tikzpicture}
  
  \caption{\footnotesize{\textit{Left}. Momentum configuration of a second-order mode (solid green) and a scalar mode (dashed black) inducing a third-order tensor mode (solid blue). The second-order perturbation is itself induced by a scalar and a first-order tensor perturbation (solid black). \textit{Right}. Geometrical configuration of the two-point function resulting from a second-order perturbation coupled to a first-order scalar perturbation with a first-order tensor mode. In this configuration the two scalar modes depend on the angle between the incoming second-order mode and the outgoing third-order mode, making the dependence on the coordinate $u$ non-trivial.}}
  \label{fig:secondorderfirstorderscalar2point}
\end{figure}

In this section, we utilize the solution given by Eq.~\eqref{GWeqsolpsi2psi1}, which describes third-order gravitational waves sourced by second-order scalar perturbations coupled to first-order scalars. The corresponding two-point function is given by
\begin{align}
    \begin{split}
            \langle h_{\lambda ^{\prime}}(\mathbf{k^{\prime}})h^{\Psi ^{(2)}\Psi}_{\lambda }(\mathbf{k}) \rangle &= \frac{6}{(2\pi)^3} \int {\rm d} ^3 \mathbf{q} \int {\rm d} ^3 \mathbf{p} \, T_h(\eta k \conf )\sum _{\lambda _1}\eps ^{ij}_{\lambda}(\mathbf{k})^* p_ip_j Q^{\lambda _1}_{\pi ,st}(\mathbf{p},\mathbf{q})  I^{\Psi ^{(2)}\Psi}(\eta , k,p,q) \\
            &\times \langle h^{\lambda \conf} _{\mathbf{k \conf}} \mathcal{R} _{\mathbf{k}-\mathbf{p}}h_{\mathbf{q}}^{\lambda _1} \mathcal{R} _{\mathbf{p}-\mathbf{q}}\rangle \, .
    \end{split}
\end{align}
The four-point function can be broken down as
\begin{equation}\label{stthirdorder4point}
     \langle h^{\lambda \conf} _{\mathbf{k \conf}}h_{\mathbf{q}}^{\lambda _1} \mathcal{R} _{\mathbf{p}-\mathbf{q}} \mathcal{R} _{\mathbf{k}-\mathbf{p}}\rangle = \delta ^{\lambda \conf \lambda _1} \delta (\mathbf{k \conf}+ \mathbf{q}) \delta (\mathbf{k}-\mathbf{q}) P^{\lambda \conf} _h (k \conf) P_{\mathcal{R}} (|\mathbf{p}-\mathbf{q}|) \, ,
\end{equation}
so that we can integrate over $\mathbf{q}$ to arrive at 
\begin{align}
    \begin{split}
            \langle h_{\mathbf{k^{\prime}},\lambda ^{\prime}}h^{(3)}_{\mathbf{k},\lambda } \rangle &= \frac{6}{(2\pi)^3}  \int {\rm d} ^3 \mathbf{p} \,   T_h(\eta k \conf )\sum _{\lambda _1}\eps ^{ij}_{\lambda}(\mathbf{k})^* p_ip_j Q^{\lambda _1}_{\pi ,st}(\mathbf{p},\mathbf{k})  I^{\Psi ^{(2)}\Psi}(\eta , k,p,q=k) \\
            &\times \delta ^{\lambda \conf \lambda _1} \delta (\mathbf{k \conf}+ \mathbf{k})  P^{\lambda \conf} _h (k \conf) P_{\mathcal{R}} (|\mathbf{p}-\mathbf{k}|) \, .
    \end{split}
\end{align}
After switching to spherical coordinates, we can integrate the polarisation functions over the azimuthal direction  
\begin{equation}
    \int {\rm d}\phi _p \,\eps ^{ij}_{\lambda}(\mathbf{k})^* p_ip_j Q_{\lambda _1}^{\Pi^{(2)}_{st}}(\mathbf{p},\mathbf{q}=\mathbf{k}) = \frac{\pi}{2}k^4v^4 \left (1-\frac{(1-u^2+v^2)^2}{4v^2} \right )^2 \delta ^{\lambda \lambda _1} \, ,
\end{equation}
where we have switched to $v$ and $u$ coordinates. As opposed to the source term made of first-order perturbations, the $u$ dependence is now non-trivial (see Fig.~\ref{fig:secondorderfirstorderscalar2point}) and this is the case for all the source terms involving second-order perturbations. 

The power spectrum can now be obtained and is given by
\begin{align}
    \begin{split}
        P^{(13)}_{R/L} &= \frac{3}{2^3\pi ^2}k^3 \int _0^{\infty} {\rm d} v  \int _{|1-v|}^{1+v} {\rm d}u \, uv \, k^4   \left (v^2-\frac{(1-u^2+v^2)^2}{4} \right )^2  \mathcal{I}^{h^{(2)}\Psi}(x , k,v,u)\\
        &\times P^{(11)}_{R/L} (k) P_{\mathcal{R}} (ku) \, , 
    \end{split}
\end{align}
where we have defined
\begin{equation}
     \mathcal{I}^{\Psi ^{(2)}\Psi}(x,k,v,u) = T_h(x)  I^{\Psi ^{(2)}\Psi}(x,k,v,u) \, ,
\end{equation}
which takes into account the sub-Hubble evolution of the tensor mode we have correlated with. Finally, the dimensionless power spectrum is given by
\begin{align}\label{pspsi2psi1}
    \begin{split}
         \mathcal{P}^{(13)}_{\Psi ^{(2)}\Psi}(\eta ,k) &= \frac{3}{4} \left ( \mathcal{P}^{(11)}_R (\eta, k) +  \mathcal{P}^{(11)}_L (\eta, k) \right ) \int _0^{\infty} {\rm d} v  \int _{|1-v|}^{1+v} {\rm d} u \, \frac{v}{u^2} \left (v^2-\frac{(1-u^2+v^2)^2}{4} \right )^2   \\
         &\times k^4 \mathcal{I}^{\Psi ^{(2)}\Psi}(x,k,v,u)  P_{\mathcal{R}} (ku) \, .
    \end{split}
\end{align}
We empahsize once again that the kernel and the primordial scalar power spectrum exhibit angular dependence, making the polar integral (the integral over $u$) non-trivial to evaluate—unlike the case of third-order waves sourced purely by first-order contributions, where the integration is trivial. The kernel $ I^{\Psi ^{(2)}\Psi}(x,k,v,u)$ was defined in Eq.~\eqref{kernelpsi2psi1} and we now give $f^{\Psi^{(2)}\Psi}(\barx,k,v,u)$ in terms of $v$ and $u$
\begin{align}\label{fpsi2psi1final}
    \begin{split}
        &f^{\Psi^{(2)}\Psi}(\barx,k,v,u) = \avant ^2 \bigg ( \frac{(1+3w)^2}{3(1+w)} \barx ^2 \frac{\partial}{\partial \barx}T_{\Psi} (\barx u )\frac{\partial}{\partial \barx} I^{\Psi ^{(2)}}_{st}(\barx , k ,v,u) \\
        &+  \frac{2(1+3w)}{3(1+w)} \barx \bigg (  \frac{\partial}{\partial \barx}T_{\Psi} (\barx u )  I^{\Psi ^{(2)}}_{st}(\barx  , k ,v,u) + \frac{1}{k^4v^4} \frac{\partial}{\partial \barx}T_{\Psi} (\barx u ) f^{\Pi ^{(2)}}_{st}(\barx , k, v,u)  \\
        &+ T_{\Psi} (\barx u ) \frac{\partial}{\partial \barx} I^{\Psi ^{(2)}}_{st}(\barx , k ,v,u) \bigg  )  + \frac{2(5+3w)}{3(1+w)} \bigg (T_{\Psi} (\barx u ) I^{\Psi ^{(2)}}_{st}(\barx  , k ,v,u) \\
        &+ \frac{1}{k^4v^4}T_{\Psi} (\barx u ) f^{\Pi ^{(2)}}_{st}(\barx , k, v,u)   \bigg ) \bigg ) \, ,
    \end{split}
\end{align}
with
\begin{align}
    \begin{split}
        f^{\Pi ^{(2)}}_{st}(\barx , k, v,u)  &= 2k^2(1+3u^2-3v^2) T_h(\barx) T_{\Psi}(\barx u)-12 k^2   T_h(\barx) T_{\Psi}(\barx u) \\
        &+12\frac{k^2}{\barx}  T_h(\barx) \frac{\partial}{\partial \barx}T_{\Psi}(\barx u)+6(w-1)k^2u^2T_h(\barx) T_{\Psi}(\barx u) \\
        &+6k^2 (1+u^2-v^2) T_h(\barx) T_{\Psi}(\barx u) + 6k^2(1-u^2+v^2)  T_h(\barx) T_{\Psi}(\barx u)\, ,
    \end{split}
\end{align}
and
\begin{align} \label{psi2nested}
    \begin{split}
        I^{\Psi ^{(2)}}_{st}(\barx , k ,v,u) &= \frac{\sqrt{w}}{\barx ^{1+b}k^2v} \int _0^{\barx v} {\rm d}  \tilde{y} \, \, \tilde{y}^{3+b} \big ( j_{b+1}(\tilde{y}\sqrt{w})y_{b+1}(v\barx\sqrt{w}) - y_{b+1}(\tilde{y}\sqrt{w})\\
        &\times j_{b+1}(v\barx\sqrt{w}) \big ) \Sigma ^{st}(\tilde{y},v,u) \, .
    \end{split}
\end{align}
We have defined $\Sigma ^{st}(\tilde{y},v,u)$ in Eq.~\eqref{Sigmapsi2} and it now reads
\begin{align}
    \begin{split}
        \Sigma ^{st}(\tilde{y},v,u) &= \frac{2}{3(1+3w)v^3\tildey ^3} \bigg (6\tildey \frac{\partial}{\partial \tildey} T_h\left (\frac{\tildey}{v}\right ) \bigg ( (3+v^2-3u^2w)\tildey T_{\Psi}\left (\frac{\tildey u}{v}\right ) - 6uv\frac{\partial}{\partial \tildey} T_{\Psi}\left (\frac{\tildey u}{v}\right ) \bigg ) \\
        &+  T_h\left (\frac{\tildey}{v}\right ) \bigg [v(1+3w) (-3-2v^2+3(u^2+v^2)w)\tildey ^3 T_{\Psi}\left (\frac{\tildey u}{v}\right ) + 6u \bigg ( \big[6v^2\\
        & + (3-3u^2w+v^2(2+3w))\tildey ^2\big] \frac{\partial}{\partial \tildey} T_{\Psi}\left (\frac{\tildey u}{v}\right ) -6uv\tildey \frac{\partial ^2}{\partial \tildey ^2}T_{\Psi}\left (\frac{\tildey u}{v}\right )  \bigg ) \bigg] \bigg ) \, ,
    \end{split}
\end{align}
where we have used the relations shown in Eq.~\eqref{yrelations}.

\subsection{The power spectrum of second-order vectors coupled to first-order scalars: 
\texorpdfstring{$\mathcal{P}^{(13)}_{B^{(2)} \Psi}$}{P(13)B2Psi}}
In this subsection, we compute the contribution coming from second-order vectors coupled to first-order scalars. This amounts to substituting in the solution Eq.~\eqref{GWeqsolb2psi1} to the two point function,
\begin{align}
    \begin{split}
            \langle h_{\lambda ^{\prime}}(\mathbf{k^{\prime}})h^{B^{(2)\Psi}}_{\lambda }(\mathbf{k}) \rangle &= \frac{6}{(2\pi)^3} \int {\rm d} ^3 \mathbf{q} \int {\rm d} ^3 \mathbf{p} \, T_h(\eta k \conf )\sum _{r}\sum _{\lambda _1} \bigg ( Q^{B^{(2)}\Psi}_{\lambda ,r}(\mathbf{k},\mathbf{p})Q^{B^{(2)}_{st1}}_{r,\lambda _1}(\mathbf{p},\mathbf{q})  \\
            &\times I^{B^{(2)}\Psi}_1(\eta , k,p,q)+Q^{B^{(2)}\Psi}_{\lambda ,r}(\mathbf{k},\mathbf{p})Q^{B^{(2)}_{st2}}_{r,\lambda _1}(\mathbf{p},\mathbf{q}) I^{B^{(2)}\Psi}_2(\eta , k,p,q) \bigg )\\
            &\times \langle h^{\lambda \conf} _{\mathbf{k \conf}}h_{\mathbf{q}}^{\lambda _1} \mathcal{R} _{\mathbf{p}-\mathbf{q}} \mathcal{R} _{\mathbf{k}-\mathbf{p}}\rangle \, .
    \end{split} 
\end{align}
Inserting the expression for the four-point function given in Eq.~\eqref{stthirdorder4point}, we arrive to
    \begin{align}
    \begin{split}
            \langle h_{\lambda ^{\prime}}(\mathbf{k^{\prime}})h^{B^{(2)\Psi}}_{\lambda }(\mathbf{k}) \rangle &= \frac{6}{(2\pi)^3} \int {\rm d} ^3 \mathbf{q} \int {\rm d} ^3 \mathbf{p} \, T_h(\eta k \conf )\sum _{r}\sum _{\lambda _1} \bigg ( Q^{B^{(2)}\Psi}_{\lambda ,r}(\mathbf{k},\mathbf{p})Q^{B^{(2)}_{st1}}_{r,\lambda _1}(\mathbf{p},\mathbf{q}=\mathbf{k})  \\
            &\times I^{B^{(2)}\Psi}_1(\eta , k,p,q=k)+Q^{B^{(2)}\Psi}_{\lambda ,r}(\mathbf{k},\mathbf{p})Q^{B^{(2)}_{st2}}_{r,\lambda _1}(\mathbf{p},\mathbf{q}=\mathbf{k}) I^{B^{(2)}\Psi}_2(\eta , k,p,q=k) \bigg )\\
            &\times \delta ^{\lambda ^{\prime} \lambda _1} \delta (\mathbf{k}+\mathbf{k^{\prime}}) P^{(11)}_{\lambda ^{\prime}}(k)P_{\mathcal{R}}(|\mathbf{p}-\mathbf{k}|) \, .
    \end{split}
\end{align}
We can now integrate over $\mathbf{q}$ and perform the azimuthal integral of the wave-vector $\mathbf{p}$ over the polarisation functions
\begin{subequations}\label{evalpolb2psi1}
    \begin{align}
        \int {\rm d} \phi_p \, \sum _{r}\sum _{\lambda _1}Q^{B^{(2)}\Psi}_{\lambda ,r}(\mathbf{k},\mathbf{p})Q^{B^{(2)}_{st1}}_{r,\lambda _1}(\mathbf{p},\mathbf{q}=\mathbf{k}) &=-\frac{k^4\pi}{32v^2} a^{B^{(2)}\Psi}(v,u)b^{B^{(2)}\Psi}(v,u)\delta ^{\lambda \lambda _1} \, , \\ 
        \int {\rm d} \phi_p \, \sum _{r}\sum _{\lambda _1}Q^{B^{(2)}\Psi}_{\lambda ,r}(\mathbf{k},\mathbf{p})Q^{B^{(2)}_{st2}}_{r,\lambda _1}(\mathbf{p},\mathbf{q}=\mathbf{k}) &=-\frac{k^2\pi}{32v^2} a^{B^{(2)}\Psi}(v,u)c^{B^{(2)}\Psi}(v,u)\delta ^{\lambda \lambda _1} \, ,
    \end{align}
\end{subequations}
where we have further defined
\begin{subequations}
    \begin{align}
        a^{B^{(2)\Psi}}(v,u) &= (-1+u-v)(1+u-v)(-1+u+v)(1+u+v) \, , \\
        b^{B^{(2)\Psi}}(v,u) &= -3+u^4+2v^2+v^4-2u^2(v^2-1) \, , \\
        c^{B^{(2)\Psi}}(v,u) &= 1+u^4+6v^2+v^4-2u^2(1+v^2) \, .
    \end{align}
\end{subequations}
The power spectrum is then given by
\begin{align}
    \begin{split}
           P^{(13)}_{R/L} &= \frac{3}{(2\pi) ^2}k^3P^{(11)}_{R/L}(k)  \int _0^{\infty} {\rm d} v  \int _{|1-v|}^{1+v} {\rm d}u \, uv\bigg ( -\frac{k^4}{32v^2} a^{B^{(2)\Psi}}(v,u)b^{B^{(2)\Psi}}(v,u)  \\
            &\times \mathcal{I}^{B^{(2)}\Psi}_1(x,k,v,u) -\frac{k^2}{32v^2} a^{B^{(2)\Psi}}(v,u)c^{B^{(2)\Psi}}(v,u) \mathcal{I}^{B^{(2)}\Psi}_2(x,k,v,u) \bigg ) P_{\mathcal{R}}(ku) \, ,
    \end{split}
\end{align}
with
\begin{equation}
     \mathcal{I}^{B^{(2)}\Psi}_{\textit{i}}(x,k,v,u) = T_h(x)  I^{B ^{(2)}\Psi}_{\textit{i}}(x,k,v,u) \, .
\end{equation}
Finally, the dimensionless power spectrum corresponding to iGWs sourced by second-order vectors coupled to first-order scalar perturbations is given by:
\begin{align}\label{finalpsb2psi1}
    \begin{split}
        \mathcal{P}^{(13)}_{B ^{(2)}\Psi}(\eta ,k) &=- \frac{3}{64} \left ( \mathcal{P}^{(11)}_R (\eta, k) +  \mathcal{P}^{(11)}_L (\eta, k) \right ) \int _0^{\infty} {\rm d} v  \int _{|1-v|}^{1+v} {\rm d} u \, \frac{1}{vu^2} \bigg ( a^{B^{(2)\Psi}}(v,u)b^{B^{(2)\Psi}}(v,u)  \\
            &\times k^4\mathcal{I}^{B^{(2)}\Psi}_1(x,k,v,u) + a^{B^{(2)\Psi}}(v,u)c^{B^{(2)\Psi}}(v,u)k^2\mathcal{I}^{B^{(2)}\Psi}_2(x,k,v,u) \bigg )  P_{\mathcal{R}} (ku) \, .
    \end{split}
\end{align}
The two kernel's $I^{B ^{(2)}\Psi}_{\textit{i}}(x,k,v,u)$, defined in Eq.~\eqref{kernelb2psi1} are time integrals over the functions
\begin{subequations} \label{fb2psi1final}
    \begin{align}
        \begin{split}
            &f^{B^{(2)\Psi}}_{\textit{i}}(\overline{x},k,p,q) = \avant ^2 \bigg ( \frac{2}{v^2k} I^{B^{(2)}}_{sti}(\overline{x} ,k,v,u) \frac{\partial}{\partial \barx}T_{\Psi}(\barx u) + \frac{2}{v^2k}\frac{\partial}{\partial \barx}I^{B^{(2)}}_{sti}(\overline{x} ,k,v,u) \\
            &\times T_{\Psi}(\barx u) + \frac{8}{v^2k\barx (1+3w)} I^{B^{(2)}}_{sti}(\overline{x} ,k,v,u) T_{\Psi}(\barx u) +\frac{\barx (1+3w)}{6k(1+w)} I^{B^{(2)}}_{sti}(\overline{x} ,k,v,u) T_{\Psi}(\barx u) \\
            &+  \frac{\barx ^2(1+3w)^2}{12k(1+w)}  I^{B^{(2)}}_{sti}(\overline{x} ,k,v,u) \frac{\partial}{\partial \barx} T_{\Psi}(\barx u)   \bigg ) \, .
        \end{split}
    \end{align}
\end{subequations}
The expression for the nested kernels ($i=1,2$) are
\begin{equation}
    I^{B^{(2)}}_{sti}(\overline{x} ,k,v,u) = \frac{1}{(kv)^{3+2b}\barx^{2(1+b)}} \int _0^{\barx v} {\rm d}  \tilde{y} \, \, \tilde{y}^{2(1+b)} f^{B^{(2)}}_{sti}(\tildey , k, v,u) \, ,
\end{equation}
with
\begin{subequations}
    \begin{align}
        f^{B^{(2)}}_{st1}(\tildey , k, v,u) &= -4 T_h\left (\frac{\tilde{y}}{v} \right )T_{\Psi} \left (\frac{\tilde{y}u}{v} \right )  \, ,
        \\
        \begin{split}
         f^{B^{(2)}}_{st2}(\tildey , k, v,u) &= 2k^2(-1+v^2-u^2)T_h\left (\frac{\tilde{y}}{v} \right )T_{\Psi} \left (\frac{\tilde{y}u}{v} \right )+4k^2T_h\left (\frac{\tilde{y}}{v} \right )T_{\Psi} \left (\frac{\tilde{y}u}{v} \right ) \\
         &-\frac{4v^2k^2}{\tildey} T_h\left (\frac{\tilde{y}}{v} \right )\frac{\partial}{\partial \tildey}T_{\Psi} \left (\frac{\tilde{y}u}{v}  \right )+2(1-w)u^2k^2T_h\left (\frac{\tilde{y}}{v} \right )T_{\Psi} \left (\frac{\tilde{y}u}{v} \right ) \, .
        \end{split}
    \end{align}
\end{subequations}

\subsection{The power spectrum of second-order tensors coupled to first-order scalars: 
\texorpdfstring{$\mathcal{P}^{(13)}_{h^{(2)} \Psi}$}{P(13)h2Psi}}
The last term to consider is the two-point function involving third-order iGWs sourced by second-order tensors and first-order scalars. The equation of motion for these iGWs was derived in Eq.~\eqref{GWeqsolh2psi1}. So we have
\begin{align}
    \begin{split}
            \langle h_{\lambda ^{\prime}}(\mathbf{k^{\prime}})h^{h^{(2)\Psi}}_{\lambda }(\mathbf{k}) \rangle &= \frac{6}{(2\pi)^3} \int {\rm d} ^3 \mathbf{q} \int {\rm d} ^3 \mathbf{p} \, T_h(\eta k \conf )\sum _{\sigma ,\lambda _1}\eps^{ij}_{\lambda}(\mathbf{k})^* \eps _{ij}^{\sigma}(\mathbf{p})\eps ^{mn}_{\sigma}(\mathbf{p})^*\eps _{mn}^{\lambda _1}(\mathbf{q})  I_{st}^{h^{(2)}\Psi}(\eta , k,p,q) \\
            &\times \langle h^{\lambda \conf} _{\mathbf{k \conf}}h_{\mathbf{q}}^{\lambda _1} \mathcal{R} _{\mathbf{p}-\mathbf{q}} \mathcal{R} _{\mathbf{k}-\mathbf{p}}\rangle \, ,
    \end{split}
\end{align}
and following the same steps outlined in the previous subsections, we arrive at 
\begin{align}
    \begin{split}
            \langle h_{\mathbf{k^{\prime}},\lambda ^{\prime}}h^{(3)}_{\mathbf{k},\lambda } \rangle &= \frac{6}{(2\pi)^3}  \int {\rm d} ^3 \mathbf{p}  T_h(\eta k \conf )\sum _{\sigma ,\lambda _1}\eps ^{ij}_{\lambda}(\mathbf{k})^* \eps _{ij}^{\sigma}(\mathbf{p})\eps ^{mn}_{\sigma}(\mathbf{p})^*\eps _{mn}^{\lambda _1}(\mathbf{k})  I_{st}^{h^{(2)}\Psi}(\eta , k,p,q=k) \\
            &\times \delta ^{\lambda \conf \lambda _1} \delta (\mathbf{k \conf}+ \mathbf{k})  P^{\lambda \conf} _h (k \conf) P_{\mathcal{R}} (|\mathbf{p}-\mathbf{k}|) \, .
    \end{split}
\end{align}
Contracting the polarisation tensors and integrating over the azimuthal direction, the only non-trivial contribution happens when $\lambda =\lambda _1$, which leads to 
\begin{align}
    \begin{split}
            \langle h_{\mathbf{k^{\prime}},\lambda ^{\prime}}h^{(3)}_{\mathbf{k},\lambda } \rangle &= \frac{6}{(2\pi)^2}  \int _0^{\infty} {\rm d}  p \int _0^{\pi} {\rm d}  \theta _p \, p^2 \sin \theta _p \left (\cos{\left (\frac{\theta _p}{2}\right )}^8 + \sin{\left (\frac{\theta _p}{2}\right )}^8 \right )  \\
            &\times T_h(\eta k \conf )I^{h^{(2)}\Psi}(\eta , k,p,q=k) \delta ^{\lambda \conf \lambda } \delta (\mathbf{k \conf}+ \mathbf{k})  P^{\lambda \conf} _h (k \conf) P_{\mathcal{R}} (|\mathbf{p}-\mathbf{k}|) \, .
    \end{split}
\end{align}
Switching to the appropriate $v$ and $u$ coordinates, the power spectrum now reads
\begin{align}
    \begin{split}
        P^{(13)}_{R/L}(\eta , k) &= \frac{6}{(2\pi)^2}k^3 \int _0^{\infty} {\rm d} v  \int _{|1-v|}^{1+v} {\rm d} u \, \frac{u}{256v^3} \big [ \left (u^2-(v+1)^2 \right )^4 + \left (u^2-(v-1)^2 \right )^4  \big ] \\
        &\times \mathcal{I}^{h^{(2)}\Psi}(x,k,v,u)P^{(11)}_{R/L} (k) P_{\mathcal{R}} (ku) \, ,
    \end{split}
\end{align}
with the kernel defined as
\begin{equation}\label{kernelh2psi1}
      \mathcal{I}^{h^{(2)}\Psi}(x,k,v,u) = T_h(x)  I^{h^{(2)}\Psi}(x , k,v,u) \, .
\end{equation}
Finally, the dimensionless power spectrum reads
\begin{align}\label{psh2psi1}
    \begin{split}
         \mathcal{P}^{(13)}_{h^{(2)}\Psi}(\eta ,k) &= \frac{3}{128} \left ( \mathcal{P}^{R}_h (\eta, k) +  \mathcal{P}^{L}_h (\eta, k) \right ) \int _0^{\infty} {\rm d} v  \int _{|1-v|}^{1+v} {\rm d} u \, \frac{1}{v^3u^2} \bigg [ \left (u^2-(v+1)^2 \right )^4 \\
         &+ \left (u^2-(v-1)^2 \right )^4  \bigg ]  \mathcal{I}^{h^{(2)}\Psi}(x,k,v,u)  P_{\mathcal{R}} (ku) \, .
    \end{split}
\end{align}
The kernel $I^{h^{(2)}\Psi}(x , k,v,u)$, defined in Eq.~\eqref{kernelh2psi1}, is an integral over the function $f^{h^{(2)\Psi}}(\overline{x},k,v,u)$, which was previously defined in Eq.~\eqref{ffuntionh2psi1}. In our new coordinates it is given by
\begin{align} \label{fh2psi1final}
    \begin{split}
        &f^{h^{(2)\Psi}}(\overline{x},k,v,u) = \left ( \frac{3+3w}{5+3w} \right )^2 \bigg ( -8k^2 T_{\Psi} ( \barx u) T_h(\barx) T_{\Psi} ( \barx u)  +4k^2(1+u^2-v^2) \\
        &\times  T_{\Psi} ( \barx u) T_h(\barx) T_{\Psi} ( \barx u) + 4 \frac{k^2}{\barx}\frac{\partial}{\partial \barx} T_{\Psi} ( \barx u) T_h(\barx) T_{\Psi} ( \barx u) \\
        &-4(1-w)k^2u^2 T_{\Psi} ( \barx u) T_h(\barx) T_{\Psi} ( \barx u) -8v^2k^2I^{h^{(2)}}_{st}(\barx,k,v,u)T_{\Psi} ( \barx u) \\
        &+8\frac{k^2}{\barx} I^{h^{(2)}}_{st}(\barx,k,v,u)\frac{\partial}{\partial \barx}T_{\Psi} ( \barx u) + 4(w-1)k^2u^2 I^{h^{(2)}}_{st}(\barx,k,v,u)T_{\Psi} ( \barx u) \\
        &+4k^2(-1+u^2-v^2)I^{h^{(2)}}_{st}(\barx,k,v,u)T_{\Psi} ( \barx u)  \bigg ) \, .
    \end{split}
\end{align}
Finally, the nested kernels can be expressed as 
\begin{align}
    \begin{split}
        I^{h^{(2)}}_{st}(\barx , k ,v,u) &= \frac{v^{b-2}}{\barx ^{b}k^2} \int _0^{\barx v} {\rm d}  \tilde{y} \, \, \tilde{y}^{2+b} \big ( j_b(\tilde{y})y_b(v\barx) - y_b(\tilde{y}) j_b(v\barx) \big )f^{h^{(2)}}_{st}(\tilde{y},k,v,u) \, ,
    \end{split}
\end{align}
with
\begin{align}
    \begin{split}
    f^{h^{(2)}}_{st}(\tilde{y} ,k, v,u)&=-2k^2T_h\left (\frac{\tilde{y}}{v} \right ) T_{\Psi}\left (\frac{\tilde{y}u}{v} \right )+k^2(1+u^2-v^2) T_h\left (\frac{\tilde{y}}{v} \right ) T_{\Psi}\left (\frac{\tilde{y}u}{v} \right ) \\
    &+ 2\frac{v^2k^2}{\tilde{y}}T_h\left (\frac{\tilde{y}}{v} \right ) T_{\Psi}\left (\frac{\tilde{y}u}{v} \right )-k^2u^2(1-w)T_h\left (\frac{\tilde{y}}{v} \right ) T_{\Psi}\left (\frac{\tilde{y}u}{v} \right )\, .
    \end{split}
\end{align}

\section{Results in a radiation dominated universe for a peaked input power spectrum}\label{sec5}
In the previous section, we derived expressions for the various contributions to the power spectrum arising from the correlation between third-order iGWs and first-order GWs. Our results, so far, hold for a general constant equation of state $w \neq  -1/3$. We now proceed to derive the spectral density of these GWs in a RD universe, setting $w=1/3$ or $b=0$ henceforth. The spectral density of the iGWs, at the time of creation, defined in Eq.~\eqref{spectraldensitydef}, consists of taking the time average of the power spectrum arising from the different contributions
\begin{equation}\label{spectraldensityallcontributions}
    \Omega ^{(13)} (\eta,k) =\frac{1}{9}x^2 \left ( \overline{\mathcal{P}^{(13)}_{\Psi \Psi h}(\eta,k)} +   \overline{\mathcal{P}^{(13)}_{\Psi ^{(2)}\Psi}(\eta,k)} + \overline{\mathcal{P}^{(13)}_{B ^{(2)}\Psi}(\eta,k)} + \overline{\mathcal{P}^{(13)}_{h ^{(2)}\Psi}(\eta,k)}\right )  \, .
\end{equation}
It is related to the present day spectral density, $\Omega (\eta _0,k)$, by a dilution factor, $d=1.62\times10^{-5}$, such that $\Omega (\eta _0,k)h^2=d\times\Omega (\eta,k)$, where $h$ is the reduced Hubble’s constant \cite{Domenech:2023fuz}. 

Eq.~\eqref{spectraldensityallcontributions} shows that we need to take the time average of the power spectra we have calculated. Since the time dependence is entirely contained within the kernels, our task now is to compute their time-averaged values, which we will do in this section. As we will see, terms involving nested kernels will need to be computed numerically; however, some terms can still be evaluated analytically. Hence, where possible, we will split kernels into an analytical and a numerical part.   

The explicit forms of the transfer functions at first order in a RD universe read
\begin{equation}
    T_h(x)=\frac{\sin x}{x}, \quad T_{\Psi}(x)=\frac{9}{x^2} \left (\frac{\sqrt{3}}{x}\sin{\frac{x}{\sqrt{3}}} - \cos{\frac{x}{\sqrt{3}}} \right )\, .
\end{equation}
Additionally, we can now provide the Green's function used for the third-order iGWs, as defined in the solution Eq.~\eqref{GWsol3rdorder},
\begin{equation}
    G^h_k(\eta , \bareta) = \frac{\sin (x - \barx)}{k} \Theta (\eta - \bareta) \, .
\end{equation}
Hence, the kernel we have defined for each contribution, $\mathcal{I}^i(x ,k,v,u)$ (here $i$ refers to the type of source), can be expanded as
\begin{align}\label{kernelspit}
    \begin{split}
        \mathcal{I}^i(x ,k,v,u)&=T_h(x)I^i(x,k,v,u) = \frac{\sin x}{x^2k^2} \int _0^{x } {\rm d}  \barx \, \barx \sin (x-\barx) f^i(\barx,k,v,u) \\
        &=\frac{\sin ^2x}{x^2k^2}\int _0^{x} {\rm d} \barx \, \barx \cos\barx f^i(\barx,k,v,u) - \frac{\sin x \cos x}{x^2k^2}\int _0^{x} {\rm d} \barx \, \barx \sin\barx f^i(\barx,k,v,u) \, .
    \end{split}
\end{align}
As mentioned previously, we need to take the time average of these kernels, and they encapsulate all the time dependence. Moreover, we are interested in these waves at present times, that is, when the iGWs are deep inside the horizon and the kernels have stabilized to a constant value. This allows us to take the limit $\eta\rightarrow \infty$ or $x\rightarrow \infty$ (for some fixed value of $k$). As a result, all remaining time dependence resides in the prefactors of the integrals.   Using the oscillation averages $\overline{\sin ^2x} = 1/2$ and $\overline{\sin x \cos x}=0$, we conclude that the second term above does not contribute to the SGWB. Accordingly, what we need to calculate is 
\begin{equation}\label{finalkernel}
    x^2\overline{\mathcal{I}^i(x\rightarrow\infty ,k,v,u)} = \frac{1}{2k^2}\int _0^{x \rightarrow \infty} {\rm d} \barx \, \barx \cos\barx f^i(\barx,k,v,u) \, ,
\end{equation}
where we recall the functions $f^i(\barx,k,v,u)$ depend on the source term and are given in Eqs.~\eqref{sshfunctionsfinal}, \eqref{fpsi2psi1final}, \eqref{fb2psi1final}, and \eqref{fh2psi1final}. We note that Ref.~\cite{Chen:2022dah} examines the impact of third-order contributions on the IR part of the iGW power spectrum, whereas here we examine the impact on the spectral density. The two terms in Eq.~\eqref{kernelspit} contribute to the power spectrum but only the first term contributes to the spectral density. The kernels and nested kernels involve integrals of products of highly oscillatory and decaying functions. As a result, the analytical solutions naturally involve trigonometric integral functions, defined as
\begin{equation}
    Si(z) = \int _0^z \frac{\sin t}{t} dt \, , \quad Ci(z) = -\int _z^{\infty} \frac{\cos t}{t} dt \, .
\end{equation}
Importantly, both have finite limits at infinity: $\lim_{z\to\infty} Si(z) = \pm\frac{\pi}{2}$ and $\lim_{z\to\infty} Ci(z) =0$. As we will see, the nested kernels depend on products of these trigonometric integrals, making it impossible to derive closed-form analytical expressions for terms containing nested kernels. Consequently, we will evaluate such kernels numerically\footnote{We do this using a fourth-order Runge-Kutta method and have tested our code by comparing numerical kernels for SIGWs and scalar-tensor iGWs to the known analytical ones.}.

\subsection{Kernel computations}
In this subsection, we compute the kernels needed throughout our work. Firstly, we look at the kernels coming from the pure first-order contribution and then move on to the second-order kernels, which involve numerical computations.

There are three kernels we need to compute for the contribution made of the scalar-scalar-tensor interaction. With the formalism established, our next step is simply to substitute in Eq.~\eqref{sshfunctions} into Eq.~\eqref{finalkernel}. This leads to
\begin{subequations}
    \begin{align}
        \begin{split}
            x^2\overline{\mathcal{I}^{ssh}_{s,1} (x\rightarrow \infty , k, v)} &= -\frac{1}{315v^6} \bigg (v^2 \left ( 648+1179v^2-739v^4\right ) +\sqrt{3}v^3 \left (1890-672v^2+11v^4 \right ) \\
            &\times \left | \frac{\sqrt{3}-v}{\sqrt{3}+v} \right | +9 \left (216-903v^2+280v^4 \right ) \log \left | \frac{v^2-3}{3} \right | \bigg ) \, ,
        \end{split}
        \\
        \begin{split}
            x^2\overline{\mathcal{I}^{ssh}_{s,2}(x\rightarrow \infty , v)} &=\frac{1}{30v^6k^2} \bigg (3v^2 \left ( 6+11v^2\right ) -2\sqrt{3}v^3 \left (v^2-15 \right )\left | \frac{\sqrt{3}-v}{\sqrt{3}+v} \right | \\
            &+18 \left (3-5v^2 \right ) \log \left | \frac{v^2-3}{3} \right | \bigg ) \, ,
        \end{split}
        \\
        \begin{split}
            x^2\overline{\mathcal{I}^{ssh}_{s,4}(x\rightarrow \infty , v)} &=\frac{1}{120v^6k^2} \bigg (6v^2 \left ( 62v^2-3\right ) +\sqrt{3}v^3 \left (285-13v^2 \right ) \left | \frac{\sqrt{3}-v}{\sqrt{3}+v} \right | \\
            &-9 \left (6+65v^2+5v^4 \right ) \log \left | \frac{v^2-3}{3} \right | \bigg ) \, .
        \end{split}
    \end{align}
\end{subequations}
We also observe that these kernels remain well-behaved and finite in the limit $v\rightarrow\sqrt{3}$, exhibiting no logarithmic divergence. A similar behaviour was found for second-order scalar-tensor induced gravitational waves, as shown in Refs.~\cite{Bari:2023rcw} and \cite{Picard:2023sbz}. 

We shift our focus to computing the kernels from source terms that couple second-order perturbations, which will involve nested kernels. For second-order scalars coupled to first-order fluctuations, the function we have to integrate over is Eq.~\eqref{fpsi2psi1final}, which depends on the nested kernel defined in Eq.~\eqref{psi2nested} 
\begin{align}
    \begin{split}
        I^{\Psi ^{(2)}}_{st}(\barx , k ,v,u)&= \frac{1}{k^2v^3\barx ^3}   \int _0^{v\barx} {\rm d} \tildey \, \, \tildey \Bigg ( 3(-v\barx+\tildey)\cos \left ( \frac{v\barx-\tildey}{\sqrt{3}} \right ) +\sqrt{3}(3+v\barx\tildey) \sin \left ( \frac{v\barx-\tildey}{\sqrt{3}} \right )  \Bigg ) \\
        &\times\Sigma ^{st}(\tilde{y},v,u)  \, , 
    \end{split}
\end{align}
in a RD universe. We further split this up as
\begin{align}
    \begin{split}
        &I^{\Psi ^{(2)}}_{st}(\barx , k ,v,u)=  \frac{1}{k^2v^3\barx ^3}  \Bigg ( 3\cos \left ( \frac{v\barx}{\sqrt{3}} \right ) I^{\Psi ^{(2)}}_{c2}(\barx ,v,u) -3v\barx \cos \left ( \frac{v\barx}{\sqrt{3}} \right ) I^{\Psi ^{(2)}}_{c1}(\barx ,v,u) \\
        &-3\sqrt{3} \cos \left ( \frac{v\barx}{\sqrt{3}} \right ) I^{\Psi ^{(2)}}_{s1}(\barx ,v,u) -\sqrt{3}v\barx  \cos \left ( \frac{v\barx}{\sqrt{3}} \right ) I^{\Psi ^{(2)}}_{s2}(\barx ,v,u) \\
        & + 3\sqrt{3} \sin \left ( \frac{v\barx}{\sqrt{3}} \right ) I^{\Psi ^{(2)}}_{c1}(\barx ,v,u)+\sqrt{3} v\barx \sin \left ( \frac{v\barx}{\sqrt{3}} \right ) I^{\Psi ^{(2)}}_{c2}(\barx ,v,u)  \\
        &+3 \sin \left ( \frac{v\barx}{\sqrt{3}} \right ) I^{\Psi ^{(2)}}_{s2}(\barx ,v,u) -  3v\barx \sin \left ( \frac{v\barx}{\sqrt{3}} \right ) I^{\Psi ^{(2)}}_{s1}(\barx ,v,u) \Bigg ) \, ,
    \end{split}
\end{align}
where we have defined
\begin{subequations}
    \begin{align}
        I^{\Psi ^{(2)}}_{c1}(\barx ,v,u)&= \int _0^{v\barx} {\rm d} \tildey \, \cos \left ( \frac{\tildey}{\sqrt{3}} \right ) \tildey \,  \Sigma ^{st}(\tilde{y},v,u)  \, ,
        \\
        I^{\Psi ^{(2)}}_{s1}(\barx ,v,u)&=  \int _0^{v\barx} {\rm d} \tildey \, \sin \left ( \frac{\tildey}{\sqrt{3}} \right ) \tildey \, \Sigma ^{st}(\tilde{y},v,u) \, ,
        \\
        I^{\Psi ^{(2)}}_{c2}(\barx ,v,u)&= \int _0^{v\barx} {\rm d} \tildey \, \cos \left ( \frac{\tildey}{\sqrt{3}} \right ) \tildey ^2 \,  \Sigma ^{st}(\tilde{y},v,u)  \, ,
        \\
        I^{\Psi ^{(2)}}_{s2}(\barx ,v,u)&= \int _0^{v\barx} {\rm d} \tildey \, \sin \left ( \frac{\tildey}{\sqrt{3}} \right ) \tildey ^2 \,  \Sigma ^{st}(\tilde{y},v,u) \, .
    \end{align}
\end{subequations}
The analytical results for $I^{\Psi ^{(2)}}_{c1}(\barx ,v,u)$, $I^{\Psi ^{(2)}}_{s1}(\barx ,v,u)$, $I^{\Psi ^{(2)}}_{c2}(\barx ,v,u)$ and $I^{\Psi ^{(2)}}_{s2}(\barx ,v,u)$ in a RD universe can be found in Eqs.~\eqref{Ic1RDfinal}, \eqref{Is1RDfinal}, \eqref{Ic2RDfinal} and \eqref{Is2RDfinal} respectively. These functions then need to be substituted back into Eq.~\eqref{fpsi2psi1final} and then integrated over in Eq.~\eqref{finalkernel}. Unfortunately, there is no analytical solutions for these functions. However, not every term in Eq.~\eqref{fpsi2psi1final} depends on these nested kernels, and we are able to compute analytical expressions for the terms that contain no nested kernels. This means we split up $ f^{\Psi ^{(2)}\Psi}(\barx,k,v,u)$ into an analytical part and numerical part:
\begin{equation}
    f^{\Psi ^{(2)}\Psi}(\barx,k,v,u) = f_a^{\Psi ^{(2)}\Psi}(\barx,k,v,u) + f_n^{\Psi ^{(2)}\Psi}(\barx,k,v,u) \, ,
\end{equation}
where
\begin{subequations}
    \begin{align}
        \begin{split}
            f_a^{\Psi ^{(2)}\Psi}(\barx,k,v,u)& =   \frac{24 \sin \barx}{k^2 u^4 v^4 \barx^7}  \sin\left(\frac{u \barx}{\sqrt{3}}\right) \Big(-\sqrt{3} u \barx (-18 + (1 + u^2 - 3 v^2) \barx^2) \cos\left(\frac{u \barx}{\sqrt{3}}\right)  \\
            &+ 3 (-18 + (1 + 3 u^2 - 3 v^2) \barx^2) \sin\left(\frac{u \barx}{\sqrt{3}}\right) \Big) \, ,
        \end{split}
        \\
        \begin{split}\label{finalkernelpsi2}
            f_n^{\Psi ^{(2)}\Psi}(\barx,k,v,u)& = \frac{4}{3 u^3 \bar{x}^2}  \bigg(6 u \bar{x} \cos\left(\frac{u \bar{x}}{\sqrt{3}}\right) + \sqrt{3} \big(-6 + u^2 \bar{x}^2\big) \sin\left(\frac{u \bar{x}}{\sqrt{3}}\right)\bigg) \frac{\partial}{\partial \barx} I^{\Psi ^{(2)}}_{st}(\barx , k ,v,u) \\
            &+\frac{4}{\sqrt{3} u \bar{x}} \sin\left(\frac{u \bar{x}}{\sqrt{3}}\right)
            I^{\Psi ^{(2)}}_{st}(\barx , k ,v,u) \, .
        \end{split}
    \end{align}
\end{subequations}
The analytical contribution once integrated is
\begin{align}\label{analyticalkernelpsi2psi1}
    \begin{split}
        x^2&\overline{\mathcal{I}_a^{\Psi^{(2)}\Psi}(x\rightarrow\infty ,k,v,u)}= \frac{1}{30 k^4 u^4 v^4} \Bigg( 
        84 u^4 + u^2 (984 - 360 v^2) + (828 + 540 u^2 - 540 v^2) \log 3 \\
        &+ \big [(-828 - 540 u^2 + 540 v^2 + 
        \sqrt{3} (-630 u - 80 u^3 + 6 u^5 + 270 u v^2 - 30 u^3 v^2) \big ] 
        \log \left |\sqrt{3} + u \right | \\
        &+ \big [-828 - 540 u^2 + 540 v^2 + 
        \sqrt{3} (630 u + 80 u^3 - 6 u^5 - 270 u v^2 + 30 u^3 v^2) \big ] 
        \log \left| \sqrt{3} - u \right| \Bigg) \, .
    \end{split}
\end{align}

For the kernels involving second-order vectors there is no `analytical' part and so we do everything numerically. In a RD universe, Eq.~\eqref{fb2psi1final} becomes
\begin{align}\label{kernelb2rd}
    \begin{split}
        f_{\textit{i}}^{B^{(2)\Psi}}(\barx,k,v,u) &=\frac{1}{3 k u^3 v^2 \bar{x}^4} \bigg( 6 u \bar{x} \big(4 + v^2 \bar{x}^2\big) \cos\left (\frac{u \bar{x}}{\sqrt{3}}\right ) + \sqrt{3} \big(-24 + \bar{x}^2 \big(-6 v^2  \\
        &+ u^2 \big(8 + v^2 \bar{x}^2\big)\big)\big) \sin\left(\frac{u \bar{x}}{\sqrt{3}} \right ) \bigg)
        I^{B^{(2)}}_{sti}(\overline{x} ,k,v,u)\\
        &+ \frac{1 }{k u^3 v^2 \bar{x}^3} 8 \sqrt{3} \sin\left(\frac{u \bar{x}}{\sqrt{3}}\right)-8 u \bar{x} \cos\left(\frac{u \bar{x}}{\sqrt{3}}\right) \frac{\partial}{\partial \barx} I^{B^{(2)}}_{sti}(\overline{x} ,k,v,u) \, ,
    \end{split}
\end{align}
and expressions for the nested kernels can be found in Eq.~\eqref{RDnestedb2psi11} for $f_1^{B^{(2)\Psi}}(\barx,k,v,u)$ and Eq.~\eqref{RDnestedb2psi12} for $f_2^{B^{(2)\Psi}}(\barx,k,v,u)$.

Finally, for the background generated by second-order tensors coupled to first-order scalars, we can similarly split up the kernel Eq.~\eqref{fh2psi1final} into an analytical contribution and a numerical contribution. We define the analytical and numerical parts as
\begin{subequations}
    \begin{align}
        \begin{split}
        f_a^{h^{(2)\Psi}}(\barx,k,v,u) &= \frac{48 k^2 \sin \barx }{u^6 \barx^9}\left( u \barx \cos\left(\frac{u \barx}{\sqrt{3}}\right) - \sqrt{3} \sin\left(\frac{u \barx}{\sqrt{3}}\right) \right) \bigg ( u \barx (-18 + (-3 + u^2 - 3 v^2)\barx^2) \\
        &\times  \cos\left(\frac{u \barx}{\sqrt{3}}\right) + 3 \sqrt{3} (6 + (1 - u^2 + v^2)\barx^2) \sin\left(\frac{u \barx}{\sqrt{3}}\right) \bigg ) \, , 
        \end{split}
        \\
        \begin{split} \label{finalkernelh2}
         f_n^{h^{(2)\Psi}}(\barx,k,v,u) &= \frac{12 k^2}{u^3 \barx^5} \bigg ( ( u \barx \left(18 + (3 - u^2 + 9 v^2) \barx^2 \right) \cos\left(\frac{u \barx}{\sqrt{3}}\right) + 3 \sqrt{3} \left(-6 + (-1 + u^2 - 3 v^2) \barx^2 \right) \\
         &\sin\left(\frac{u \barx}{\sqrt{3}}\right) \bigg ) I^{h^{(2)}}_{st}(\barx ,k,v,u) \, .
        \end{split}
    \end{align}
\end{subequations}
Substituting $f_a^{h^{(2)\Psi}}(\barx,k,v,u)$ into Eq.~\eqref{finalkernel}, we arrive to  
\begin{align}\label{analyticalkernelh2psi1}
    \begin{split}
         x^2&\overline{\mathcal{I}_a^{h^{(2)}\Psi}(x\rightarrow\infty ,v,u)}=\frac{2}{315 u^6} \Bigg( 127 u^6 + 324 (3 + 7 v^2) \log 3 
        + 9 u^4 \big (25 - 77 v^2 + 140 \log 3 \big)  \\
        &- 54 u^2 \big (3 + 28 \log 3 + v^2 (7 + 70 \log 3) \big )  + 2 \bigg[-81 (3 + 7 v^2) + u^2 \bigg ( 189 (2 + 5 v^2)  + u \big[ 315 \sqrt{3} v^2  \\
        &+ u \big (-315 + \sqrt{3} u \{4 u^2 - 21 (4 + v^2)\}\big )\big]\bigg )\bigg ] 
        \log\left| 3 + \sqrt{3} u \right|- 2 \bigg [(81 (3 + 7 v^2) 
        + u^2 \bigg(-189 (2 \\
        &+ 5 v^2)  + u \big [ 315 \sqrt{3} v^2  + u \big (315 + \sqrt{3} u \{4 u^2 - 21 (4 + v^2)\}\big )\big]\bigg )\bigg ] 
        \log\left| -3 + \sqrt{3} u \right| \Bigg) \, .
    \end{split}
\end{align}
The nested kernel has been computed and is shown in Eq.~\eqref{RDnestedh2psi1}.
\subsection{Log-normal input power spectrum}
We are now ready to compute the spectral density of these contributions. We will look at an input peaked power spectrum for both the primordial scalar and tensor power spectra and look at the effect of varying the width of the peak. 

The peaked input power spectrum on small scales, for both scalars and tensors, is modelled by 
\begin{equation}
\label{lognorm_def}
    \mathcal{P}_{\mathcal{R} , h}= \frac{\mathcal{A}_{\mathcal{R} , h}}{\sqrt{2\pi}\sigma}\exp \left ( {-\frac{\log ^2(k/k_{\mathcal{R} , h})}{2\sigma ^2}}\right )\, ,
\end{equation}
where $\mathcal{A}_{\mathcal{R} /h}$ is the amplitude of input power spectra for scalars and tensors, respectively, $\sigma$ controls the width of the peak, and $k_{\mathcal{R} , h}$ is the location of the peak in k-space. In direct continuation of the work previously done in Ref.~\cite{Picard:2023sbz}, the primordial scalar and tensor power spectrum peaks in the centre LISA band: $f_{\mathcal{R}}=f_h=3.4$ mHz and we take $\mathcal{A}_{\mathcal{R} }=10\mathcal{A}_{ h}\approx2\times 10^{-2}$.

The integrands are calculated numerically, using the numerical integration package \texttt{vegas+}~\cite{Lepage:2020tgj}. We transform to the coordinate system $(s,t)$ to perform these integrals, where
\begin{equation}
    s=u-v \, , \quad t = u+v-1 \, .
\end{equation}
This has the advantage of making the integration region rectangular. With these variables, the integral becomes
\begin{equation}
    \int _0^{\infty} {\rm d} v  \int _{|1-v|}^{1+v} {\rm d} u = \frac{1}{2}  \int _0^{\infty} {\rm d} t  \int _{1}^{-1} {\rm d} s \, .
\end{equation}

\begin{figure}[h!]
    \centering
    \includegraphics[width=\linewidth]{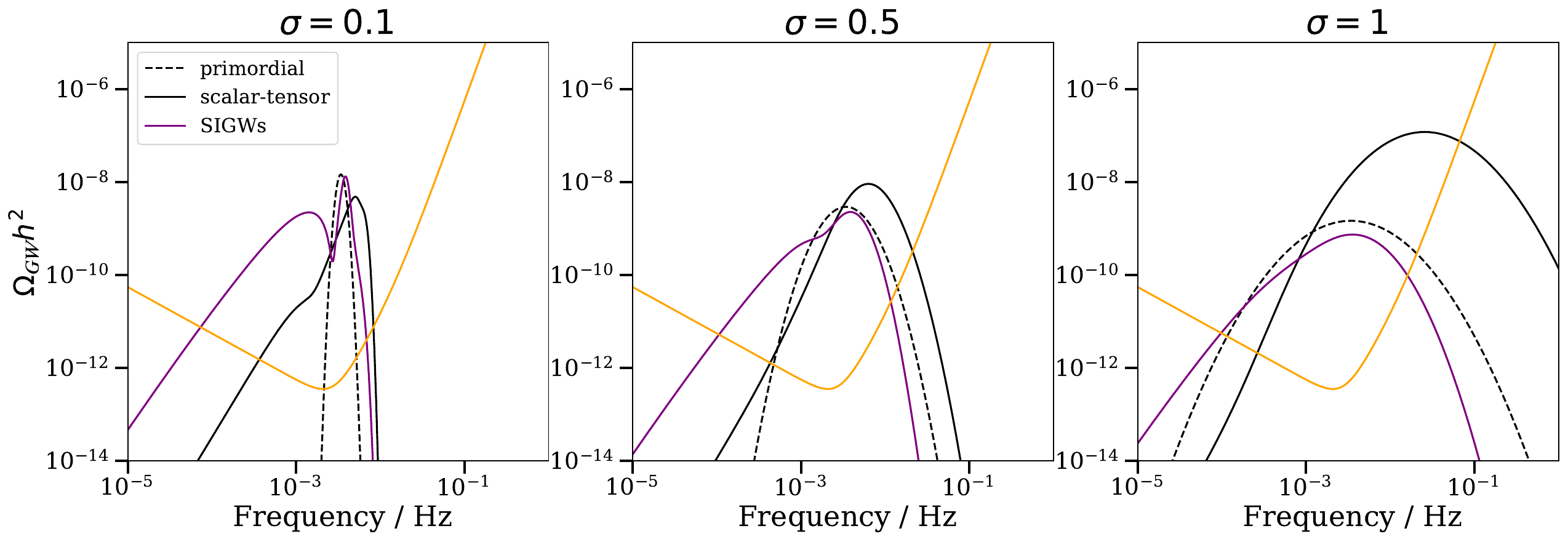}
    \caption{\footnotesize{Plot of the spectral density against frequency, where we look at SIGWs (solid purple) and scalar-tensor iGWs at second order (solid black). The dashed black lines correspond to primordial GWs and the orange curve is the LISA sensitivity \cite{Moore:2014lga, Sathyaprakash:2009xs}. Each graph corresponds to a different value of $\sigma$, increasing as we go towards the right. Regardless, of the value of $\sigma$ we would expect the higher order contributions to converge. This is not the case for the scalar-tensor contribution (as opposed to the SIGWs) which becomes bigger for increasing $\sigma$. This `unphysical' enhancement was discussed in Refs.~\cite{Bari:2023rcw, Picard:2023sbz}. Additionally, we note that the integrand for SIGWs is well behaved in the IR and UV limits.}}
    \label{fig:secondorderplots}
\end{figure}

In Fig.~\ref{fig:secondorderplots} we have plotted the (current-day) spectral density of the scalar-tensor iGWs and SIGWs for reference. 
The expressions for the time-averaged power spectrum can respectively be found in Eqs.~\eqref{powerspectrumstigws} and \eqref{powerspectrumsigws}. Notably, as the width of the primordial peak increases, the scalar-tensor contribution seems enhanced. This is not the case for SIGWs. This `unphysical-enhancement' was attributed to a divergence of the integrand in Eq.~\eqref{powerspectrumstigws} when taking certain limits. The IR limit ($k\rightarrow 0$) of the scalar-tensor integrand is finite. However, in the UV limit ($k\rightarrow\infty$) there are two distinct limits: $v\rightarrow 1$ and $u\rightarrow 0$ (or $t\rightarrow 0$ and $s\rightarrow -1$) which couples large wavelength scalar perturbations with short wavelength tensor modes and $v\rightarrow 0$ and $u\rightarrow 1$ (or $t\rightarrow 0$ and $s\rightarrow 1$) which couples short wavelength scalar perturbations with long wavelength tensor modes. Ref.~\cite{Bari:2023rcw} showed that the integrand blows up as $u^{-4}$ in the $v\rightarrow 1$ and $u\rightarrow 0$ limit. In the opposite UV limit, Ref.~\cite{Picard:2023sbz} found a $v^{-2}$ divergence. We note that, when the input primordial power spectrum is sufficiently peaked, the enhancements are not an issue since the spectrum is sufficiently suppressed away from the peak. However, as the peak of the scalar power spectrum is widened, we start integrating over more and more scales, which include the UV scales where the rest of the integrand diverges. This results in an unphysical enhancement of the observable. In App.~\ref{meshstuff} we present mesh plots of the integrands from the new numerical contributions we will consider. For reference, we have also a mesh plot of the scalar-tensor contribution in Fig.~\ref{fig:st_meshplots}. The enhancements of the integrand appear in the limits we have discussed. We would like to emphasise that, from the mesh plot, we see that only the $u^{-4}$ divergence seems to enhance the observable, which corresponds to scalar sector.

\begin{figure}[h!]
    \centering
    \includegraphics[width=\linewidth]{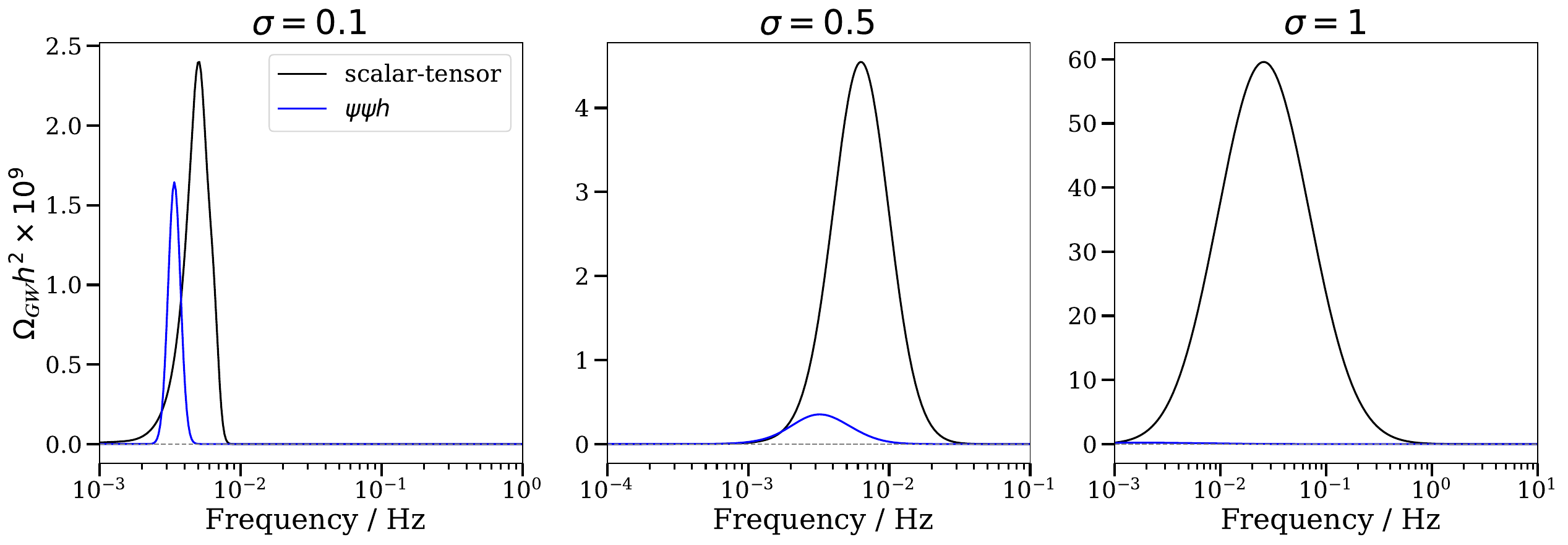}
    \caption{\footnotesize{Contribution of the scalar-scalar-tensor source term to the iGW background. We have plotted the spectral density enhanced by a factor of $10^9$ (on a linear scale) against the frequency (on a logarithmic scale). We see that these induced waves are subdominant compared to the scalar-tensor iGWs and also decrease in amplitude as $\sigma$ gets larger.}}
    \label{fig:sshvaryingsigma}
\end{figure}

\begin{figure}[h!]
    \centering
    \includegraphics[width=\linewidth]{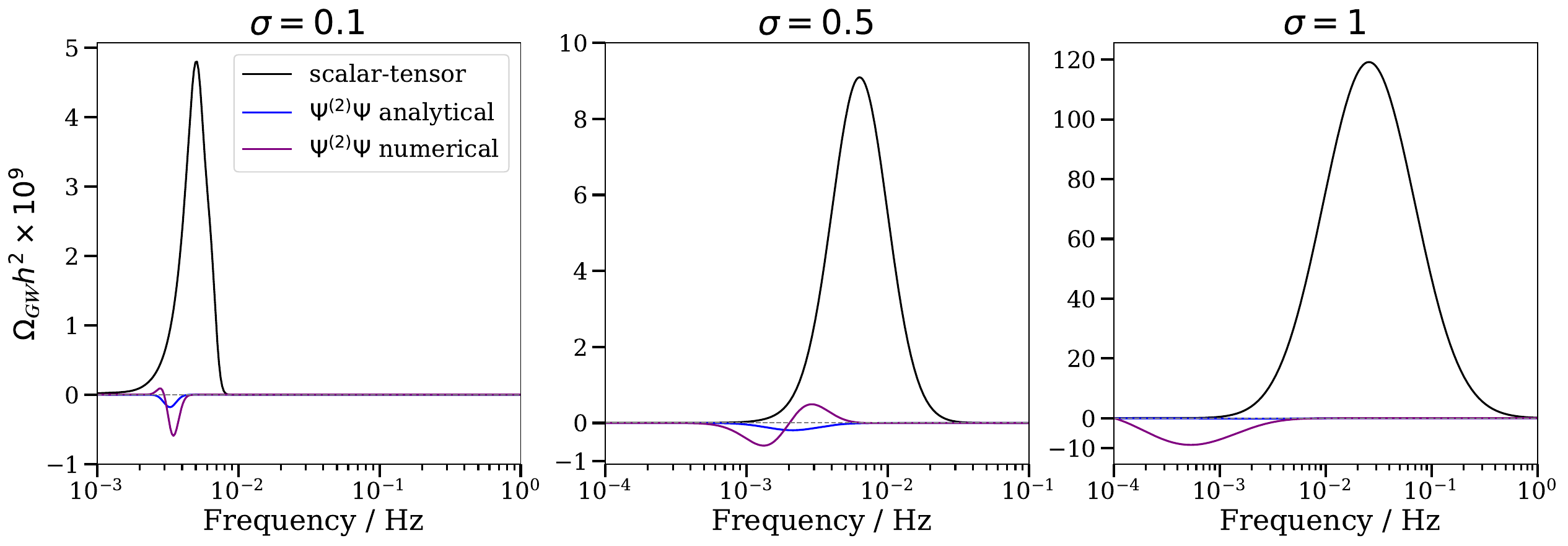}
    \caption{\footnotesize{Contribution to the iGW background originating from second-order scalar perturbations coupled to first-order scalar modes. We have plotted the numerical contribution (purple), analytical contribution (blue) and second-order scalar-tensor induced waves (black). This new contribution acts as a suppression of the signal on certain scales when $\sigma$ is small and the suppression gets larger whilst increasing the width of $\sigma$. The numerical and analytical contributions are different, with the numerical contribution dominating in magnitude. When $\sigma =1$ the spectral density is heavily suppressed for lower frequencies.}}
    \label{fig:psi2psi1varyingsigma}
\end{figure}

\begin{figure}[h!]
    \centering
    \includegraphics[width=\linewidth]{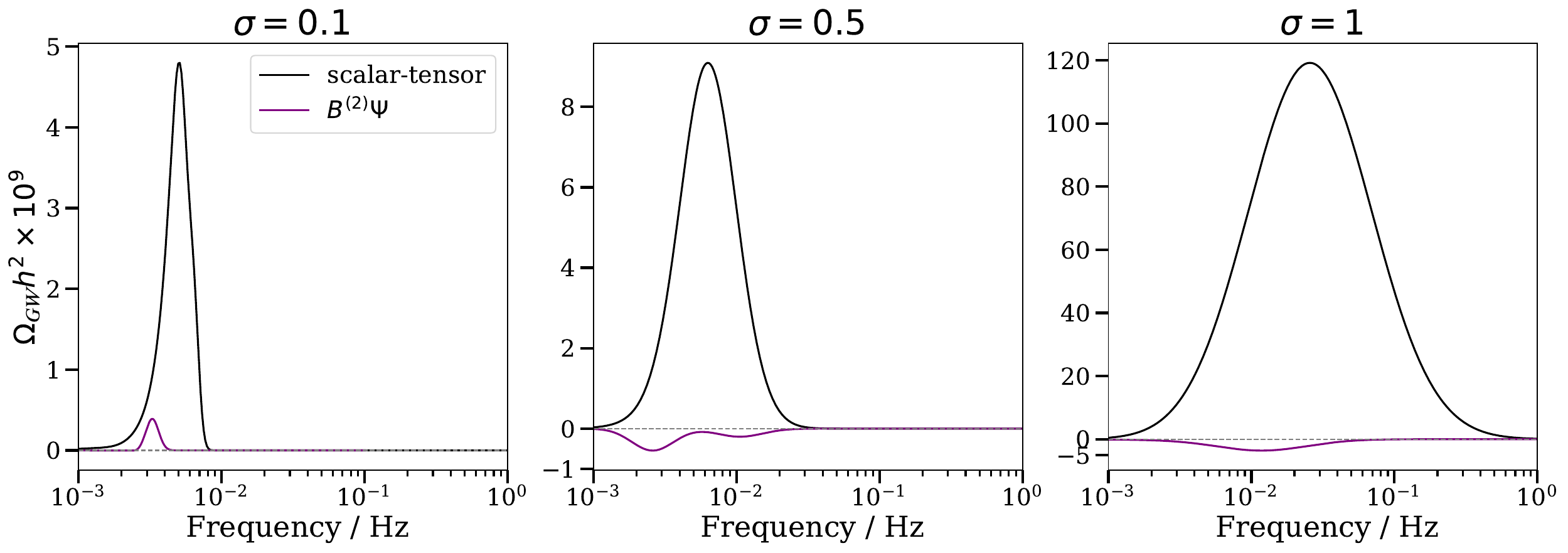}
    \caption{\footnotesize{Contribution to the iGW background originating from second-order vector perturbations coupled to first-order scalar modes (purple). For small input power spectra ($\sigma=0.1$) the contribution in positive but subdominant compared to the scalar-tensor iGWs. However, as $\sigma$ gets large the contribution becomes negative and acts as suppression to the spectral density. The suppression occurs at a different scale than the waves sourced by second-order scalar perturbations and is also subdominant in magnitude.}}
    \label{fig:b2psi1varyingsigma}
\end{figure}

\begin{figure}[h!]
    \centering
    \includegraphics[width=\linewidth]{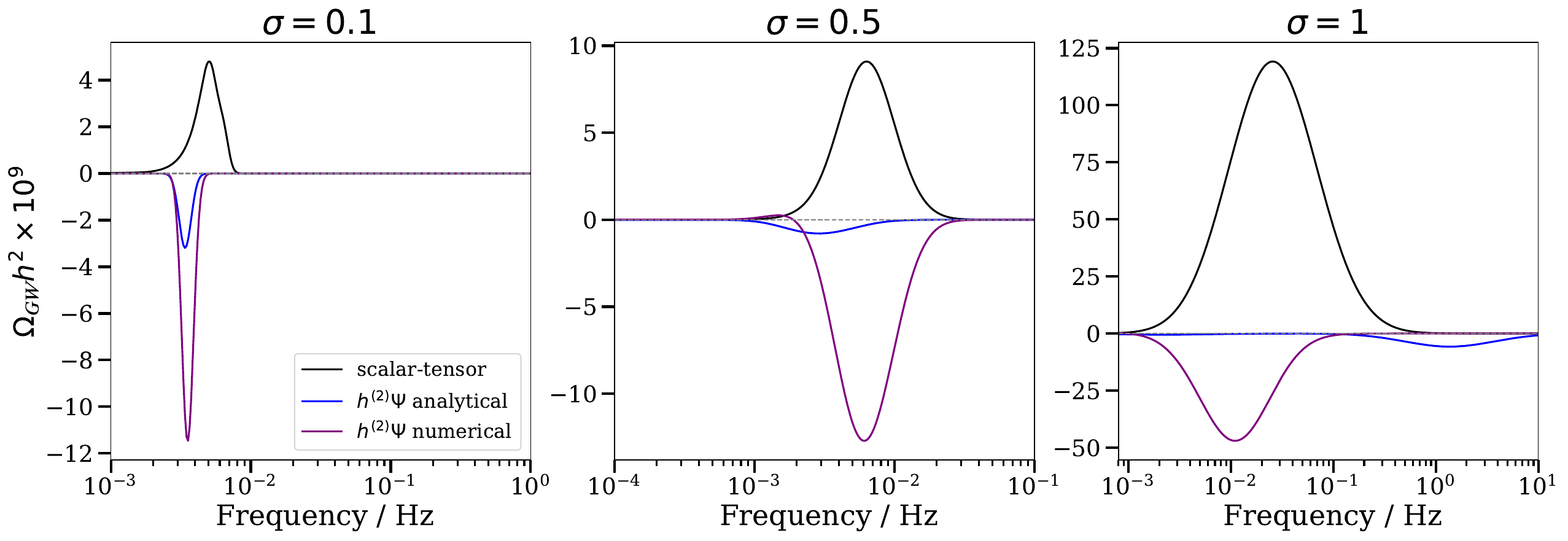}
    \caption{\footnotesize{Contribution to the iGW background originating from second-order tensor perturbations coupled to first-order scalar modes. Similarly to the previous plots, we differentiate the analytical contribution (blue) from the numerical contribution (purple). Strikingly, the magnitude of the suppression is large compared to the other contributions for both the analytical and numerical contributions, particularly from the numerical contribution.}}
    \label{fig:h2psi1varyingsigma}
\end{figure}

We now move on to computing the spectral density of the new contributions we have derived. Instead of using a logarithmic scale for the spectral density, as in Fig.~\ref{fig:secondorderplots}, we plot $\Omega_{GW} h^2 \times 10^9$ on a linear scale, since the contributions can take negative values. We first show the contributions from each individual term: Fig.~\ref{fig:sshvaryingsigma} shows the contribution coming from the scalar-scalar-tensor-induced waves, Fig.~\ref{fig:psi2psi1varyingsigma} shows the contribution coming from the second-order scalars coupled to first-order scalars, Fig.~\ref{fig:b2psi1varyingsigma} shows the contribution coming from the second-order vectors coupled to first-order scalars, and Fig.~\ref{fig:h2psi1varyingsigma} shows the contribution coming from second-order tensors and scalars. Where possible, we distinguish between contributions that we have computed numerically and analytically. 

Our results show a suppression of the scalar-tensor iGWs generated at second order from all the source terms including second-order perturbations. The contribution from pure first-order terms, i.e.~the scalar-scalar-tensor iGWs, is additive yet subdominant compared to the scalar-tensor iGWs. For iGWs generated by source terms which couple first-order scalar to second-order perturbations, we have differentiated between the `analytical contribution' (terms where we were able to compute an analytical kernel) and `numerical contribution' (terms where we were unable to compute an analytical kernel). There are two main differences between the analytical and numerical contributions: Firstly, the analytical contributions always acts as a suppression to the spectral density (i.e.~always negative or zero) whereas the numerical contributions can be positive at certain scales.  Secondly, the magnitude of the numerical contribution is always larger than its analytical counterpart. Furthermore, as $\sigma$ increases, the magnitude of both numerical and analytical contributions increase in magnitude. When this was the case for the scalar-tensor iGWs we concluded that this was due to how the integrand behaves in the UV limit and so we shall carry out a similar analysis here. For terms involving analytical expressions, we can track the limits of the integrands in the UV and IR limits, however, this is not possible for the terms where we have integrated terms numerically. Hence, when the kernels were computed numerically, we look at the mesh plots in App.~\ref{meshstuff} and try to draw some conclusions by comparing with the scalar-tensor mesh plot in Fig.~\ref{fig:st_meshplots}. For the third-order contributions, we have plotted the logarithm of the absolute values of the integrand since they can be positive and negative and then we can deduce if there is a positive or negative divergence by looking at the observable\footnote{There are ways to plot on logarithmic scales negative and positive values (for example by using \texttt{symlog}) however, the data spans lots of scales in magnitude and we found that the most informative way was to plot the absolute value of the data.}. In our analysis, we have kept the width of the primordial spectra equal, however, we insist that since we are only integrating over the primordial scalar power spectrum, if there is a divergence, it will come from there. 

For terms involving second-order scalars, we find that the analytical contribution is well behaved in the IR limit and in the UV limit. A potential problematic limit would be when $u=\sqrt{3}$ , then the integrand diverges as $v^{-3}$ and then integrating over $v=0$ would cause a divergence. However, this is not a region we integrate over (this corresponds to $s=\sqrt{3}\approx1.7$). To study the numerical behaviour of the integrand, we look at Fig.~\ref{fig:psi2psi1um_meshplots}. From the mesh plots we see that the kernel does get enhanced as $v\rightarrow0$ and $u\rightarrow 1$, however, this enhancement does not blow up as the width of the primordial scalar power spectrum gets larger. We are therefore unable to conclude that the contribution which involves second-order scalars diverges. 

The contribution coming from terms involving second-order vectors is fully numerical, the corresponding mesh plots of the integrand are shown in Fig.~\ref{fig:b2psi1um_meshplots}. Both kernels exhibit an enhancement in the limit of $t\rightarrow 0$ and $s\rightarrow -1$. We also notice diagonals in the mesh plots of the kernels which seem to be enhanced, which could correspond to constant values of $u$ entering a resonance (for example, a logarithmic divergence like in the case of SIGWS). However, we have plotted constant lines of multiples of $u$ and cannot come to the conclusion that these lines corresponds to a logarithmic resonance. Nevertheless, we see that as the primordial peak of the scalar spectrum gets wider, the enhancement in the UV limit stands out. We conclude from these mesh plots that this term diverges negatively in $v\rightarrow1$ and $u\rightarrow 0$ limit, at a slower rate than the scalar-tensor iGWS.

The analytical term coming from second-order tensors is finite in the IR and UV limit where $v\rightarrow 0$ $u\rightarrow1$. However, the over UV limit which couples large wavelength scalar modes with short wavelength tensor modes tends to $-u^{-2}$. We recall that scalar-tensor iGWs also had a similar divergence in this limit but as $u^{-4}$, hence we cannot conclude any `cancellation' coming from these terms. From the numerical mesh plots in Fig.~\ref{fig:h2psi1um_meshplots}, we can see an enhancement in the same limit which seems similar in magnitude to the one occurring in scalar-tensor iGWS, however, we cannot conclude its of magnitude $-u^{-4}$.

Parity violation in the primordial tensor power spectrum \cite{Obata:2016tmo,Bartolo:2017szm,Bartolo:2018elp} leaves observational signatures for scalar-tensor iGWs as was shown in Ref.~\cite{Bari:2023rcw} for Gaussian initial scalar conditions and also for non-Gaussian initial conditions in Ref.~\cite{Picard:2024ekd}. We note that for scalar-tensor iGWs, the tensor power spectrum appears in the integrand and is coupled to the polarisation tensors in a way which depends on parity. In the case of third-order iGWs the tensor power spectrum appears outside the integrand and the way in which it couples to the polarisation tensors is independent of the parity (i.e.~we get the same result whether $\lambda = R$ or $\lambda = L$), making the dependence on the parity of the initial spectrum trivial. For similar reasons, we do not consider peaks at different scales. 

We would like to clarify that the results presented are independent of any tunable parameters in the \texttt{vegas+}~\cite{Lepage:2020tgj} integration package used in our analysis. In particular, the final plots remain unchanged when varying the sampling strategy, the number of iterations (\texttt{nitn}), or the number of integrand evaluations per iteration (\texttt{neval}). Furthermore, we checked that our results are insensitive to integrating over more scales (i.e. changing the upper limit of the $t$ integrand in the (s,t) coordinates system) and also changing the lower limit of the $t$ integrand from $0$ to a value infinitely close to $0$ or to the value of $1$.

\begin{figure}[htbp]
    \centering
    \begin{subfigure}[b]{0.7\textwidth}
        \includegraphics[width=\linewidth]{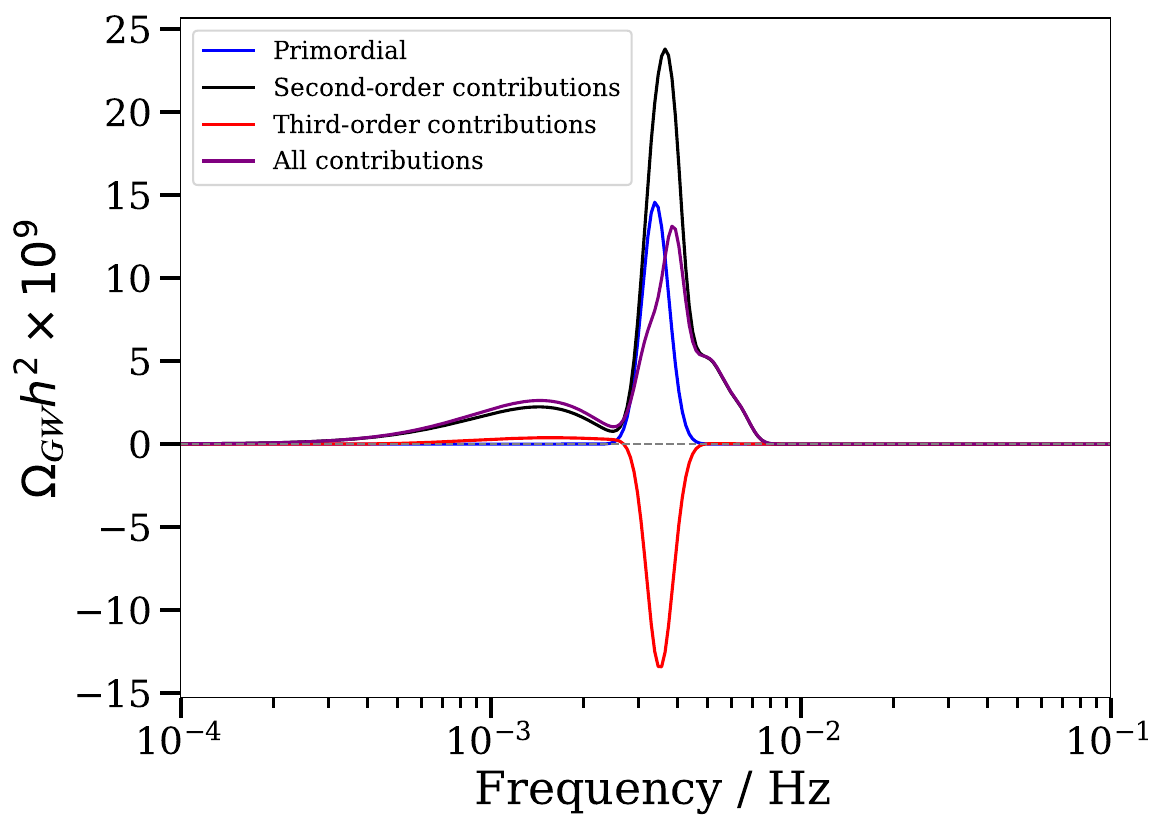}
    \end{subfigure}

    \vspace{0.5cm}
    
    \begin{subfigure}[b]{0.4\textwidth}
        \includegraphics[width=\linewidth]{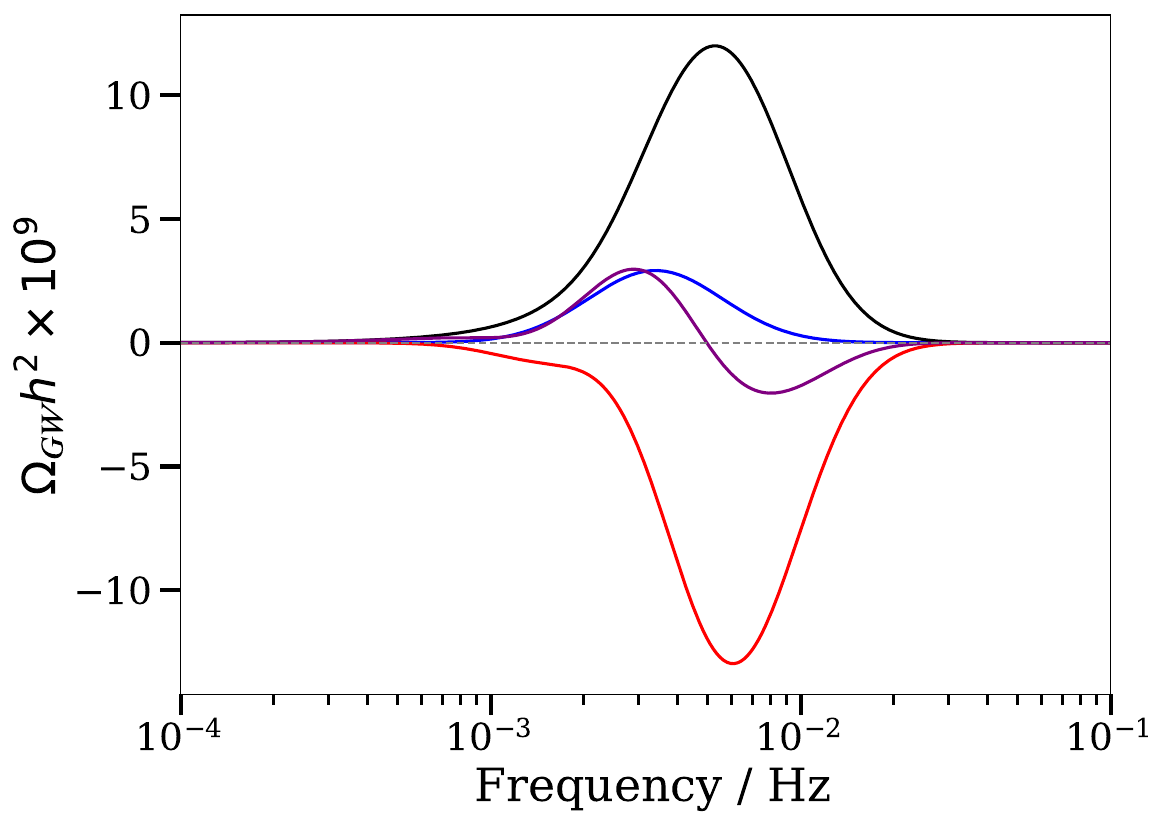}
    \end{subfigure}
    \begin{subfigure}[b]{0.4\textwidth}
        \includegraphics[width=\linewidth]{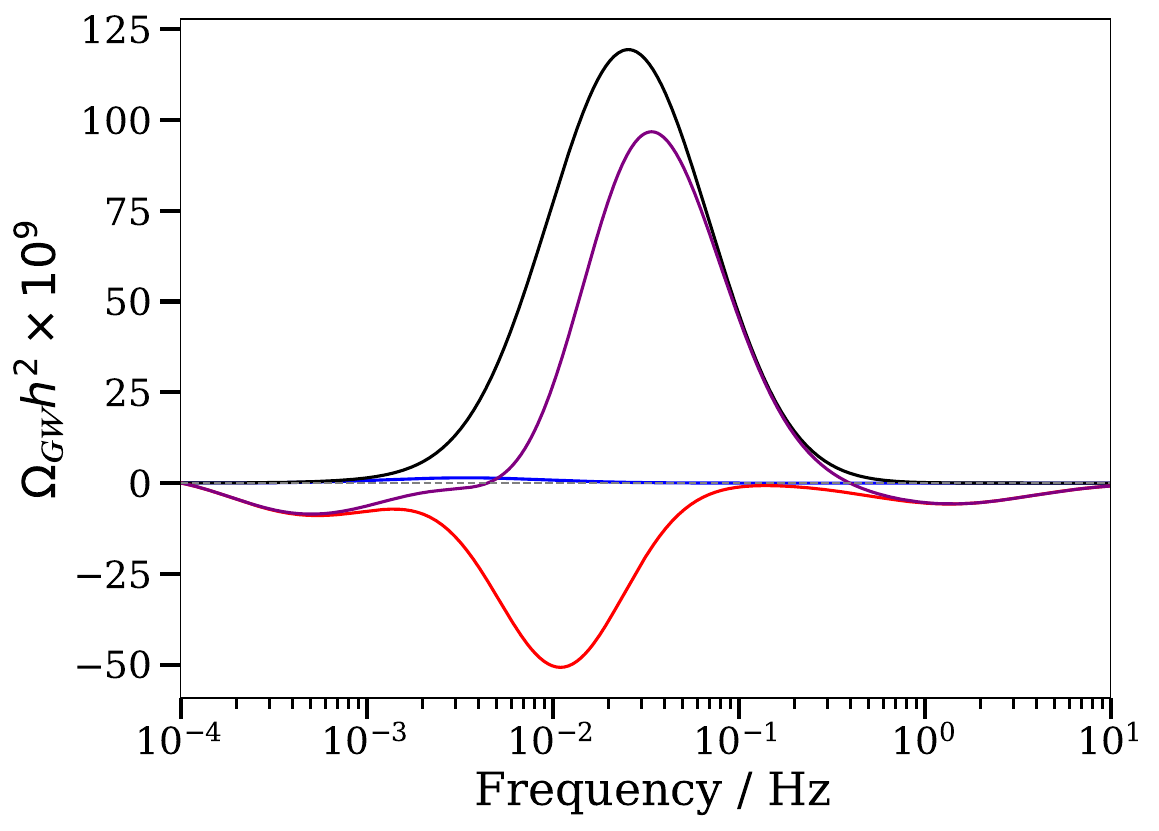}
    \end{subfigure}
   
    \caption{\footnotesize{Plot of the spectral  density today for tensor modes coming from inflation (blue), second-order contributions (black), third-order contributions (red) and the total contributions (purple) for input log-normal power spectra with $\sigma=0.1$ corresponding to the top plot, $\sigma=0.5$ bottom left plot and $\sigma=1$ bottom right plot. The inclusion of third-order tensor modes drastically reduces the spectra around peak scales. As $\sigma$ gets larger, more scales are suppressed. For values of $\sigma >0.1$, the observable becomes negative, which is unphysical. Although this is partially due to the divergence, including the auto-correlations of third-order iGWs would add a positive contribution and could potentially make the overall observable positive.}}
    \label{fig:allcontirbutions}
\end{figure}

Finally, we present all contributions together in Fig.~\ref{fig:allcontirbutions} for different values of $\sigma$. We stress that only the case $\sigma = 0.1$ should be regarded as realistic, since in this regime the input power spectra are sufficiently peaked. As shown earlier, the integrands for third-order and scalar–tensor iGWs exhibit divergent behaviour in the UV limit. Including third-order tensor modes correlated with linear tensor perturbations suppresses the spectral density for all $\sigma$ values. The scale at which this suppression occurs depends on $\sigma$; for $\sigma = 0.1$ it coincides with the peak, whereas for larger $\sigma$ the suppression extends to a broader range of scales, and for $\sigma = 1$ it affects the entire spectrum. Although we attribute the appearance of negative spectral densities in part to unphysical enhancements at small scales, we emphasise that a negative contribution is not inherently problematic. Correlations can be negative which might indicate that we should also include the auto-correlation of third-order iGWs, which would yield a positive contribution, much like the auto-correlation of second-order iGWs, which is always positive\footnote{This was shown in Ref.~\cite{Zhou:2021vcw} for the case of first-order scalar-perturbations inducing second and third-order perturbations.}. Put differently, the observable (the spectral density) receives contributions from various correlations in our analysis: $\langle h^{(1)} h^{(1)} \rangle$, $\langle h^{(2)} h^{(2)} \rangle$, and $\langle h^{(1)} h^{(3)} \rangle$. If the $\langle h^{(3)} h^{(3)} \rangle$ term were also included, and the total spectral density then remained positive for all $\sigma$ values, the negative contribution from $\langle h^{(1)} h^{(3)} \rangle$ would not be problematic.

\newpage
\section{Conclusion}
In this work, we set out to compute correlation of third-order iGWs with primordial tensor modes produced during inflation. These contribute at the same order in perturbation theory as scalar-tensor iGWs which were the focus of previous research in Refs.~\cite{Chang:2022vlv,Bari:2023rcw, Picard:2023sbz,Yu:2023lmo} for peaked input power spectra. In this regard, there had already been some work on certain sources of third-order iGWs in the IR limit, which found that the absolute value of the power-spectrum was suppressed \cite{Chen:2022dah}. However, since the integrands of scalar-tensor iGWs showed problematic behaviour in their UV limits, which lead to enhancements of the observable \cite{Bari:2023rcw, Picard:2023sbz} for wider input peaks, we set out to compute all the contributions involving scalar-tensor interactions up to third-order at all scales. 

We derived expressions for all the source terms of third-order sourced gravitational waves which give rise to non-trivial correlations with first-order tensor modes. The expressions provided of the time-averaged power spectra of these source terms are valid for all values of the equation of state parameter $w>-\tfrac{1}{3}$. Finally, we computed the spectral density of these terms in a radiation dominated universe with an input log-normal power spectrum for the first-order curvature and tensor perturbations. For very peaked spectra, we found that these new contributions suppress the spectral density, particularly around peak scales. As the peak of the input spectra gets wider, the suppression happens over more scales. However, some of the suppression was due to the divergent nature of the integrands. We showed that some of the source terms diverge in the UV limit, but we cannot claim that these divergences cancel out the ones seen in the scalar-tensor iGWs. Indeed, source terms made of second-order perturbations gave rise to nested kernels which made it impossible to determine analytically how these terms behave. Numerically, our results show that there is no cancellation of the divergence.

Furthermore, we argued that although the suppression of the spectral density happens in part due to the divergent nature of the integrand, we think it is also because we are not including the auto-correlation of third-order tensor modes. Additionally, for the third-order terms, which we could only compute numerically due to the nested integrals, taking the UV limits of the nested kernels may allow for an analytical evaluation. This, in turn, could clarify the nature of the divergence. It would also be worth investigating how the spectral density looks over epochs dominated by other components (like considering a period of early matter domination)\footnote{It will also be interesting to compute the second-order scalar-tensor contribution and the third-order terms we have computed in a different gauge than the conformal Newtonian gauge. If the divergence does not occur in a different gauge, this will show that the gauge we are using are not suitable for this computation. We thank P.~Bari and G.~Dom\`{e}nech for raising this point.}. Finally, it was shown in Ref.~\cite{Picard:2024ekd} that inclusion of scalar primordial non-Gaussianity can leave an imprint on the spectral density of scalar-tensor iGWs, and so one should check whether this is the case for the new source terms we have computed. We leave these tasks for future work.

\acknowledgments
RP is thankful for useful discussions with S{\' e}bastien Paine and Chris Clarkson. Furthermore, the authors are grateful for comments on a draft of this paper from Pritha Bari and Guillem Dom\`{e}nech. This research utilised Queen
Mary’s Apocrita HPC facility, supported by QMUL Research-IT~\cite{king_2017_438045}. 
RP is funded by STFC grant ST/P000592/1. LEP is supported by
a Royal Society funded post-doctoral position. DJM acknowledges financial support from the STFC under grant ST/X000931/1.


\appendix
\section{Einstein equations}
In this Appendix we present all the Einstein equations we need to  build up our work up to third-order. Ultimately, the fully simplified source terms of the third-order GW equation depend on $\Psi(\eta ,k)$, $h_{\lambda}(\eta ,k)$, $\Psi^{(2)}(\eta ,k)$, $B_r^{(2)}(\eta ,k)$ and $h^{(2)}_{\lambda}(\eta ,k)$, so in this Appendix we show how to relate these to other perturbation variables. We present the relevant Einstein equations order by order. 

All equations have been found with the help of the $xPand$ package \cite{Pitrou:2013hga}.
\subsection{Background}
At the background level, the two Friedmann equations are
\begin{equation}
    \mathcal{H}^2 =\frac{a^2}{3M_{Pl}^2} \rho, \quad \mathcal{H}' = -\frac{a^2\rho}{6M_{Pl}^2}(1+3w) \, .
\end{equation}
There is also an energy conservation equation
\begin{equation}
    \rho ^{\prime} + 3\mathcal{H}\left (\rho + P \right )=0 \, .
\end{equation}
The matter content of the universe is modelled as a barotropic fluid, by $w$,
\begin{equation}
    P=w\rho \, ,
\end{equation}
and we set $w$ to be constant, so that $w=c_s^2$, where $c_s^2$ is the speed of sound defined by
\begin{equation}
    c_s^2 = \frac{P^{\prime}}{\rho ^{\prime}} \, .
\end{equation}
As a result ($w\neq -\frac{1}{3}$)
\begin{equation}
    a(\eta) \propto \eta ^{\frac{2}{1+3w}} \quad \text{and} \quad \hubble =\frac{a\conf}{a}= \frac{2}{\eta(1+3w)} \, .
\end{equation}

\subsection{first-order} \label{firstorderEQ}
We work with adiabatic perturbations of the fluid, meaning that entropy perturbations are zero at all orders. Pressure and density perturbations are subsequently linked by a simple relationship
\begin{equation}
    \delta P = c_s^2 \delta \rho \, . 
\end{equation}
First-order density perturbations and velocity perturbations source GWs at second and third-order. The evolution equation for the first-order density perturbation, $\delta \rho$, comes from the time-time component of the Einstein equation
\begin{equation}
     \delta \rho = -2\rho \Phi + \frac{M_{Pl}^2}{a^2}\left (-6\mathcal{H}\Psi ^{\prime} +2\nabla ^2 \Psi \right )\, .
\end{equation}
Taking the divergence of the $0-i$ component of the Einstein equations we get an evolution equation for $v$, the first-order velocity perturbation,
\begin{equation}
     v=-\frac{2}{3(1+w)\mathcal{H}} \left ( \Phi +\frac{\Psi ^{\prime}}{\mathcal{H}} \right ) \, .
\end{equation}
From the pure spatial part of the Einstein equations we can extract two equations: the symmetric trace-free part, which for a perfect fluid in the absence of anisotropic stress implies $\Phi = \Psi$, and the trace, which after some simplification reads
\begin{equation} 
     \Psi ^{\prime \prime} +3 \mathcal{H}\left (1+c_s^2  \right )\Psi ^{\prime} + c_s^2 k^2\Psi  =0 \, .
\end{equation}
The above equation of motion can be solved in Fourier space and has the solution\footnote{Our Fourier conventions are in the following Appendix.}
\begin{equation} \label{psievolutioneq}
    \Psi(\eta ,k) = \avant \mathcal{R} _{\mathbf{k}} \frac{2^{2+b}\Gamma \left (\frac{5}{2} +b\right )}{w^{\frac{1+b}{2}}x^{1+b}\sqrt{\pi}}j_{b+1}(x\sqrt{w}) \equiv  \avant \mathcal{R} _{\mathbf{k}} T_{\Psi}(\sqrt{w}x) \,
\end{equation}
where we recall the parameter $b$ is defined in Eq.~\eqref{defb}, $w>0$ and the function $j_n(x)$ is the spherical Bessel function of the first kind. 

Finally, the transverse trace-free part of the spatial component of the Einstein equations gives us an equation of motion for the first-order tensor modes
\begin{equation}
    h_{ij}^{\prime \prime} + 2\mathcal{H}h_{ij}^{'} - \nabla ^2h_{ij} = 0 \, .
\end{equation}
Similarly, this can be solved in Fourier space,
\begin{equation}\label{hevolutioneq}
    h_{\lambda} (\eta , k) = h_{\mathbf{k}}^{\lambda} \frac{j_b(x)}{x^b} \equiv h_{\mathbf{k}}^{\lambda} T_h(x) \, ,
\end{equation}
$h_{\mathbf{k}}^{\lambda}$ being the primordial value of the perturbation.
\subsection{second-order}\label{appsecondorder}
Here, we present the second-order equations needed throughout our work. We have simplified them using the first-order equations presented in the previous sub-section.

at second order, the second-order pressure becomes gauge dependent, however in the Newtonian gauge it can be related to familiar quantities \cite{Carrilho:2015cma}
\begin{equation}
    \delta P^{(2)} = c_s^2 \delta \rho ^{(2)} + \frac{2}{\rho ^{\prime}} \delta \rho \delta P^{\prime} - \left (\frac{P^{\prime \prime}}{\rho ^{\prime ^2}} - \frac{P^{\prime}\rho ^{\prime \prime}}{\rho ^{\prime 3}} \right )\delta \rho ^2 - \frac{2c_s^2}{\rho ^{\prime}} \delta \rho \delta \rho ^{\prime} \, .
\end{equation}
In the case of a constant equation of state then
\begin{equation}
    \delta P^{(2)} = c_s^2 \delta \rho ^{(2)} \, .
\end{equation}

From the time-time component of the second-order Einstein equations, we get a relationship between $\rho ^{(2)}$ and $\Phi^{(2)}$, $\Psi ^{(2)}$ and over first-order perturbations
\begin{align}
    \begin{split}
        \frac{3(1+w)a^2\hubble^2}{M_{Pl}^2}\rho^{(2)}=& -3 (1 + w) \hubble ^2  h_{m n}' \left ( h^{m n'} + 8\hubble h^{m n}  \right )   -18 (1 + w) \mathcal{H}^4 \left( \Phi^{(2)} - 4 \Psi^2 \right)  \\
        & - 18 (1 + w) \mathcal{H}^3 \Psi^{(2)'} 
        - 16 \mathcal{H}  \partial_m \Psi' \partial^m \Psi - 8 \partial_m \Psi' \partial^m \Psi'  + \mathcal{H}^2 \bigg( 18 (1 + w) \Psi'^2
        \\
        &  + 48 (1 + w) \Psi \nabla ^2 \Psi + 6(1+w) \nabla^2 \Psi^{(2)}  + 2(5+9w) \partial_m \Psi \partial^m \Psi   
        \\
        &  
        - 12 (1+w)   h^{m n} \partial_n \partial_m \Psi + 12 (1 + w)  h^{m n}\nabla ^2 h_{m n} \bigg ) 
        \\
        & - 3 (1 + w) \mathcal{H}^2 \big( 
        2 \partial_n h_{m c} 
        - 3 \partial_c h_{m n} \big)\partial^c h^{m n} \, .
    \end{split}
\end{align}
Similarly, from the $0-i$ component, an equation for $v_i^{(2)}$
\begin{align}
\begin{split}
    18(1+w)\hubble ^4 v_i^{(2)}& = -18 (1 + w) \mathcal{H}^4 B_i^{(2)}  
    - 18 (1 + w) \mathcal{H}^4 \partial_i v^{(2)}
    + 12 \mathcal{H}^3 \bigg( 4 h_{i m} \partial^m \Psi 
    \\
    & - \partial_i \Phi^{(2)} - 2 \Psi \partial_i \Psi \bigg)  + 16 \nabla ^2 \Psi \partial_i \Psi ^{\prime}
    + 16 \mathcal{H} \bigg( \nabla ^2 \Psi \partial_i \Psi - 3 \Psi ^{\prime} \partial_i \Psi ^{\prime} \bigg)
    \\
    & + 3 \mathcal{H}^2 \bigg( \nabla ^2  B_i^{(2)} 
    + 8 h_{im}^{\prime} \partial^m \Psi + 8 h_{i m} \partial ^m \Psi'
    + 8 h^{m n} ( \partial_n h_{im}^{\prime} - \partial_i h_{mn}^{\prime}) 
    \\
    & -  4h^{mn '} \partial_i h_{m n} 
    + 24 \Psi ^{\prime} \partial_i \Psi  + 40 \Psi \partial _i \Psi ^{\prime} + 4 \partial_i \Psi^{(2)'}\bigg) \, .
\end{split}
\end{align}
Finally, we present the spatial part of the Einstein tensor
\begin{equation}
    \begin{split}
        G_{ij}^{(2)} &= h_{ij}^{\prime \prime (2)} + 2 \mathcal{H}h_{ij}^{\prime (2)} - h_{ij}^{(2)} (2\mathcal{H}^2 + 4\mathcal{H}^{\prime}) -\nabla ^2 h_{ij}^{(2)} + \delta _{ij} (\Phi ^{(2)} + \Psi ^{(2)})(2\mathcal{H}^2 + 4\mathcal{H}^{\prime}) \\
        &+2\delta _{ij} \Phi ^{\prime (2)}\mathcal{H} +4 \delta _{ij} \Psi ^{\prime (2)}\mathcal{H} + 2 \delta _{ij} \Psi ^{\prime \prime (2)} + \delta _{ij} \nabla ^2 \Phi ^{(2)} - \delta _{ij} \nabla ^2 \Psi ^{(2)}  \\
        &- \partial _i \partial _j \Phi ^{(2)} + \partial _i \partial _j \Psi ^{(2)} - 4 h_i^{m\prime}h_{jm}^{\prime} +  \delta _{ij} \left ( 3h_{mk}^{\prime} h^{mk \prime} + 4 h^{mk}h_{mk}^{\prime \prime} +8 \mathcal{H} h^{mk}h_{mk}^{\prime}  \right ) \\
        &- \Phi \left (4 h_{ij}^{\prime \prime} + 8 \mathcal{H}h_{ij}^{\prime} - 8\mathcal{H}^2 h_{ij} - 16\mathcal{H}^{\prime} h_{ij} \right ) - \delta _{ij} \left (8\mathcal{H}^2 + 16 \mathcal{H}^{\prime} \right ) \Phi ^{(2)} - 16 \delta _{ij} \mathcal{H} \Phi \Phi ^{\prime} \\
        &+\Phi ^{\prime} \left (8 \mathcal{H}h_{ij} - 2h_{ij}^{\prime} \right ) + \delta _{ij} \left ( 16 \mathcal{H}^{\prime} - 8\mathcal{H}^2 \right )\Phi \Psi - 8 \delta _{ij} \mathcal{H} \Psi \Phi ^{\prime} + 2 h_{ij}^{\prime} \Psi ^{\prime} + 24 \mathcal{H} h_{ij} \Psi ^{\prime} \\
        &-16 \delta _{ij} \mathcal{H} \Phi \Psi ^{\prime} - 4 \delta _{ij} \Phi ^{\prime} \Psi ^{\prime} + 2 \delta _{ij} \Psi ^{\prime 2} + 12h_{ij} \Psi ^{\prime \prime} - 8 \delta _{ij} \Phi \Psi ^{\prime \prime} - 4 \Psi \nabla ^2 h_{ij} + 4 h_i^m \partial _m \partial _j \Psi \\
        &+ \left ( 4 h_{ij} - 4\delta _{ij} \Phi \right )\nabla ^2 \Phi - \left ( 8 h_{ij} + 4\delta _{ij} \Psi \right ) \nabla ^2 \Psi + 4 h_j^m \partial _m \partial _i \Psi - 2 \partial _k h_{ij} \partial ^k \Phi \\
        & - 2 \delta _{ij} \partial _k \Phi \partial ^k \Phi - 6 \partial _k h_{ij} \partial ^k \Psi - 4 \delta _{ij} \partial _k \Psi \partial ^k \Psi + 4 h^{mk} \partial _m \partial _k h_{ij} - 4 h^{mk} \delta _{ij} \partial _m \partial _k \Phi \\
        & - 4 \partial _m h_{jk} \partial ^k h_i^m + 4 \partial _k h_{jm} \partial ^ k h_i^m - 4h^{mk}\delta _{ij} \nabla ^2 h_{mk} + 2 \delta _{ij} \partial _k h_{md} \partial ^d h^{mk} \\ 
        & - 3\delta _{ij} \partial _d h_{mk} \partial ^d h^{mk} + 2 \partial ^m \Phi \partial _i h_{jm} + 2 \partial ^m \Psi \partial _i h_{jm} - 4 h^{mk} \partial _i \partial _k h_{jm} + 2 \partial _i h^{mk} \partial _j h_{mk} \\
        &+2 \partial ^m \Psi \partial _j h_{im} +2 \partial ^m \Phi \partial _j h_{im} + 2\partial _i \Phi \partial _j \Phi - 2\partial _i \Psi \partial _j \Phi - 2\partial _i \Phi \partial _j \Psi +6 \partial _i \Psi \partial _j \Psi \\ 
        &-4h^{mk} \partial _j \partial _k h_{im} + 4 h^{mk} \partial _j \partial _i h_{mk} + 4\Phi \partial _j \partial _i \Phi + 4 \Psi \partial _j \partial _i \Psi
    \end{split}\, ,
\end{equation}
and the energy-momentum tensor
\begin{eqnarray}
    T_{ij}^{(2)} &=& a^2 \bigg ( \delta _{ij} \left [ \frac{1}{2}P^{(2)} -2c_s^2\delta \rho \Psi -w\rho \Psi ^{(2)} \right]  + 2c_s^2\delta \rho  h_{ij} +(1+w)\rho \partial _i v \partial _j v \nonumber \\
    &+&  w\rho h_{ij}^{(2)}\bigg )\, .
\end{eqnarray}
We have left these unsimplified since from these two definitions we can see what perturbations contribute to the GW background at second order (Eq.~\eqref{GW2eq}), the second-order induced vectors (Eq.~\eqref{eqinducedvectorsrealspace}) and the induced second-order scalars (Eqs.~\eqref{eqtracerealspace} and \eqref{eqanisrealspace}). 

\subsection{third-order}\label{appthirdorderEinstein}
In this sub-section we give the spatial part of the third-order Einstein equations. After the perturbations have been separated into scalar, vector and tensor degrees of freedom there are approximately 500 terms. Hence, we have disregarded some terms, namely: pure first-order tensor contributions, second-order perturbations coupled to first-order tensors, terms that couple quadratic first-order tensor perturbations to a first-order scalar fluctuation and terms made of cubic first-order scalars. The first two terms are subdominant and not the focus of this work and the last two terms disregarded do not contribute to the correlation we are interested in. The third-order Einstein tensor is  
\begin{equation}
    \begin{split}\label{thirdordereinstein}
        G_{ij}^{(3)} &= h_{ij}^{(3)''} + 2h_{ij}^{(3)'} \mathcal{H} - 2h_{ij}^{(3)} \mathcal{H}^2 - 4h_{ij}^{(3)} \mathcal{H}' - 6h_{ij}^{(2)''} \Phi - 12h_{ij}^{(2)'} \mathcal{H} \Phi + 12h_{ij}^{(2)} \mathcal{H}^2 \Phi + 24h_{ij}^{(2)} \mathcal{H}' \Phi \\
        &+ 24h_{ij}^{''} \Phi^2+ 48h_{ij}^{'} \mathcal{H} \Phi^2- 48h_{ij} \mathcal{H}^2 \Phi^2- 96h_{ij} \mathcal{H}' \Phi^2- 3h_{ij}^{(2)'} \Phi'+ 12h_{ij}^{(2)} \mathcal{H} \Phi'+ 24h_{ij}^{'} \Phi \Phi' \\
        &- 96h_{ij} \mathcal{H} \Phi \Phi' + 3h_{ij}^{(2)'} \Psi'+ 36h_{ij}^{(2)} \mathcal{H} \Psi'- 12h_{ij}^{'} \Phi \Psi'- 144h_{ij} \mathcal{H} \Phi \Psi'- 36h_{ij} \Phi' \Psi'+ 12h_{ij}^{'} \Psi \Psi' \\
        &+ 144h_{ij} \mathcal{H} \Psi \Psi'+ 24h_{ij} \Psi'^2+ 18h_{ij}^{(2)} \Psi''- 72h_{ij} \Phi \Psi''+72h_{ij} \Psi \Psi''- 24\Psi^2 \nabla^2 h_{ij} \\
        &- 6\Psi \nabla^2 h_{ij}^{(2)}- \nabla^2 h_{ij}^{(3)}+ 6h_{ij}^{(2)} \nabla^2 \Phi - 24h_{ij} \Phi \nabla^2 \Phi + 24h_{ij} \Psi \nabla^2 \Phi - 6h_{ij} \Phi \nabla^2 \Phi^{(2)}  \\
        &+ 6h_{jm}^{(2)} \nabla^m \nabla_i \Psi + 48h_{jm} \Psi \nabla^m \nabla_i \Psi + 6h_{im}^{(2)} \nabla^m \nabla_j \Psi + 48h_{im} \Psi \nabla^m \nabla_j \Psi \\
        &+ 12\Phi \nabla^m h_{ij} \nabla_m \Phi - 12\Psi \nabla^m h_{ij} \nabla_m \Phi - 3\nabla^m h_{ij}^{(2)} \nabla_m \Phi - 12h_{ij} \nabla^m \Phi \nabla_m \Phi - 12h_{ij} \nabla^m \Psi \nabla_m \Phi \\
        &- 72\Psi \nabla^m h_{ij} \nabla_m \Psi - 9\nabla^m h_{ij}^{(2)} \nabla_m \Psi - 60h_{ij} \nabla^m \Psi \nabla_m \Psi + 6\mathcal{H} \Phi \nabla_i B_j^{(2)} \\
        &+ \frac{3}{2}\Phi' \nabla_i B_j^{(2)} + \frac{3}{2}\Psi' \nabla_i B_j^{(2)} + 3\Phi \nabla_i B_j^{(2)'} - 12\Phi \nabla^m \Phi \nabla_i h_{jm}- 12h_{ij}^{(2)} \nabla^2 \Psi - 96h_{ij} \Psi \nabla^2 \Psi \\
        &+ 12\Psi \nabla^m \Phi \nabla_i h_{jm} + 24\Psi \nabla^m \Psi \nabla_i h_{jm} + 3\nabla^m \Phi \nabla_i h_{jm}^{(2)} + 3\nabla^m \Psi \nabla_i h_{jm}^{(2)} - 3B_j^{(2)'} \nabla_i \Psi \\
        &- 6 B_j^{(2)} \mathcal{H} \nabla_i \Psi  + 12 h_{jm} \nabla^m \Phi \nabla_i \Psi  + 36 h_{jm} \nabla^m\Psi \nabla_i \Psi  - 3 B_j^{(2)} \nabla_i \Psi'  + 6 \mathcal{H} \Phi \nabla_j B_i^{(2)}  \\
        & + \frac{3}{2} \Phi' \nabla_j B_i^{(2)} + \frac{3}{2} \Psi' \nabla_j B_i^{(2)}  + 3 \Phi \nabla_j B_i^{(2)'}  - 12 \Phi \nabla^m \Phi \nabla_j h_{im}  + 12 \Psi \nabla^m \Phi \nabla_j h_{im}  \\
        & + 24 \Psi \nabla^m \Psi \nabla_j h_{im}  + 3 \nabla^m \Phi \nabla_j h_{im}^{(2)}  + 3 \nabla^m \Psi \nabla_j h_{im}^{(2)} + 6 \Psi \nabla_j \nabla_i \Psi^{(2)} \\
        & + 3 \nabla_i \Phi^{(2)} \nabla_j \Phi  - 3 \nabla_i \Psi^{(2)} \nabla_j \Phi  + 3 \nabla_i \Phi \nabla_j \Phi^{(2)}  - 3 \nabla_i \Psi \nabla_j \Phi^{(2)}  - 3 B_i^{(2)'} \nabla_j \Psi  - 6 B_i^{(2)} \mathcal{H} \nabla_j \Psi  \\
        & + 12 h_{im} \nabla^m \Phi \nabla_j \Psi  + 36 h_{im} \nabla^m \Psi \nabla_j \Psi  - 3 \nabla_i \Phi^{(2)} \nabla_j \Psi  + 9 \nabla_i \Psi^{(2)} \nabla_j \Psi  - 3 B_i^{(2)} \nabla_j \Psi'  \\
        & - 3 \nabla_i \Phi \nabla_j \Psi^{(2)} + 9 \nabla_i \Psi \nabla_j \Psi^{(2)}  + 6 \Phi^{(2)} \nabla_j \nabla_i \Phi  + 6 \Phi \nabla_j \nabla_i \Phi^{(2)}  + 6 \Psi^{(2)} \nabla_j \nabla_i \Psi \\
        &+ \text{(trace part)}\delta _{ij} + (hhh,\, \Psi \Psi \Psi, \, hh\Psi,\, \Psi ^{(2)}h, \, \Phi ^{(2)}h, \, B ^{(2)}h, \, h ^{(2)}h, \,\text{terms})   \, ,
    \end{split} 
\end{equation}
whilst the third-order energy-momentum tensor reads
\begin{equation}
    \begin{split}\label{thirdorderenergymomentum}
        a^{-2}T_{ij}^{(3)} &=  2 h_{ij}^{(3)} P + 6 h_{ij}^{(2)} \delta P  + 6 h_{ij} P^{(2)} + h_{ij} P^{(3)} - 6 h_{ij} P^{(2)} \Psi - 6 h_{ij} \delta P \Psi^{(2)}- 2 h_{ij} P \Psi^{(3)} \\
        &   + 3 B^{(2)}_{j} P \partial_{i} v + 3 P v^{(2)}_{j} \partial_{i} v + 3 B^{(2)}_{j} \rho \partial_{i} v + 3 v^{(2)}_{j} \rho \partial_{i} v+ 12 h_{jm} P \partial^{m} v \partial_{i} v \\
        &   + 12 h_{jm} \rho \partial^{m} v \partial_{i} v + 3 B^{(2)}_{i} P \partial_{j} v + 3 P v^{(2)}_{i} \partial_{j} v + 3 B^{(2)}_{i} \rho \partial_{j} v + 3 v^{(2)}_{i} \rho \partial_{j} v  \\
        &+ 12 h_{im} P \partial^{m} v \partial_{j} v + 12 h_{im} \rho \partial^{m} v \partial_{j} v + 6 \delta P \partial_{i} v \partial_{j} v + 6 \delta \rho \partial_{i} v \partial_{j} v - 24 P \Psi \partial_{i} v \partial_{j} v \\
        &  - 24 \rho \Psi \partial_{i} v \partial_{j} v + 3 P \partial_{i} v^{(2)} \partial_{j} v + 3 \rho \partial_{i} v^{(2)} \partial_{j} v + 3 P \partial_{i} v \partial_{j} v^{(2)} + 3 \rho \partial_{i} v \partial_{j} v^{(2)}  \, .
    \end{split}
\end{equation}

\section{Induced second-order scalars, vectors and tensors in Fourier space: solutions} \label{secondorderApp}
In this Appendix we present the equations of motion for the induced scalars, vectors and tensors at second order in perturbation theory. It is useful to switch to Fourier space, and so first we give our conventions and define the operators we need to extract scalar, vector and tensorial degrees of freedom from the spatial part of the Einstein equations.

A scalar perturbation can be expressed as a Fourier integral
\begin{equation}
    S(\eta , \mathbf{x}) = \int \frac{d^3\mathbf{k}}{(2\pi)^{\frac{3}{2}}} S(\eta , \mathbf{k}) e^{i\mathbf{k}\cdot \mathbf{x}} \, .
\end{equation}
We need to extract the symmetric trace free part of the spatial Einstein equations. This can be done via an operator $D^{ij}(\mathbf{x})$, defined in real space as acting on a term $S_{ij}(\mathbf{\eta ,x^{\prime}})$ as
\begin{equation}\label{scalarextractor}
    D^{ij}(\mathbf{x})S_{ij}(\mathbf{\eta ,x^{\prime}}) = \int \frac{{\rm d} ^3 \mathbf{k^{\prime}}}{(2\pi)^{\frac{3}{2}}} e^{i\mathbf{k^{\prime}}\cdot \mathbf{x}} \int \frac{{\rm d} ^3 \mathbf{x^{\prime}}}{(2\pi)^{\frac{3}{2}}} \left (-k^{\prime i}k^{\prime j}+\frac{1}{3}k^{\prime 2}\right ) e^{-i\mathbf{k^{\prime}}\cdot \mathbf{x^{\prime}}} S_{ij}(\mathbf{\eta ,x^{\prime}}) \, .
\end{equation}

A vector can also be expressed as a Fourier integral, where it decouples into its two parities ($r=+,-$):
\begin{equation}
    V_i(\eta , \mathbf{x}) = \sum _r \int \frac{d^3\mathbf{k}}{(2\pi)^{\frac{3}{2}}}  V^r(\eta , \mathbf{k}) e^r_i({\mathbf{k}}) e^{i\mathbf{k}\cdot \mathbf{x}} \, .
\end{equation}
For simplicity we work in the circular basis, so the parity vectors are defined as 
\begin{subequations}
    \begin{align}
        e^+_i({\mathbf{k}}) &= \frac{1}{\sqrt{2}} \left ( e_i(\mathbf{k}) +i \overline{e}_i(\mathbf{k}) \right ) \, ,
        \\
        e^-_i({\mathbf{k}}) &= \frac{1}{\sqrt{2}} \left ( e_i(\mathbf{k}) -i \overline{e}_i(\mathbf{k}) \right ) \, ,
    \end{align}
\end{subequations}
where $\{ e_i(\mathbf{k}) , \overline{e}_i(\mathbf{k})\}$ form a subspace perpendicular to the wave-vector $\mathbf{k}$. Useful relations used throughout this work include
\begin{equation}
    e^r_i(\mathbf{k})k^i=0 \, , \quad e^r_i(\mathbf{k})^*e_s^i(\mathbf{k}) = \delta ^r_s \, \quad \text{and} \quad e^r_i(\mathbf{k})^* = e^{-r}_i(\mathbf{k}) = e^r_i(\mathbf{-k}) \, .
\end{equation}
where $-r$ refers to the opposite parity. In order to extract the vectorial degrees of freedom of a tensor $S_{ij}(\mathbf{\eta ,x^{\prime}})$, we can construct a transverse extraction operator \cite{Lu:2007cj} $V^{ij}_a(\mathbf{x})$
\begin{equation}\label{vectorextractor}
    V^{ij}_a(\mathbf{x})S_{ij}(\mathbf{\eta ,x^{\prime}}) =-i \sum _{s=+,-} \int \frac{{\rm d} ^3 \mathbf{k^{\prime}}}{(2\pi)^{\frac{3}{2}}} e^s_a(\mathbf{k^{\prime}}) e^{i\mathbf{k^{\prime}}\cdot \mathbf{x}} \frac{1}{k^{\prime 2}} \int \frac{{\rm d} ^3 \mathbf{x^{\prime}}}{(2\pi)^{\frac{3}{2}}} k^{\prime i}e_s^j(\mathbf{k^{\prime}})^* e^{-i\mathbf{k^{\prime}}\cdot \mathbf{x^{\prime}}} S_{ij}(\mathbf{\eta ,x^{\prime}}) \, .
\end{equation}

Similarly, a tensor splits into two polarisations ($\lambda = R,L$ the so called-circular basis):
\begin{equation}
    T_{ab}(\eta , \mathbf{x}) = \sum _{\lambda} \int \frac{d^3\mathbf{k}}{(2\pi)^{\frac{3}{2}}}  T^{\lambda}(\eta , \mathbf{k}) \epsilon ^{\lambda}_{ab}({\mathbf{k}})  e^{i\mathbf{k}\cdot \mathbf{x}} \text{ .}
\end{equation}
The polarisation tensors $\epsilon ^{\lambda}_{ab}({\mathbf{k}})$ are constructed from a linear superposition of the $\{+,\times \}$ polarisation basis
\begin{align}
    \begin{split}
        \eps ^R_{ab}(\mathbf{k}) &= \frac{1}{\sqrt{2}} \left ( q^+_{ab}(\mathbf{k}) + i q^{\times}_{ab}(\mathbf{k})  \right ) \text{ ,}  \\
        \eps ^L_{ab}(\mathbf{k}) &= \frac{1}{\sqrt{2}} \left ( q^+_{ab}(\mathbf{k}) - i q^{\times}_{ab}(\mathbf{k})  \right )\,,
    \end{split}
\end{align}
where
\begin{align}
    \begin{split}
         q_{ab}^{+}(\mathbf{k})&=\frac{1}{\sqrt{2}}\left ( e_a(\mathbf{k})e_b(\mathbf{k}) - \overline{e}_a(\mathbf{k}) \overline{e}_b(\mathbf{k})\right )\text{ ,} \\
        q_{ab}^{\times}(\mathbf{k})&=\frac{1}{\sqrt{2}}\left ( e_a(\mathbf{k})\overline{e}_b(\mathbf{k}) + \overline{e}_a(\mathbf{k}) e_b(\mathbf{k})\right )\,.
    \end{split}
\end{align}
Some properties and useful relation of $\epsilon ^{\lambda}_{ab}({\mathbf{k}})$ are
\begin{equation}
    \eps _{ab}^{\lambda}(\mathbf{k})\delta ^{ab} =0, \quad \eps _{ab}^{\lambda}(\mathbf{k}) k^a = 0, \quad  \eps _{ab}^{\lambda}(\mathbf{k}) ^* \eps ^{ab,\lambda ^{\prime}}(\mathbf{k}) = \delta ^{\lambda \lambda ^{\prime}}, \quad \text{and} \quad \left ( \eps _{ab}^{\lambda}(\mathbf{k}) \right)^* = \eps _{ab}^{-\lambda}(\mathbf{k})  =\eps _{ab}^{\lambda}(-\mathbf{k})\, ,
\end{equation}
where $-\lambda$ refers to the opposite polarisation. We can also construct an operator $\Lambda ^{ij}_{ab}(\mathbf{x})$ \cite{Ananda:2006af} to extract the TT degrees of freedom of a tensor $S_{ij}(\mathbf{\eta ,x^{\prime}})$
\begin{equation}\label{tensorextractor}
    \Lambda ^{ij}_{ab}(\mathbf{x})S_{ij}(\mathbf{\eta ,x^{\prime}}) = \sum _{\lambda=R,L} \int \frac{{\rm d} ^3 \mathbf{k^{\prime}}}{(2\pi)^{\frac{3}{2}}} \eps ^{\lambda ^{\prime}}_{ab}(\mathbf{k^{\prime}}) e^{i\mathbf{k^{\prime}}\cdot \mathbf{x}}  \int \frac{{\rm d} ^3 \mathbf{x^{\prime}}}{(2\pi)^{\frac{3}{2}}}\eps _{\lambda ^{\prime}}^{ij}(\mathbf{k^{\prime}})^* e^{-i\mathbf{k^{\prime}}\cdot \mathbf{x^{\prime}}} S_{ij}(\mathbf{\eta ,x^{\prime}}) \, .
\end{equation}

Finally, we give the explicit expressions for the wave-vector $\mathbf{k}$ and the orthonormal basis $\{ e_i(\mathbf{k}) , \overline{e}_i(\mathbf{k})\}$ . Due to homogeneity and isotropy it is straightforward to parameterize vectors in spherical coordinates. Explicitly,
\begin{equation}\label{kwavevector}
    \mathbf{k}=k(\sin{\theta_k}\cos{\phi_k},\sin{\theta_k}\sin{\phi_k},\cos{\theta_k}) \, ,
\end{equation}
then we can construct the vectors  $\{ e_i(\mathbf{k}) , \overline{e}_i(\mathbf{k})\}$ as
\begin{subequations}
    \begin{align}
         \mathbf{e}(\mathbf{k})&=(\cos{\theta_k}\cos{\phi_k},\cos{\theta_k}\sin{\phi_k},-\sin{\theta_k})\, , \\
         \overline{\mathbf{e}}(\mathbf{k})&=(-\sin{\phi_k},\cos{\phi_k},0)\,.
    \end{align}
\end{subequations}
It is straightforward to construct the wave-vector $\mathbf{p}$ and its corresponding orthonormal basis, the reader can refer to Ref.~\cite{Picard:2023sbz} for explicit expressions. 
\subsection{Induced second-order scalars}\label{AppSOscalars}
In Sec.~\ref{inducedscalar} we presented two equations of motion for the second-order scalars, in this subsection we solve them in Fourier space.

The second-order anisotropic stress equation sourced by first-order tensor and scalar fluctuations, Eq.~\eqref{eqanisrealspace}, reads
\begin{equation} \label{anis}
    \Phi ^{(2)}(\eta , \mathbf{k}) - \Psi ^{(2)}(\eta , \mathbf{k}) = \frac{1}{k^4} \Pi ^{(2)}_{st} (\eta , \mathbf{k}) \, ,
\end{equation}
with
\begin{align}\label{anisinducedst}
    \begin{split}
        \Pi ^{(2)}_{st} (\eta , \mathbf{k}) & = \left (\frac{3+3w}{5+3w} \right ) \sum _{\lambda _1 }  \int \frac{d^3\mathbf{p}}{(2\pi)^{\frac{3}{2}}}  \text{ } Q_{\lambda _1}^{\Psi ^{(2)} _{st}}(\mathbf{k},\mathbf{p})f^{\Pi ^{(2)}}_{st}(\eta ,k,p,|\mathbf{k}-\mathbf{p}|) h^{\lambda _1}_{\mathbf{p}}\mathcal{R} _{\mathbf{k}-\mathbf{p}} \, .
    \end{split}
\end{align}
In the above we factored out the primordial values and defined for compactness
\begin{align}\label{polinducedscalars}
    Q_{\lambda _1}^{\Psi ^{(2)} _{st}}(\mathbf{k},\mathbf{p}) &= k^ik^j\epsilon_{ij}^{\lambda _1}(\mathbf{p}),
\end{align}
and
\begin{align}\label{fpistscalar}
    \begin{split}
        f^{\Pi ^{(2)}}_{st}(\eta ,k,p,|\mathbf{k}-\mathbf{p}|) &= -12 \left (k^ip_i-\frac{2}{3}k^2 \right ) T_h(\eta p) T_{\Psi}(\eta|\mathbf{k}-\mathbf{p}|)-12 p^2  T_h(\eta p)T_{\Psi}(\eta|\mathbf{k}-\mathbf{p}|)  \\
        &+6(1+3w)\hubble T_h(\eta p) T_{\Psi}\conf(\sqrt{w}\eta|\mathbf{k}-\mathbf{p}|)+6(w-1)|\mathbf{k}-\mathbf{p}|^2 T_h(\eta p) T_{\Psi}(\eta|\mathbf{k}-\mathbf{p}|) \\
        &- 12 (k^ip_i-p^2)T_h(\eta p) T_{\Psi}(\eta|\mathbf{k}-\mathbf{p}|) + 12 \, k^ip_i T_h(\eta p)T_{\Psi}(\eta|\mathbf{k}-\mathbf{p}|) \, .
    \end{split}
\end{align}

The second-order trace equation for scalars, Eq.~\eqref{eqtracerealspace}, in Fourier space is given by
\begin{align}
    \begin{split}
        &3\Psi ^{\prime \prime (2)}(\eta, \mathbf{k}) + 3(2+3w)\hubble \Psi ^{\prime (2)}(\eta, \mathbf{k}) + (3w+1)k^2 \Psi ^{(2)}(\eta, \mathbf{k}) + 3\hubble \Phi ^{\prime (2)}(\eta, \mathbf{k}) \\
        &- k^2 \Phi ^{(2)}(\eta, \mathbf{k}) =  \mathcal{S}^{\Psi ^{(2)}}_{st}(\eta , \mathbf{k})
    \end{split}
\end{align}
where
\begin{align}
    \begin{split}
        \mathcal{S}^{\Psi ^{(2)}}_{st}(\eta , \mathbf{k}) & = \left (\frac{3+3w}{5+3w} \right ) \sum _{\lambda _1 } \int \frac{d^3\mathbf{p}}{(2\pi)^{\frac{3}{2}}} Q_{\lambda _1}^{\Psi ^{(2)} _{st}}(\mathbf{k},\mathbf{p})f^{\Psi ^{(2)}}_{st}(\eta , p,|\mathbf{k}-\mathbf{p}|) h^{\lambda _1}_{\mathbf{p}}\mathcal{R} _{\mathbf{k}-\mathbf{p}} \, ,
    \end{split}
\end{align}
\begin{equation}
    f^{\Psi ^{(2)}}_{st}(\eta , p,|\mathbf{k}-\mathbf{p}|) = -2  (1-3w)T_h(\eta p)T_{\Psi} (\eta |\mathbf{k}-\mathbf{p}|) \, .
\end{equation}
We can now solve for $\Psi ^{(2)} (\eta , \mathbf{k})$ by making the substitution in Eq.~(\ref{anis}) to get rid of $\Phi ^{(2)} (\eta , \mathbf{k})$. This leaves us with
\begin{align} \label{eqscalarfourier}
    \begin{split}
        &\Psi ^{\prime \prime (2)}(\eta, \mathbf{k}) + 3(1+w)\hubble \Psi ^{\prime (2)}(\eta, \mathbf{k}) +wk^2 \Psi ^{(2)}(\eta, \mathbf{k}) = \frac{1}{3}  \mathcal{S}^{st}(\eta , \mathbf{k}) - \frac{\hubble}{k^4} \timeder   \Pi ^{(2)}_{st} (\eta , \mathbf{k})    \\
        &+ \frac{1}{3k^2} \Pi ^{(2)}_{st} (\eta , \mathbf{k})= \sum _{\lambda _1 , \lambda _2} \int \frac{d^3\mathbf{p}}{(2\pi)^{\frac{3}{2}}} Q_{\lambda _1}^{\Psi ^{(2)} _{st}}(\mathbf{k},\mathbf{p})\Sigma ^{st} (\eta , p,|\mathbf{k}-\mathbf{p}|)h^{\lambda _1}_{\mathbf{p}}\mathcal{R} _{\mathbf{k}-\mathbf{p}} \, ,
    \end{split}
\end{align}
where for convenience we defined
\begin{equation}\label{Sigmapsi2}
        \Sigma ^{st} (\eta , p,|\mathbf{k}-\mathbf{p}|) =\frac{1}{3}f^{\Psi ^{(2)}}_{st}(\eta , p,|\mathbf{k}-\mathbf{p}|) - \frac{\hubble}{k^4}\timeder  f^{\Pi ^{(2)}}_{st}(\eta ,k , p,|\mathbf{k}-\mathbf{p}|) + \frac{1}{3k^2}f^{\Pi ^{(2)}}_{st}(\eta ,k , p,|\mathbf{k}-\mathbf{p}|) \, .
\end{equation}
Eq.~\eqref{eqscalarfourier} can be solved via Green's method. The Green's function is given by (with initial conditions $\Psi^{(2)}(\eta _i ,k)=0$ and $\partial _{\eta}\Psi^{(2)}(\eta _i ,k)=0$ at some initial conformal time $\eta _i$)
\begin{equation}
    G^s_k(\eta,\bareta) = \frac{(k\bareta)(k\eta)\sqrt{w}}{k} \left (j_{b+1}(k\bareta\sqrt{w})y_{b+1}(k\eta\sqrt{w}) - y_{b+1}(k\bareta\sqrt{w})j_{b+1}(k\eta\sqrt{w}) \right ) \Theta (\eta - \bareta) \, ,
\end{equation}
valid for $w>0$ and $b$ is defined in Eq.~\eqref{defb}. The solution to the equation of motion for $\Psi ^{(2)}(\eta , \mathbf{k})$ is then given by
\begin{equation}\label{psiinducedst}
    \Psi ^{(2)}_{st}(\eta , \mathbf{k}) = \left (\frac{3+3w}{5+3w} \right )\sum _{\lambda _1 }\int \frac{d^3\mathbf{p}}{(2\pi)^{\frac{3}{2}}} Q_{\lambda _1}^{\Psi ^{(2)} _{st}}(\mathbf{k},\mathbf{p}) I^{\Psi ^{(2)}}_{st}(\eta ,p,|\mathbf{k}-\mathbf{p}|)h_{\mathbf{p}}^{\lambda _1} \mathcal{R} _{\mathbf{k}-\mathbf{p}} \, ,
\end{equation}
where we have further defined
\begin{equation} \label{kernelinducedstscalars}
    I^{\Psi ^{(2)}}_{st}(\eta ,p,|\mathbf{k}-\mathbf{p}|)= \int _{\eta _i}^{\infty} {\rm d}  \bareta \, \left ( \frac{a(\bareta)}{a(\eta)} \right )^{\frac{3(1+w)}{2}} G^s_k(\eta , \bareta) \Sigma ^{st}(\bareta , p,|\mathbf{k}-\mathbf{p}|) \, .
\end{equation}
\subsection{Induced second-order vectors}\label{AppSOvectors}
In Sec. \ref{inducedvector} we presented an equation that describes how vectors are sourced as second-order in perturbation theory.  Eq.~(\ref{eqinducedvectorsrealspace}) in Fourier spaces reads
\begin{align} \label{inducedvectorequation}
    \begin{split}
         B_r^{\prime (2)}(\eta , \mathbf{k}) + 2\mathcal{H}B_r^{(2)}(\eta , \mathbf{k}) =   \mathcal{S}^{B ^{(2)}_{st}}_{r}(\eta , \mathbf{k})\, ,
    \end{split}
\end{align}
where the source terms read
\begin{align}
    \begin{split}
        \mathcal{S}^{B ^{(2)}_{st}}_{r}(\eta , \mathbf{k}) & = \left ( \frac{3+3w}{5+3w} \right ) \sum _{\lambda _1} \frac{i}{k^2}\int \frac{d^3\mathbf{p}}{(2\pi)^{\frac{3}{2}}} \bigg ( Q^{B ^{(2)}_{st1}} _{r,\lambda _1 } (\mathbf{k},\mathbf{p}) f^{B ^{(2)}}_{st1}(\eta , p ,|\mathbf{k}-\mathbf{p}| ) + Q^{B ^{(2)}_{st2}} _{r,\lambda _1 } (\mathbf{k},\mathbf{p})  \\
    &\times f^{B ^{(2)}}_{st2}(\eta , p ,|\mathbf{k}-\mathbf{p}| ) \bigg ) h^{\lambda _1}_{\mathbf{p}} \mathcal{R}_{\mathbf{k}-\mathbf{p}} \, .
    \end{split}
\end{align}
We have defined
\begin{subequations}
    \begin{align}
       \label{pol2ndvecst1}
        Q^{ B^{(2)}_{st1}}_{r,\lambda _1 }(\mathbf{k},\mathbf{p}) &= (k^2-k^ip_i)k^me^j_r(\mathbf{k})^*\eps_{mj}^{\lambda _1} (\mathbf{p}) + e^j_r(\mathbf{k})^*p_jk^ik^m\eps_{im}^{\lambda _1} (\mathbf{p}) \, ,
        \\ \label{pol2ndvecst2}
        Q^{ B^{(2)}_{st2}}_{r,\lambda _1 }(\mathbf{k},\mathbf{p}) &= k^ie^j_r(\mathbf{k})^*\eps_{ij}^{\lambda _1} (\mathbf{p}) \, ,
    \end{align}
\end{subequations}
and
\begin{subequations}
    \begin{align}
        \label{fun2ndvecst1}
        f^{B^{(2)}}_{st1}(\eta , p,|\mathbf{k}-\mathbf{p}|) & = -4 T_h(\eta p)T_\Psi( \eta |\mathbf{k}- \mathbf{p}|) \, ,
        \\ \label{fun2ndvecst2}
        \begin{split} 
            f^{B^{(2)}}_{st2}(\eta , p,|\mathbf{k}-\mathbf{p}|) &=  4(\mathbf{p}\cdot \mathbf{k}-p^2)T_h(\eta p)T_\Psi (\eta |\mathbf{k}-\mathbf{p}|) +4p^2T_h(\eta p)T_\Psi (\eta |\mathbf{k}-\mathbf{p}|) \\
            &-2(1+3w)\hubble T_h(\eta p)T^{\prime}_{\Psi}  (\eta |\mathbf{k}-\mathbf{p}|) + 2(1-w)|\mathbf{k}-\mathbf{p}|^2T_h(\eta p)T_\Psi (\eta |\mathbf{k}-\mathbf{p}|) \, .
        \end{split}
    \end{align}
\end{subequations}
The general solution to Eq.~\eqref{inducedvectorequation} can be found via Green's method (with initial condition $B^{(2)}(\eta_i,k)=0$)
\begin{equation} \label{greensvectors}
    G^v(\eta , \bareta) = \left( \frac{\bareta}{\eta} \right )^{2(1+b)} \Theta(\eta - \bareta) \, .
\end{equation}
The Green's function for induced vectors does not depend on $k$ as for the case of induced scalars or tensors, i.e. there is no propagating mode. The solution to the equation, is then
\begin{align}\label{2ndordervecst}
    \begin{split}
        B_{r}^{(2)st}(\eta , \mathbf{k}) &= \left ( \frac{3+3w}{5+3w} \right ) \sum _{\lambda _1} \frac{i}{k^2}\int \frac{d^3\mathbf{p}}{(2\pi)^{\frac{3}{2}}} \bigg ( Q^{B ^{(2)}_{st1}} _{r,\lambda _1 } (\mathbf{k},\mathbf{p}) I^{B^{(2)}}_{st1}(\eta,p,|\mathbf{k}-\mathbf{p}|) + Q^{B ^{(2)}_{st2}} _{r,\lambda _1 } (\mathbf{k},\mathbf{p})  \\
            &\times I^{B^{(2)}}_{st2}(\eta,p,|\mathbf{k}-\mathbf{p}|) \bigg ) h^{\lambda _1}_{\mathbf{p}} \mathcal{R}_{\mathbf{k}-\mathbf{p}} \, ,
    \end{split}
\end{align}
where
\begin{subequations}
    \begin{align}
          I^{B^{(2)}}_{st1}(\eta,p,|\mathbf{k}-\mathbf{p}|) &=  \int _{\eta _i}^{\infty} {\rm d}  \bareta \,  G^v(\eta , \bareta) f^{B^{(2)}}_{st1}(\overline{\eta},  p,|\mathbf{k}-\mathbf{p}|) \label{kernelvectorst1}  \, ,
          \\
          I^{B^{(2)}}_{st2}(\eta,p,|\mathbf{k}-\mathbf{p}|) &=  \int _{\eta _i}^{\infty} {\rm d}  \bareta \,  G^v(\eta , \bareta) f^{B^{(2)}}_{st2}(\overline{\eta},  p,|\mathbf{k}-\mathbf{p}|) \label{kernelvectorst2}  \, .
    \end{align}
\end{subequations}

\subsection{Induced second-order tensors}\label{AppSOtensors}
In this subsection we list expressions regarding the second-order induced tensors. 
Eq.~\eqref{GW2eq} in Fourier space can be expressed as
\begin{equation}\label{GW2eqfourier}
    h_{\lambda}^{(2)\prime \prime} (\eta , \mathbf{k}) + 2\mathcal{H}h_{\lambda}^{(2)\prime} (\eta , \mathbf{k}) +k^2h_{\lambda}^{(2)} (\eta , \mathbf{k}) =  4\left (S_{\lambda}^{h^{(2)}_{ss}}(\eta , \mathbf{k})+ S_{\lambda}^{h^{(2)}_{st}}(\eta , \mathbf{k}) \right )\,,
\end{equation}
The Green's function for arbitrary $w$ is and say same initial conditions as third-order
\begin{equation}\label{Greenstensor}
    G^h_k(\eta,\bareta)= \frac{(k\eta)(k\bareta)}{k} \left ( y_b(k\bareta)j_b(k\eta) - y_b(k\eta)j_b(k\bareta) \right ) \Theta (\eta - \bareta) \, .
\end{equation}
where we recall $b$ is defined in Eq.~\eqref{defb}.

For the SIGWs the solution to the equation of motion Eq.~\eqref{GW2eqfourier} is (see Refs.~\cite{Espinosa:2018eve,Kohri:2018awv} for details)
\begin{equation} 
    h^{(2)ss}_{\lambda}(\eta , \mathbf{k}) = 4\int \frac{{\rm d} ^3 \mathbf{p}}{(2\pi)^{\frac{3}{2}}} Q_{\lambda}^{h^{(2)}_{ss}}(\mathbf{k},\mathbf{p})I^{h^{(2)}}_{ss}(\eta , p , |\mathbf{k}-\mathbf{p}|) \mathcal{R} _{\mathbf{p}}\mathcal{R} _{\mathbf{k}-\mathbf{p}} \,,
\end{equation}
with 
\begin{equation}
    Q_{\lambda}^{h^{(2)}_{ss}}(\mathbf{k},\mathbf{p})= \eps ^{ab}_{\lambda}(\mathbf{k}) ^*  p_{a}p_{b} \, ,
\end{equation}
\begin{equation}
    I^{h^{(2)}}_{ss}(\eta , p , |\mathbf{k}-\mathbf{p}|) = \left (\frac{3+3w}{5+3w} \right )^2 \int _{\eta _i}^{\infty} {\rm d}  \bareta \, \frac{a(\bareta)}{a(\eta)} G^h_k(\eta , \bareta)f^{h^{(2)}}_{ss}(\bareta , p , |\mathbf{k}-\mathbf{p}|) \, ,
\end{equation}
and
\begin{align}
    \begin{split}
    f^{h^{(2)}}_{ss}(\bareta , p , |\mathbf{k}-\mathbf{p}|)&=\frac{2}{3(1+w)} \bigg [ \bigg ((T_{\Psi}(\eta |\mathbf{k}-\mathbf{p}|) + \frac{T^{\prime}_{\Psi}(\eta |\mathbf{k}-\mathbf{p}|)}{\mathcal{H}}\bigg )  \bigg ( T_{\Psi}(\eta p) \\
    &+ \frac{T^{\prime}_{\Psi}(\eta p)}{\mathcal{H}} \bigg )  \bigg ] 
    +T_{\Psi}(\eta |\mathbf{k}-\mathbf{p}|)T_{\Psi}(\eta p)\, .
    \end{split}
\end{align}
During RD the dimensionless time-averaged power spectrum is
\begin{align}\label{powerspectrumsigws}
    \begin{split}
        \overline{\mathcal{P}_{h^{(2)}_{ss}}(\eta , k)} &= 8 \int _0^{\infty} {\rm d} v \int _{|1-v|}^{1+v} {\rm d} u \, \left (\frac{4v^2-(1-u^2+v^2)^2}{4vu} \right ) ^2 \overline{I^{h^{(2)}}_{ss}(x\rightarrow \infty , v , u)^2} \\
        &\times \mathcal{P}_{\mathcal{R}}(kv)\mathcal{P}_{\mathcal{R}}(ku)  \, ,
    \end{split}
\end{align}
with 
\begin{align}
     x^2\overline{I^{h^{(2)}}_{ss}(x\rightarrow \infty , v , u)^2} &= \frac{1}{2} \left ( \frac{3(u^2+v^2-3)}{8u^3v^3}\right )^2 \bigg [\left ( -4uv+(u^2+v^2-3) \log \bigg |\frac{3-(u+v)^2}{3-(u-v)^2} \bigg |\right )^2 \nonumber \\
    &+\pi ^2 (u^2+v^2-3)^2 \Theta (v+u-\sqrt{3})  \bigg]\,.
\end{align}

For the scalar-tensor term we have \cite{Yu:2023lmo,Picard:2023sbz} 
\begin{equation} \label{solh2st}
    h^{(2)st}_{\lambda}(\eta , \mathbf{k}) = 4  \left (\frac{3+3w}{5+3w} \right ) \sum _{\lambda _1}\int \frac{{\rm d} ^3 \mathbf{p}}{(2\pi)^{\frac{3}{2}}} Q_{\lambda ,\lambda _1}^{h^{(2)}_{st}}(\mathbf{k},\mathbf{p})I^{h^{(2)}}_{st}(\eta , p , |\mathbf{k}-\mathbf{p}|) \mathcal{R} _{\mathbf{k}-\mathbf{p}}h^{\lambda _1} _{\mathbf{p}} \,,
\end{equation}
with 
\begin{equation}
    Q_{\lambda , \lambda _1}^{h^{(2)}_{st}}(\mathbf{k},\mathbf{p})= \eps ^{ab}_{\lambda}(\mathbf{k}) ^*  \eps _{ab}^{\lambda _1}(\mathbf{p}) \, ,
\end{equation}
\begin{equation} \label{h2stkernel}
    I^{h^{(2)}}_{st}(\eta , p , |\mathbf{k}-\mathbf{p}|) =\int _{\eta _i}^{\infty} {\rm d}  \bareta \, \frac{a(\bareta)}{a(\eta)} G^h_k(\eta , \bareta)f^{h^{(2)}}_{st}(\bareta , p , |\mathbf{k}-\mathbf{p}|) \, ,
\end{equation}
and
\begin{align}
    \begin{split}
    f^{h^{(2)}}_{st}(\bareta , p , |\mathbf{k}-\mathbf{p}|)&=-2p^2T_h(\eta p) T_{\Psi}( \eta|\mathbf{k}-\mathbf{p}|)-2(\mathbf{k}-\mathbf{p})\cdot \mathbf{p}\, T_h(\eta p) T_{\Psi}( \eta|\mathbf{k}-\mathbf{p}|) \\
    &+ \mathcal{H}(1+3w^2)T_h(\eta p) T_{\Psi}^{\prime}( \eta|\mathbf{k}-\mathbf{p}|)-|\mathbf{k}-\mathbf{p}|^2(1-w^2)T_h(\eta p) \\
    &\times T_{\Psi}( \eta|\mathbf{k}-\mathbf{p}|)\, .
    \end{split}
\end{align}
during RD
\begin{align}\label{powerspectrumstigws}
    \begin{split}
        \overline{\mathcal{P}_{h^{(2)}_{st} }(\eta , k)} &= \frac{1}{32} \int _0^{\infty} {\rm d} v \int _{|1-v|}^{1+v} {\rm d} u \,\frac{1}{u^2v^6}\left ( (u^2-(v+1)^2)^4 + (u^2-(v-1)^2)^4 \right )  \\
        &\times \overline{I^{h^{(2)}}_{st}(x\rightarrow \infty , v , u)^2} \left ( \mathcal{P}^R_{h}(kv) + \mathcal{P}^L_{h}(kv) \right )\mathcal{P}_{\mathcal{R}}(ku)  \, ,
    \end{split}
\end{align}
with 
\begin{align}\label{kernelscalartensor}
    \begin{split}
        x^2 \overline{I^{h^{(2)}}_{st}(x\rightarrow \infty , v , u)^2} &=  \frac{1}{1152v^2u^6} \bigg [3 \pi ^2 (u^2-3(v-1)^2)^2(u^2-3(v+1)^2)^2 \theta (v+\frac{u}{\sqrt{3}}-1)  \\
        &+ \bigg ( 4uv\left (9-9v^2 +u^2 \right )-\sqrt{3}(u^2-3(v-1)^2)(u^2-3(v+1)^2) \\
        &\times \log \bigg | \frac{\left (\sqrt{3}v-u \right )^2-3}{\left (\sqrt{3}v+u \right )^2-3} \bigg | \bigg )^2 \bigg ]
    \end{split}
\end{align}
To study the integrand of Eq.~\eqref{powerspectrumstigws} in different limits we define
\begin{equation}
    \overline{\mathcal{P}_{h^{(2)}_{st} }(\eta , k)} = \int _0^{\infty} {\rm d} v \int _{|1-v|}^{1+v} {\rm d} u \, \, \mathcal{F}_{st} (v,u) \left ( \mathcal{P}^R_{h}(kv) + \mathcal{P}^L_{h}(kv) \right )\mathcal{P}_{\mathcal{R}}(ku) \, ,
\end{equation}
with 
\begin{equation}\label{stintegrand}
     \mathcal{F}_{st} (v,u) \equiv \frac{1}{32} \frac{1}{u^2v^6}\left ( (u^2-(v+1)^2)^4 + (u^2-(v-1)^2)^4 \right )
         \overline{I^{h^{(2)}}_{st}(x\rightarrow \infty , v , u)^2}
\end{equation}

\section{Nested kernels in a RD universe} \label{nestedkernels}
In this Appendix we present the analytical results of the nested kernels in RD universe.

For the terms involving scalar perturbations, the relevant results are 
\begin{subequations}
\setlength{\jot}{1pt}
    \begin{align}
        \begin{split} \label{Ic1RDfinal}
        &I^{\Psi^{(2)}}_{c1}(\barx ,v,u) = \frac{1}{120 u^3 v^2 \barx^6} \bigg\{  
        - v \times a(v, u) \barx^6 \bigg[ \text{Ci} \left( \barx \left|1 + \frac{u - v}{\sqrt{3}} \right| \right)  
        - \log \left| \sqrt{3} + u - v \right| \bigg] \\  
        &+ v \times b(v, u) \barx^6 \bigg[ \text{Ci} \left( \barx \left|1 + \frac{-u + v}{\sqrt{3}} \right| \right)  
        - \log \left| \sqrt{3} - u + v \right| \bigg] \\  
        &+ v \times c(v, u) \barx^6 \bigg[ \text{Ci} \left( \barx \left|-1 + \frac{u + v}{\sqrt{3}} \right| \right)  
        - \log \left| -\sqrt{3} + u + v \right| \bigg] \\  
        &+ v \times d(v, u) \barx^6 \bigg[ \text{Ci} \left( \barx \left|1 + \frac{u + v}{\sqrt{3}} \right| \right)  
        - \log \left| \sqrt{3} + u + v \right| \bigg] \\  
        &+ 36 \sin \left( \frac{u \barx}{\sqrt{3}} \right) \bigg\{  
        - 3 \sqrt{3} \cos \left( \frac{v \barx}{\sqrt{3}} \right) 
        \bigg[ v^2 \barx^3 \bigg( -6 + \left( -\frac{7}{3} + u^2 + \frac{23 v^2}{9} \right) \barx^2 \bigg) \cos \barx \\  
        &+ 4 \bigg( -90 + (-15 + 15 u^2 - 11 v^2) \barx^2 + v^2 \left( -\frac{7}{6} + u^2 - \frac{2 v^2}{9} \right) \barx^4 \bigg) \sin \barx \bigg] \\  
        &- \frac{1}{3} v \barx \bigg[ 3 \barx \left( 54 + (21 - 9 u^2 + 31 v^2) \barx^2 \right) \cos \barx + ( 648 + 6 (21 - 18 u^2 + 22 v^2) \barx^2 \\  
        & + (-63 - 12 u^4 - 51 v^2 + 8 v^4 + u^2 (69 + 8 v^2)) \barx^4 ) \sin \barx \bigg] \sin \left( \frac{v \barx}{\sqrt{3}} \right)  
        \bigg\} \\  
        &+ 12 \barx \bigg[ u \bigg( 12 u^2 (-5 + u^2) - (21 + 17 u^2) v^2 + 23 v^4 \bigg) \barx^5 + u \cos \left( \frac{u \barx}{\sqrt{3}} \right) \bigg\{  2 \cos \left( \frac{v \barx}{\sqrt{3}} \right) \bigg[ -1620 \sin \barx \\  
        &+ \barx^2 \bigg( 90 (-3 + u^2) \sin \barx + v^2 \bigg( -27 \barx \cos \barx + (-198 + (-21 + 6 u^2 - 4 v^2) \barx^2) \sin \barx \bigg) \bigg) \bigg] \\  
        &+ \sqrt{3} v \barx \bigg[ \barx (54 + (21 - 3 u^2 + 31 v^2) \barx^2) \cos \barx + 2 (108 + (21 - 6 u^2 + 22 v^2) \barx^2) \sin \barx \bigg] \sin \left( \frac{v \barx}{\sqrt{3}} \right)  
        \bigg\}  
        \bigg]  
    \bigg\} ,
        \end{split}
        \\
        \begin{split} \label{Is1RDfinal}
        &I^{\Psi^{(2)}}_{s1}(\barx ,v,u) =\frac{1}{120 u^3 v^2 \barx^6} \bigg\{  
    36 \sin \left( \frac{u \barx}{\sqrt{3}} \right) \bigg[  
    \frac{1}{3} v \barx \cos \left( \frac{v \barx}{\sqrt{3}} \right)  
    \bigg( 3 \barx (54 + (21 - 9 u^2 + 31 v^2) \barx^2) \cos \barx \\  
    &+ (648 + 6 (21 - 18 u^2 + 22 v^2) \barx^2  
    + (-63 - 12 u^4 - 51 v^2 + 8 v^4 + u^2 (69 + 8 v^2)) \barx^4) \sin \barx  
    \bigg) \\  
    &- 3 \sqrt{3} \bigg(  
    v^2 \barx^3 (-6 + (-\frac{7}{3} + u^2 + \frac{23 v^2}{9}) \barx^2) \cos \barx + 4 (-90 + (-15 + 15 u^2 - 11 v^2) \barx^2 \\  
    &+ v^2 (-\frac{7}{6} + u^2 - \frac{2 v^2}{9}) \barx^4) \sin \barx  
    \bigg) \sin \left( \frac{v \barx}{\sqrt{3}} \right)  
    \bigg] \\  
    &+ 12 u \barx \cos \left( \frac{u \barx}{\sqrt{3}} \right) \bigg[  
    - \sqrt{3} v \barx \cos \left( \frac{v \barx}{\sqrt{3}} \right)  
    \bigg( \barx (54 + (21 - 3 u^2 + 31 v^2) \barx^2) \cos \barx \\  
    &+ 2 (108 + (21 - 6 u^2 + 22 v^2) \barx^2) \sin \barx  
    \bigg) -2 \bigg( 1620 \sin \barx  
    + \barx^2 \bigg( -90 (-3 + u^2) \sin \barx \\  
    &+ v^2 (27 \barx \cos \barx  
    + (198 + (21 - 6 u^2 + 4 v^2) \barx^2) \sin \barx)  
    \bigg) \bigg) \sin \left( \frac{v \barx}{\sqrt{3}} \right)  
    \bigg] \\  
    &- v \times b(v, u) \barx^6 \text{Si} \left(-\barx + \frac{(u - v) \barx}{\sqrt{3}} \right)  
    + v \times c(v, u) \barx^6 \text{Si} \left(-\barx + \frac{(u + v) \barx}{\sqrt{3}} \right) \\  
    &+ v \times a(v, u) \barx^6 \text{Si} \left(\barx + \frac{(u - v) \barx}{\sqrt{3}} \right)  
    + v \times d(v, u) \barx^6 \text{Si} \left(\barx + \frac{(u + v) \barx}{\sqrt{3}} \right)  
    \bigg\}  ,
        \end{split}
    \end{align}
\end{subequations}
with
\begin{subequations}
\setlength{\jot}{1pt}
    \begin{align}
        \begin{split}
            a( v, u) &= -90 u^2 \left(v^2+3\right)+v^2 \left(v \left(v \left(8 \sqrt{3} v+45\right)-120 \sqrt{3}\right)+90\right)+189 \\
            &+12 \sqrt{3} u^5+ 45 u^4-20 \sqrt{3} u^3 \left(v^2+3\right) \, ,
        \end{split}
        \\
        \begin{split}
            b(v, u) &= 90 u^2 \left(v^2+3\right)+v^2 \left(v \left(v \left(8 \sqrt{3} v-45\right)-120 \sqrt{3}\right)-90\right)-189 \\
            &+12 \sqrt{3} u^5-45 u^4-20 \sqrt{3} u^3 \left(v^2+3\right) \, ,
        \end{split}
        \\
        \begin{split}
            c( v, u) &=-90 u^2 \left(v^2+3\right)+v^2 \left(v \left(v \left(8 \sqrt{3} v+45\right)-120 \sqrt{3}\right)+90\right)+189 \\ 
            & -12 \sqrt{3} u^5+45 u^4+20 \sqrt{3} u^3 \left(v^2+3\right) \, ,
        \end{split}
        \\
        \begin{split}
            d( v, u) &= -90 u^2 \left(v^2+3\right)+v^2 \left(v \left(v \left(45-8 \sqrt{3} v\right)+120 \sqrt{3}\right)+90\right)+189 \\
            &+12 \sqrt{3} u^5+45 u^4-20 \sqrt{3} u^3 \left(v^2+3\right)\, .
        \end{split}
    \end{align}
\end{subequations}
Furthermore,
\begin{subequations}
    \begin{align}
    \begin{split} \label{Ic2RDfinal}
    I^{\Psi^{(2)}}_{c2}(\barx ,v,u) &= \frac{3}{10 u^3 v \barx^5} \bigg\{  
    \cos \left( \frac{v \barx}{\sqrt{3}} \right) \bigg[  
    3 u \barx \cos \left( \frac{u \barx}{\sqrt{3}} \right)  
    \bigg( \barx (-18 + (-7 + u^2 - \frac{31 v^2}{3}) \barx^2) \cos \barx \\  
    &+ 2 (-216 + (-37 + 12 u^2 - \frac{82 v^2}{3}) \barx^2) \sin \barx  
    \bigg) + \frac{1}{\sqrt{3}} \bigg(  
    3 \barx (54 + (21 - 9 u^2 + 31 v^2) \barx^2) \cos \barx \\  
    &+ (3888 + 6 (111 - 108 u^2 + 82 v^2) \barx^2  
    - (63 + 12 u^4 - 39 v^2 + 22 v^4 + u^2 (-69 + 22 v^2)) \barx^4) \\
    &\times\sin \barx  
    \bigg) \sin \left( \frac{u \barx}{\sqrt{3}} \right)  
    \bigg]- 6 v \barx \bigg[  
    \sqrt{3} u \barx \cos \left( \frac{u \barx}{\sqrt{3}} \right) 
    \bigg( -7 \barx \cos \barx + (-18 + (-\frac{8}{3} + u^2 \\  
    &- \frac{41 v^2}{9}) \barx^2) \sin \barx  
    \bigg) + \frac{1}{6} \bigg( \barx (126 + (69 - 21 u^2 + 113 v^2) \barx^2) \cos \barx \\  
    &+ 2 (162 + (24 - 27 u^2 + 41 v^2) \barx^2) \sin \barx  
    \bigg) \sin \left( \frac{u \barx}{\sqrt{3}} \right)  
    \bigg] \sin \left( \frac{v \barx}{\sqrt{3}} \right) \\  
    &+ e(v, u) \barx^5 \text{Si} \left(-\barx + \frac{(u - v) \barx}{\sqrt{3}} \right)  
    - g(v, u) \barx^5 \text{Si} \left(-\barx + \frac{(u + v) \barx}{\sqrt{3}} \right) \\  
    &+ f(v, u) \barx^5 \text{Si} \left(\barx + \frac{(u - v) \barx}{\sqrt{3}} \right)  
    + h(v, u) \barx^5 \text{Si} \left(\barx + \frac{(u + v) \barx}{\sqrt{3}} \right)  
    \bigg\}  ,
    \end{split}
    \\
    \begin{split} \label{Is2RDfinal}
    I^{\Psi^{(2)}}_{s2}(\barx ,v,u) &=\frac{3}{10 u^3 v \barx^5} \bigg\{  
        e(v, u) \barx^5 \bigg[ \text{Ci} \left( \barx \left|-1 + \frac{u - v}{\sqrt{3}} \right| \right)  
        - \log \left| -\sqrt{3} + u - v \right| \bigg] \\  
        &+ f(v, u) \barx^5 \bigg[ \text{Ci} \left( \barx \left|1 + \frac{u - v}{\sqrt{3}} \right| \right)  
        - \log \left| \sqrt{3} + u - v \right| \bigg] \\  
        &+ g(v, u) \barx^5 \bigg[ \text{Ci} \left( \barx \left|-1 + \frac{u + v}{\sqrt{3}} \right| \right)  
        - \log \left| -\sqrt{3} + u + v \right| \bigg] \\  
        &- h(v, u) \barx^5 \bigg[ \text{Ci} \left( \barx \left|1 + \frac{u + v}{\sqrt{3}} \right| \right)  
        - \log \left| \sqrt{3} + u + v \right| \bigg] \\  
        &+ \frac{1}{3} \bigg[ \sqrt{3} u v (-69 + 7 u^2 - 113 v^2) \barx^5 + u \barx \cos \left( \frac{u \barx}{\sqrt{3}} \right) \bigg\{  
        -2 \sqrt{3} v \barx \cos \left( \frac{v \barx}{\sqrt{3}} \right)   \\  
        &\times \bigg[ 63 \barx \cos \barx + \big( 162 + (24 - 9 u^2 + 41 v^2) \barx^2 \big) \sin \barx \bigg] -3 \bigg[ \barx (54 + (21 - 3 u^2 + 31 v^2) \barx^2) \cos \barx  \\  
        &
        + 2 (648 + (111 - 36 u^2 + 82 v^2) \barx^2) \sin \barx \bigg]  
        \sin \left( \frac{v \barx}{\sqrt{3}} \right) \bigg\} \\  
        &+ \sin \left( \frac{u \barx}{\sqrt{3}} \right) \bigg\{  
        3 v \barx \cos \left( \frac{v \barx}{\sqrt{3}} \right)  
        \bigg[ \barx (126 + (69 - 21 u^2 + 113 v^2) \barx^2) \cos \barx  
        + 2 (162\\  
        & + (24 - 27 u^2 + 41 v^2) \barx^2) \sin \barx \bigg] + \sqrt{3} \bigg[ 3 \barx (54 + (21 - 9 u^2 + 31 v^2) \barx^2) \cos \barx \\  
        &+ (3888 + 6 (111 - 108 u^2 + 82 v^2) \barx^2  
        - (63 + 12 u^4 - 39 v^2 + 22 v^4 \\
        &+ u^2 (-69 + 22 v^2)) \barx^4) \sin \barx \bigg]  
        \sin \left( \frac{v \barx}{\sqrt{3}} \right) \bigg\}  
        \bigg]  
    \bigg\}  ,
    \end{split}
    \end{align}
\end{subequations}
with 
\begin{subequations}
    \begin{align}
        \begin{split}
            e( v, u) &=\frac{5}{2} \left(u^2-3\right)^2 v+\frac{1}{4} \sqrt{3} \left(5 \left(u^4-2 u^2 \left(v^2+3\right)+9 v^4\right)+10 v^2+21\right)+\frac{11 v^5}{6} \\
            & -u^5+\frac{5 u^3 v^2}{3}+5 u^3-5 \left(u^2-5\right) v^3\, ,
        \end{split}
        \\
        \begin{split}
            f( v, u) &=  
            -\frac{5}{2} \left(u^2-3\right)^2 v+\frac{1}{4} \sqrt{3} \left(5 \left(u^4-2 u^2 \left(v^2+3\right)+9 v^4\right)+10 v^2+21\right)-\frac{11 v^5}{6} \\
            &+u^5-\frac{5 u^3 v^2}{3}-5 u^3+5 \left(u^2-5\right) v^3\, ,
        \end{split}
        \\
        \begin{split}
            g( v, u) &=  
            \frac{5}{2} \left(u^2-3\right)^2 v-\frac{1}{4} \sqrt{3} \left(5 \left(u^4-2 u^2 \left(v^2+3\right)+9 v^4\right)+10 v^2+21\right)+\frac{11 v^5}{6}\\
            &+u^5-\frac{5 u^3 v^2}{3}-5 u^3-5 \left(u^2-5\right) v^3\, ,
        \end{split}
        \\
        \begin{split}
            h( v, u) &=  
            \frac{5}{2} \left(u^2-3\right)^2 v+\frac{1}{4} \sqrt{3} \left(5 \left(u^4-2 u^2 \left(v^2+3\right)+9 v^4\right)+10 v^2+21\right)+\frac{11 v^5}{6} \\
            &+u^5-\frac{5 u^3 v^2}{3}-5 u^3-5 \left(u^2-5\right) v^3\, .
        \end{split}
    \end{align}
\end{subequations}

For the second-order vector nested kernels 
\begin{subequations}
    \begin{align} \label{RDnestedb2psi11}
        I^{B^{(2)}}_{st1}(\barx ,k,v,u) &=\frac{18 \sqrt{3}}{k^3 u^3 \overline{x}^3} 
        \bigg\{
        2 \sin \overline{x} \sin \left(\frac{u \overline{x}}{\sqrt{3}}\right)  
        + \overline{x} \text{Si} \left( \overline{x} - \frac{u \overline{x}}{\sqrt{3}} \right)  
        - \overline{x} \text{Si} \left( \overline{x} + \frac{u \overline{x}}{\sqrt{3}} \right)  
        \bigg\} 
        \\
        \begin{split} \label{RDnestedb2psi12}
            I^{B^{(2)}}_{st2}(\barx ,k,v,u) &= \frac{3}{ku^3 \overline{x}^5} \bigg\{  
            6 u \overline{x} \cos \left(\frac{u \overline{x}}{\sqrt{3}}\right)  
            \left(\overline{x} \cos \overline{x} + 2 \sin \overline{x} \right) \\  
            &+ 2 \sqrt{3} \bigg( -3 \overline{x} \cos \overline{x}  
            + \left(-6 + ( u^2 - 3 v^2) \overline{x}^2 \right) \sin \overline{x} \bigg)  
            \sin \left(\frac{u \overline{x}}{\sqrt{3}}\right) \\  
            &+ 3\sqrt{3} v^2 
            \overline{x}^3 \bigg[  
            \text{Si} \left(- \overline{x} + \frac{u \overline{x}}{\sqrt{3}} \right)  
            + \text{Si} \left( \overline{x} + \frac{u \overline{x}}{\sqrt{3}} \right)  
            \bigg]  
            \bigg\} .
        \end{split}
    \end{align}
\end{subequations}

And finally, for the second-order tensors
\begin{align}\label{RDnestedh2psi1}
    \begin{split}
        I^{h^{(2)}}_{st}(\barx ,k,v,u) &=\frac{1}{16 u^3 v \overline{x}^5} \bigg \{ 4 \overline{x} 
        \Big( 
        18 u v \overline{x} \cos\left(\frac{u \overline{x}}{\sqrt{3}}\right) \left(\overline{x} \cos \overline{x} + \sin \overline{x} \right) 
        + 3 \sqrt{3} v \big( -6 \overline{x} \cos \overline{x} \\
        &+ \left(-6 + (3 + u^2 - 3 v^2) \overline{x}^2\right) \sin \overline{x} \big ) 
        \sin\left(\frac{u \overline{x}}{\sqrt{3}}\right) 
        + u (-9 + u^2 + 9 v^2) \overline{x}^3 \sin(v \overline{x}) 
        \Big) \\ 
        &+ \sqrt{3} (u^2 - 3(-1 + v)^2) (u^2 - 3(1 + v)^2) \overline{x}^4 
        \bigg ( \sin(v \overline{x})
        \Big[-\text{Ci} \left( \overline{x} \left|1 + \frac{u}{\sqrt{3}} - v \right| \right) \\ 
        &+ \text{Ci} \left( \overline{x} \left|1 - \frac{u}{\sqrt{3}} + v \right| \right) 
        + \text{Ci} \left( \overline{x} \left|-1 + \frac{u}{\sqrt{3}} + v \right| \right) 
        - \text{Ci} \left( \overline{x} \left|1 + \frac{u}{\sqrt{3}} + v \right| \right) \\
        &+ \log \left| \frac{(\sqrt{3} + u)^2 - 3 v^2}{(\sqrt{3} - u)^2 - 3 v^2} \right| \Big ]  
        - \cos(v \overline{x}) \bigg [ 
        \text{Si} \left( (1 - \frac{u}{\sqrt{3}} + v) \overline{x} \right) \\
        &+ \text{Si} \left( (-1 + \frac{u}{\sqrt{3}} + v) \overline{x} \right) 
        - \text{Si} \left( (1 + \frac{u}{\sqrt{3}} + v) \overline{x} \right) 
        + \text{Si} \left( \overline{x} + \frac{u \overline{x}}{\sqrt{3}} - v \overline{x} \right) 
        \bigg ] 
        \bigg) 
        \bigg \} .
    \end{split}
\end{align}

\newpage

\section{Mesh plots of the integrands for the numerical contributions} \label{meshstuff}

\begin{figure}[h!]
    \centering
    \begin{subfigure}[b]{0.48\textwidth}
        \includegraphics[width=\linewidth]{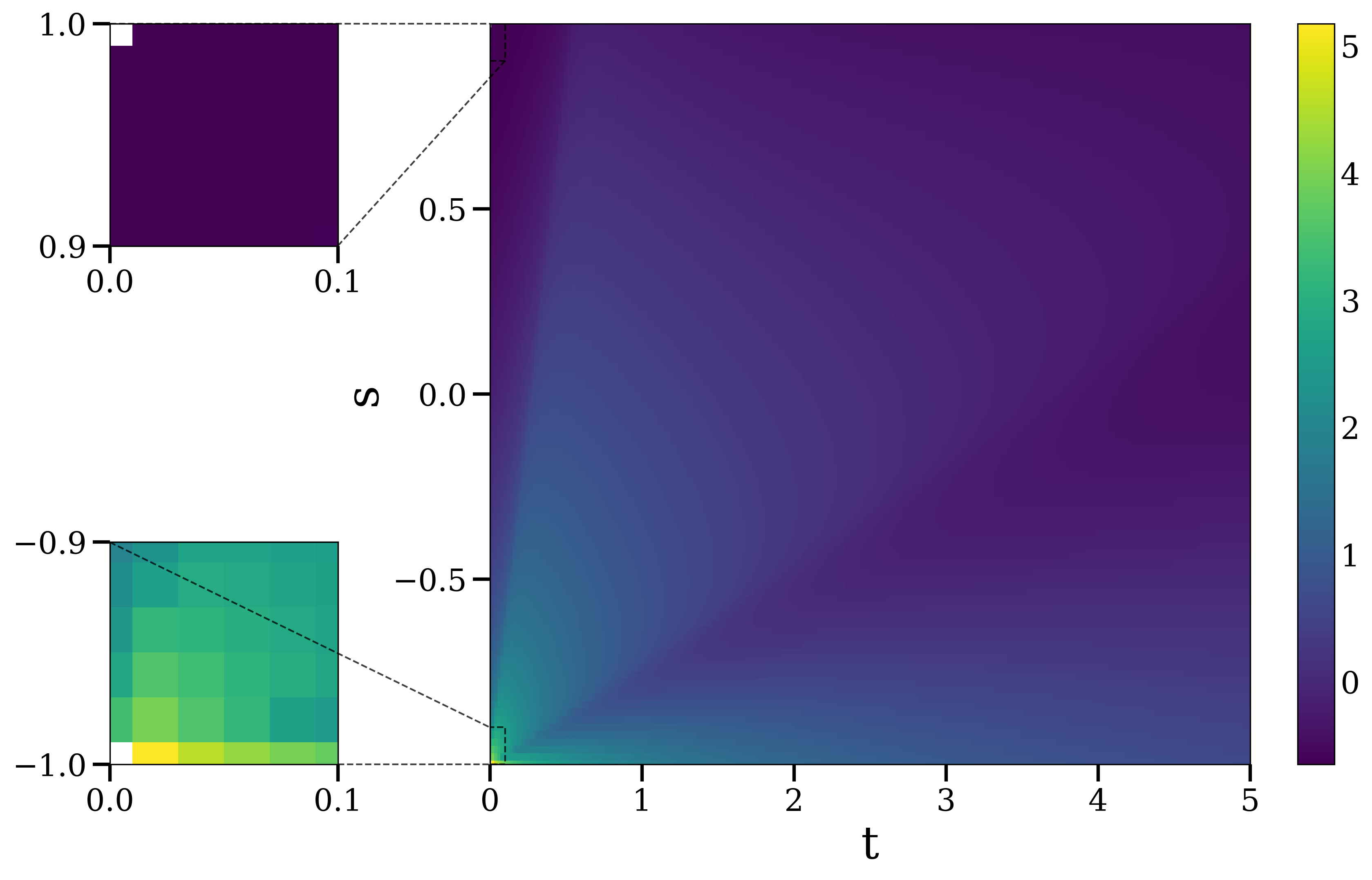}
    \end{subfigure}
    \begin{subfigure}[b]{0.48\textwidth}
        \includegraphics[width=\linewidth]{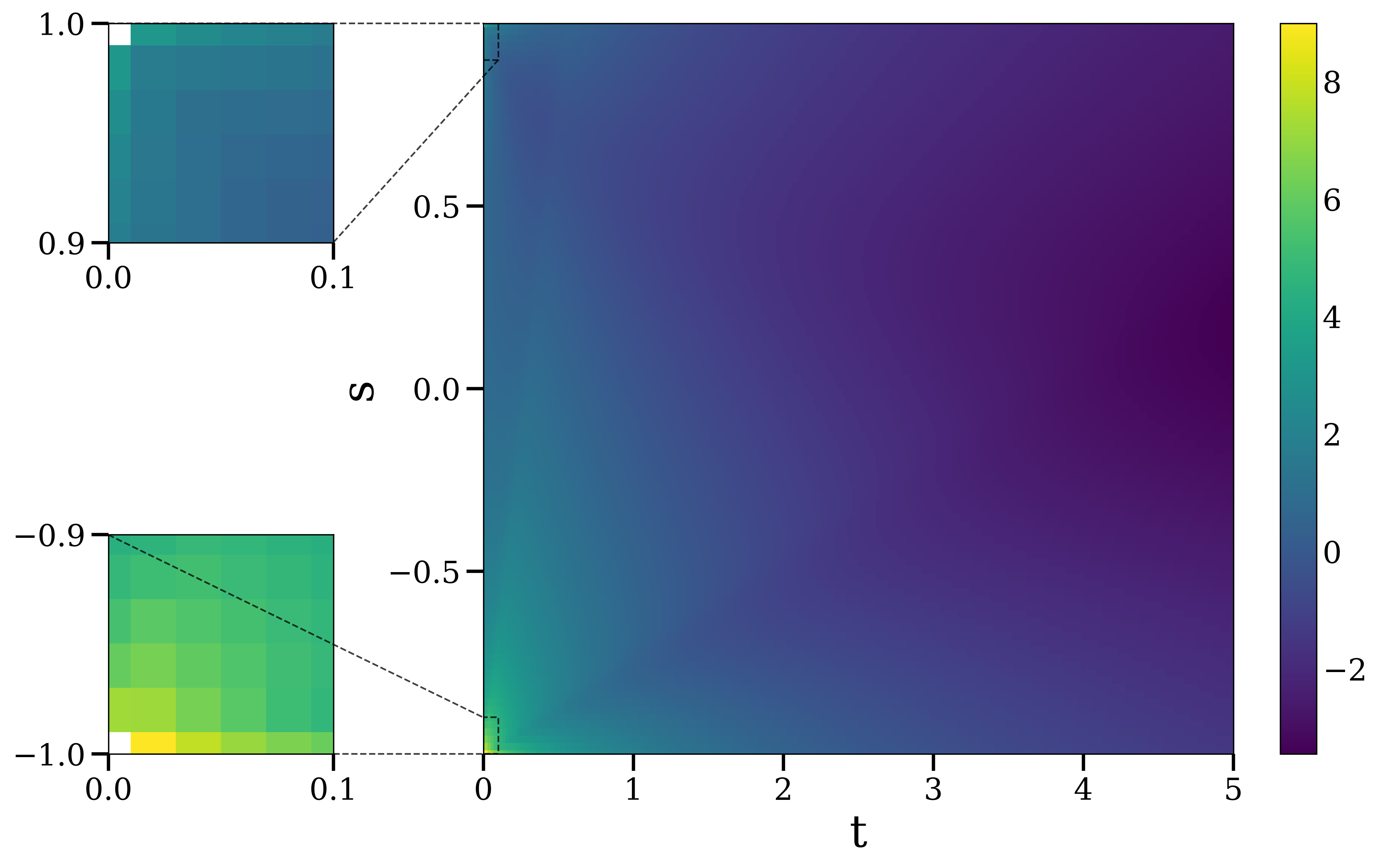}
    \end{subfigure}
    
    \begin{subfigure}[b]{0.48\textwidth}
        \includegraphics[width=\linewidth]{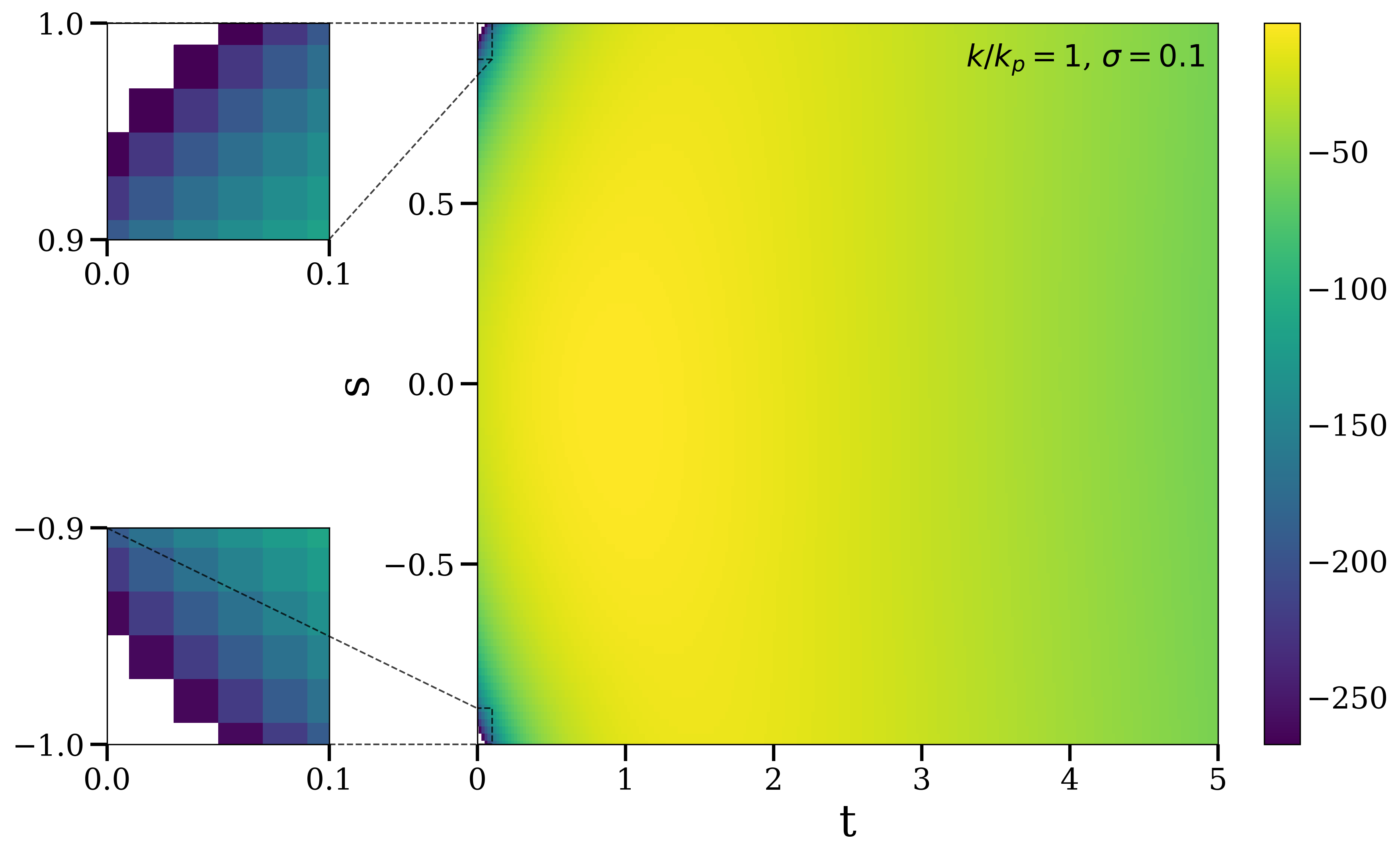}
    \end{subfigure}
    \begin{subfigure}[b]{0.48\textwidth}
        \includegraphics[width=\linewidth]{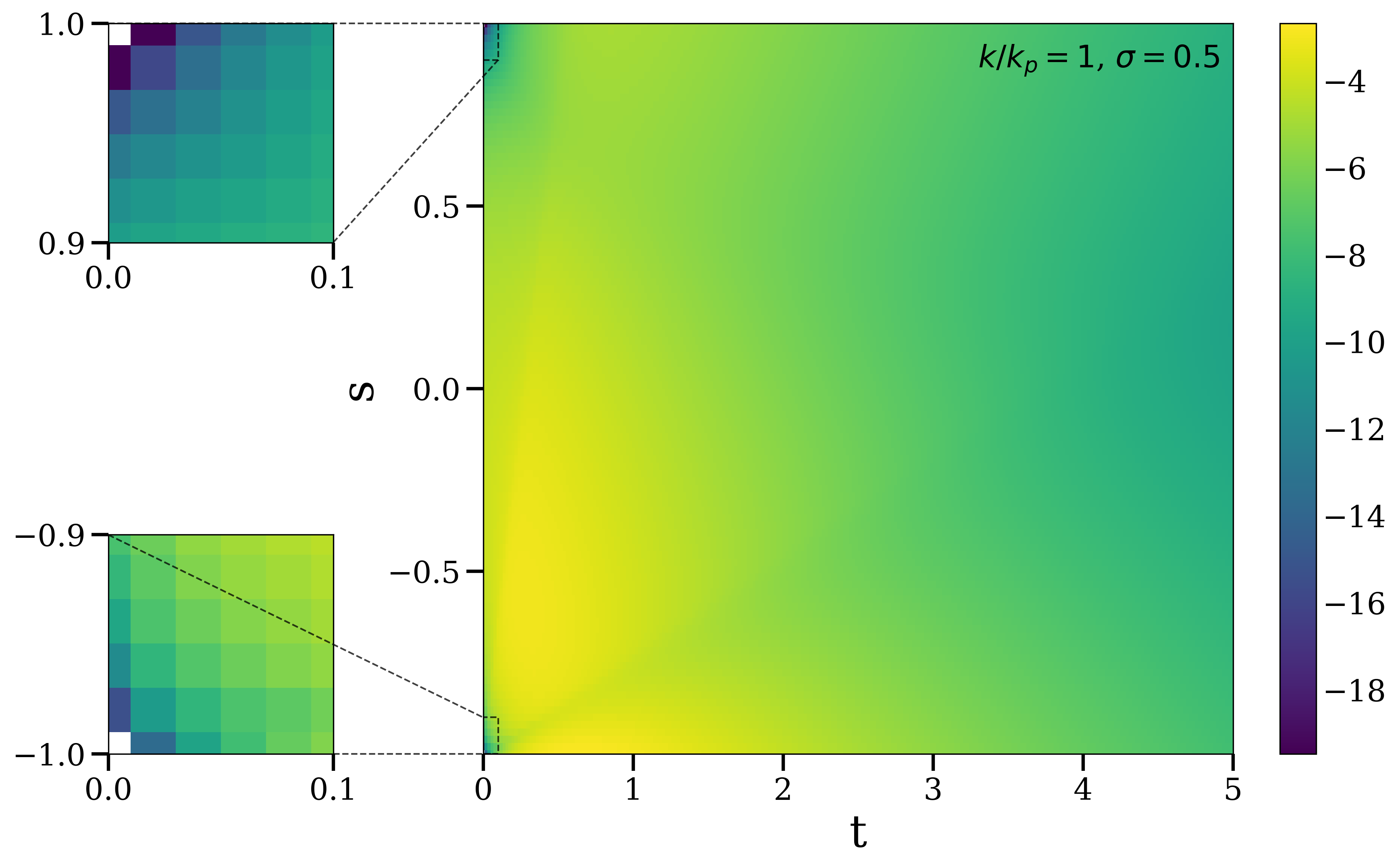}
    \end{subfigure}
    
    \begin{subfigure}[b]{0.48\textwidth}
        \includegraphics[width=\linewidth]{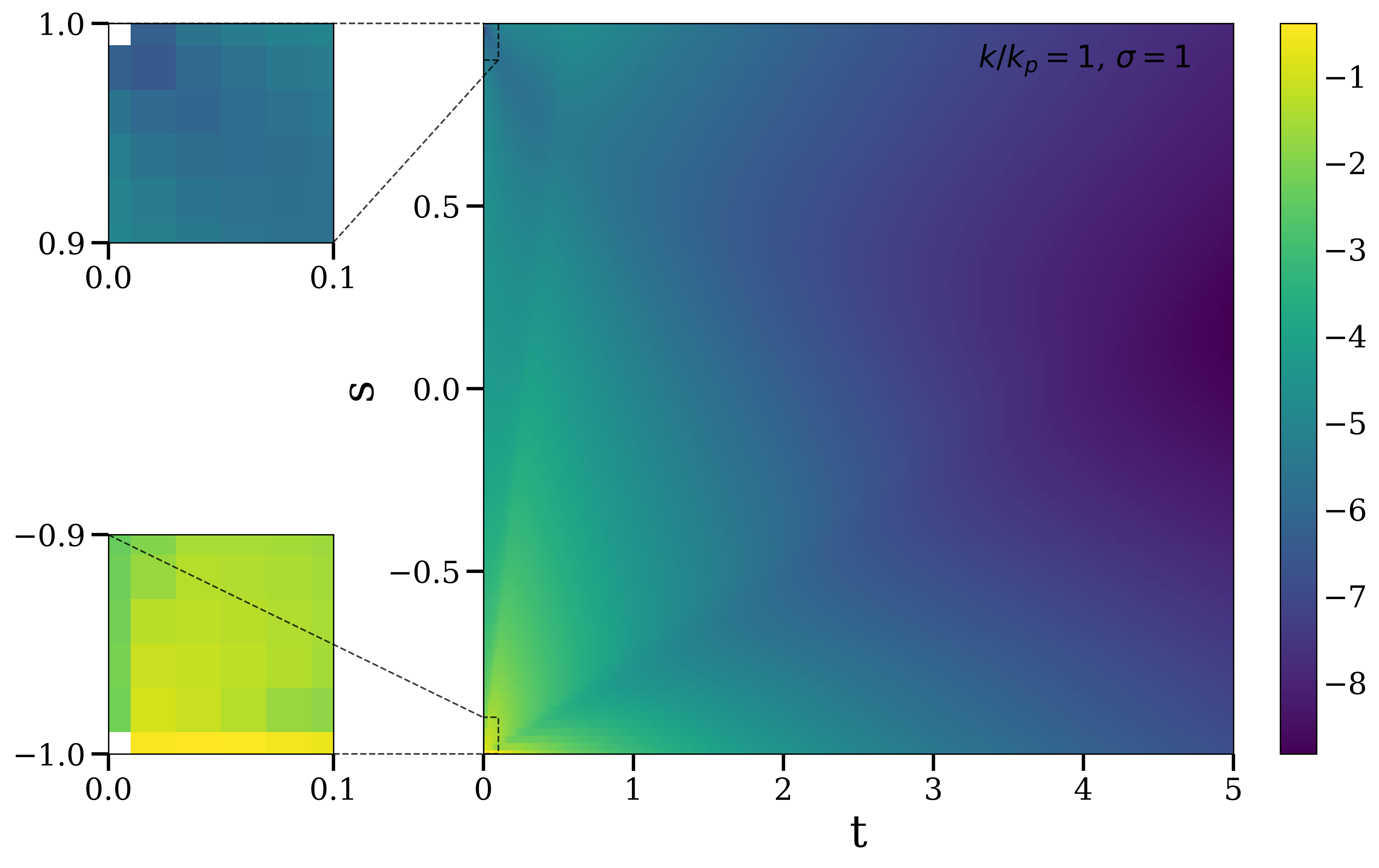}
    \end{subfigure}

    \caption{\footnotesize{Logarithmic mesh plots of the integrand of the power spectrum for the second-order scalar-tensor contribution showed in Eq.~\eqref{powerspectrumstigws} for different values of $\sigma$ in Eq.~\eqref{lognorm_def}. We note here that we do not show the integrand region beyond $t=5$ since we have established that the integrand is IR safe. (\textit{Top left}) Plot of the scalar-tensor kernel in a RD dominated universe. One of the UV limits blows up. (\textit{Top right}) Plot of the integrand without the contribution of the power spectrum, $\mathcal{F}_{st}(v,u)$ in Eq.~\eqref{stintegrand}. We see the plot blows up in the top-left corner ($v^{-2}$ divergence) and even quicker in the bottom-left corner ($u^{-4}$ divergence). (\textit{Middle left}) Plot of the integrand this time including the primordial scalar and tensor spectra, for $k=k_p$ where $k_p$ is where the primordial spectra peak. The inclusion of primordial power spectra has masked the divergence. (\textit{Middle right}) Same plot as previous but with $\sigma =0.5$. The overall numerical values of the integrand increase compared to $\sigma = 0.1$ (\textit{Bottom row}) Same plot as previous but with $\sigma =1$. We see that now the divergence as $v\rightarrow1$ and $u\rightarrow 0$ dominates in magnitude, and we conclude that it is the scalar power spectrum being too wide which causes the observable to be enhanced.}}
    \label{fig:st_meshplots}
\end{figure}

\newpage

\begin{figure}[h!]
    \centering
    \begin{subfigure}[b]{0.48\textwidth}
        \includegraphics[width=\linewidth]{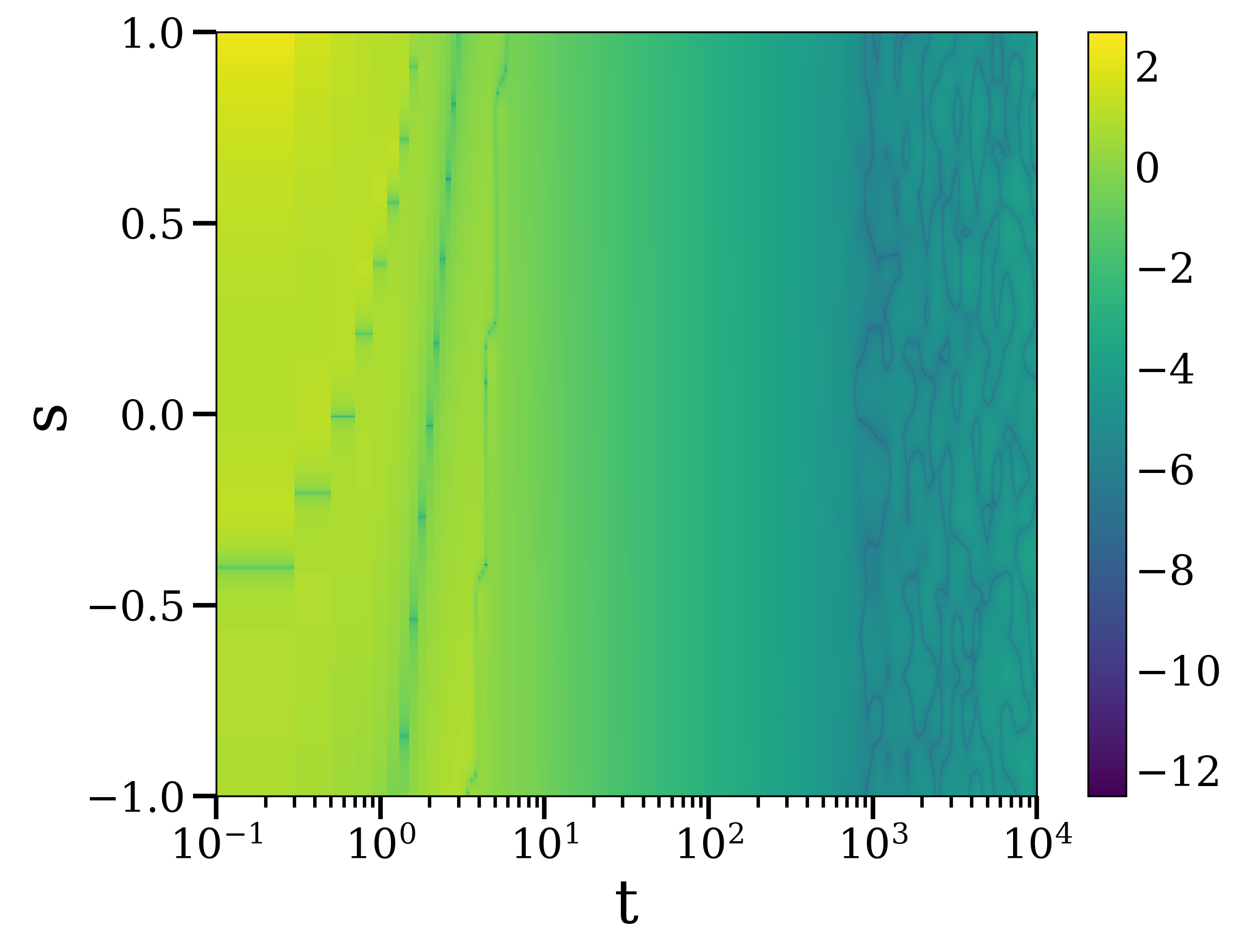}
    \end{subfigure}
    \begin{subfigure}[b]{0.48\textwidth}
        \includegraphics[width=\linewidth]{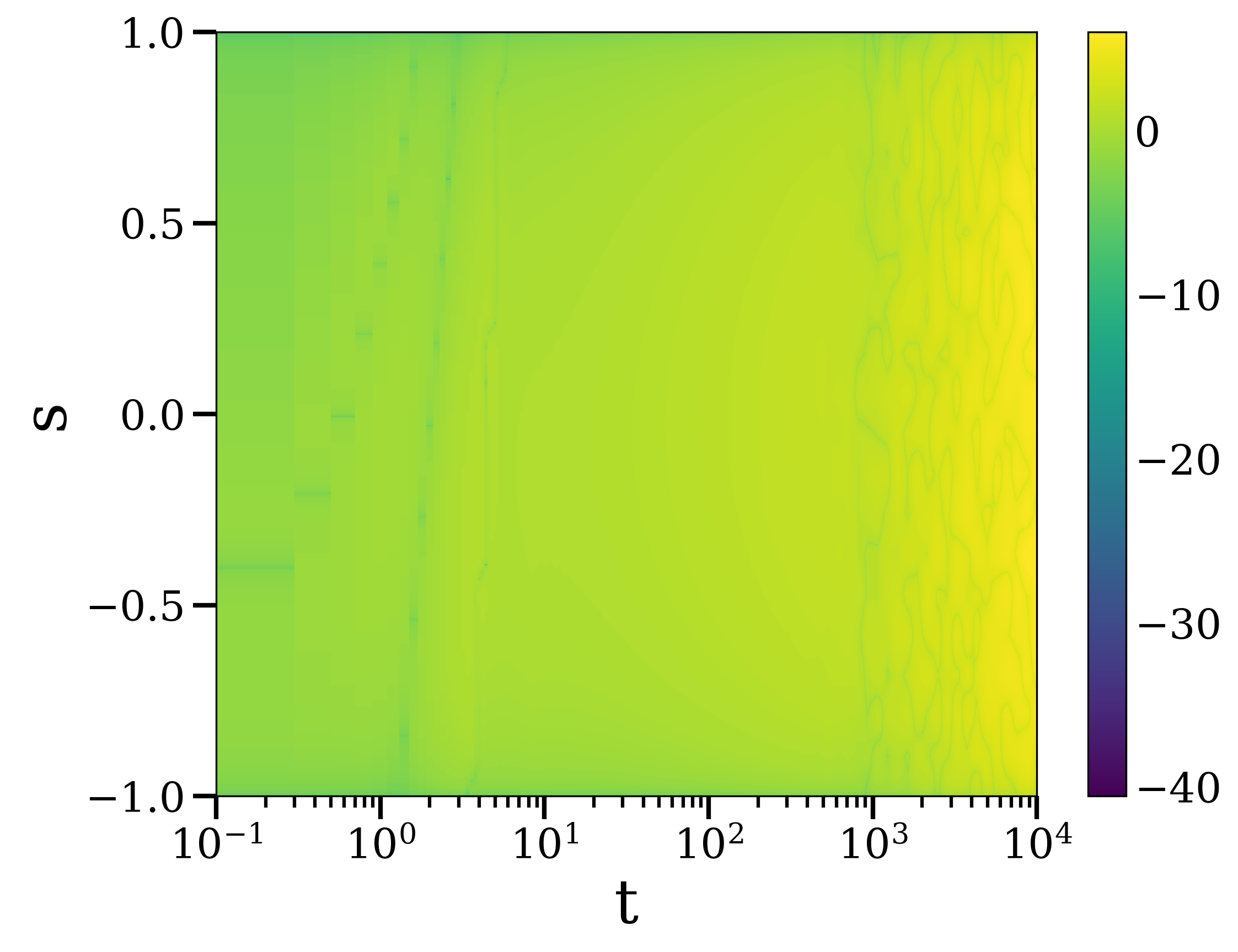}
    \end{subfigure}
    
    \begin{subfigure}[b]{0.48\textwidth}
        \includegraphics[width=\linewidth]{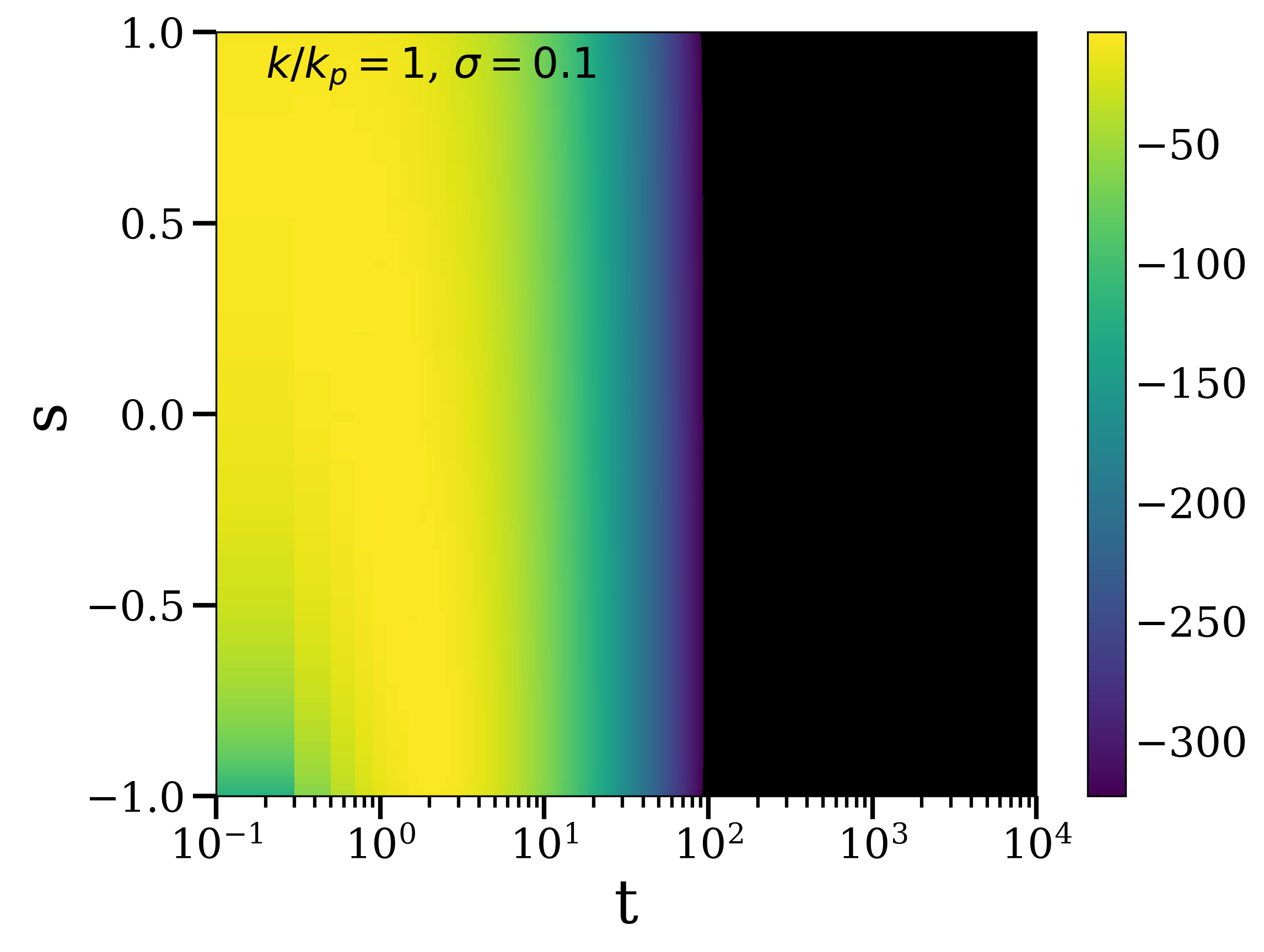}
    \end{subfigure}
    \begin{subfigure}[b]{0.48\textwidth}
        \includegraphics[width=\linewidth]{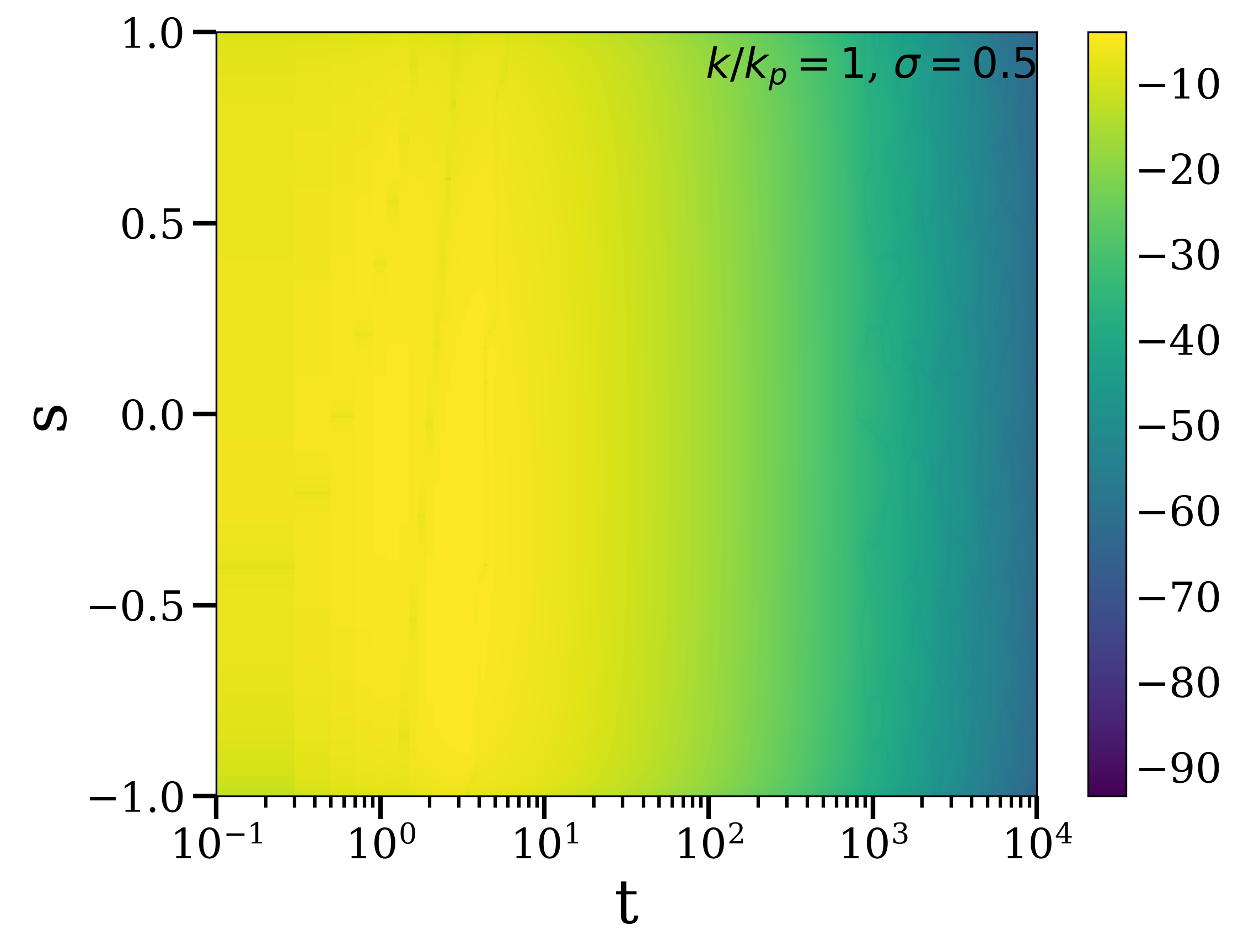}
    \end{subfigure}
    
    \begin{subfigure}[b]{0.48\textwidth}
        \includegraphics[width=\linewidth]{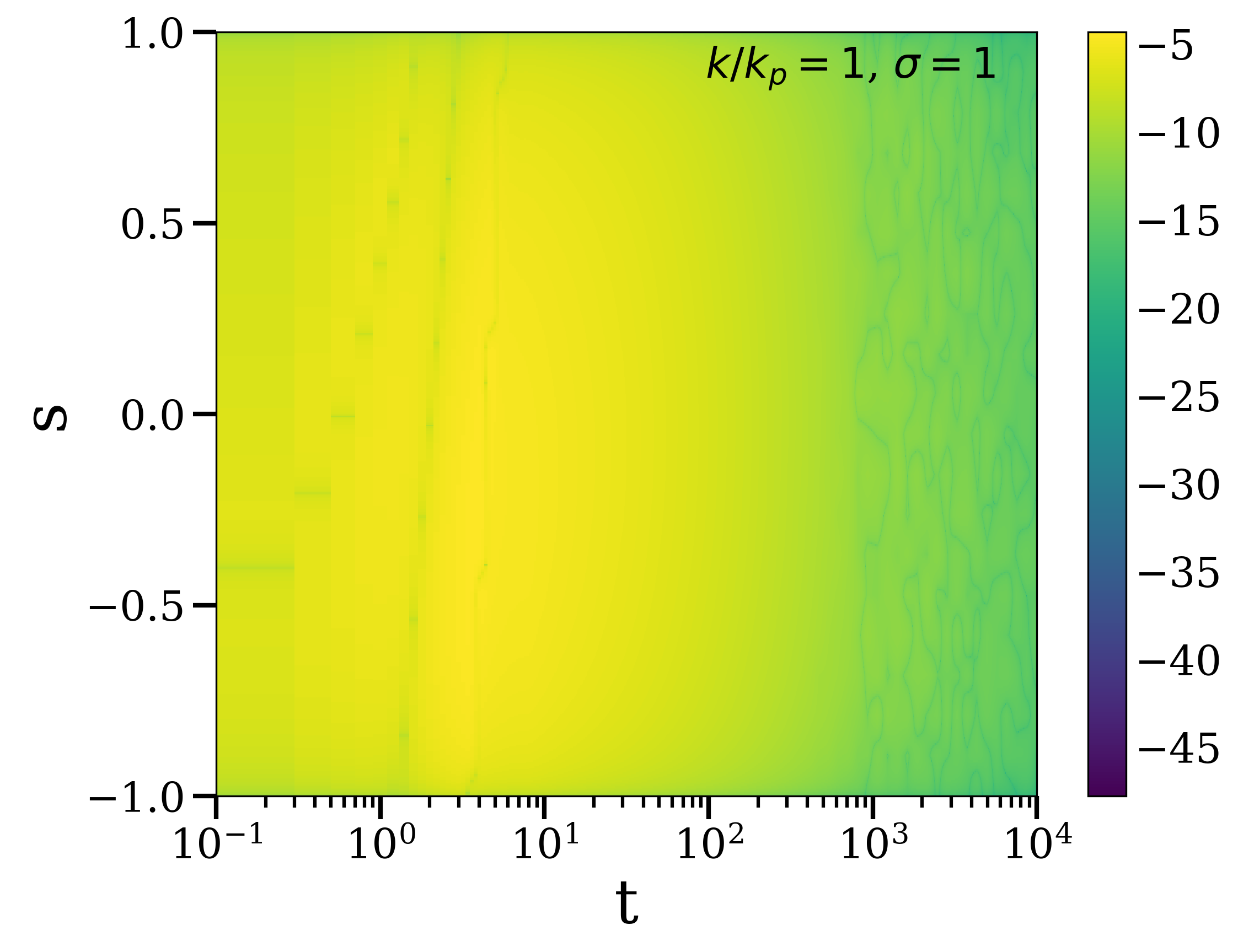}
    \end{subfigure}

    \caption{\footnotesize{Coloured mesh plots for the source term containing second-order scalars. (\textit{Top left}) Numerical kernel shown in Eq.~\eqref{finalkernelpsi2}  (\textit{Top right}) Numerical part of the integrand in Eq.~\eqref{pspsi2psi1} without the power spectra. (\textit{Middle left}) Integrand with the power spectra, with $\sigma = 0.1$ (\textit{Middle right}) $\sigma =0.5$ (\textit{Bottom row}) $\sigma =1$. Although the kernel exhibits an enhancement in the limit $v \rightarrow 1$ and $u \rightarrow 0$, we cannot conclude that there is a UV divergence as $\sigma$ increases.}}
    \label{fig:psi2psi1um_meshplots}
\end{figure}

\newpage

\begin{figure}[h!]
    \centering
    \begin{subfigure}[b]{0.48\textwidth}
        \includegraphics[width=\linewidth]{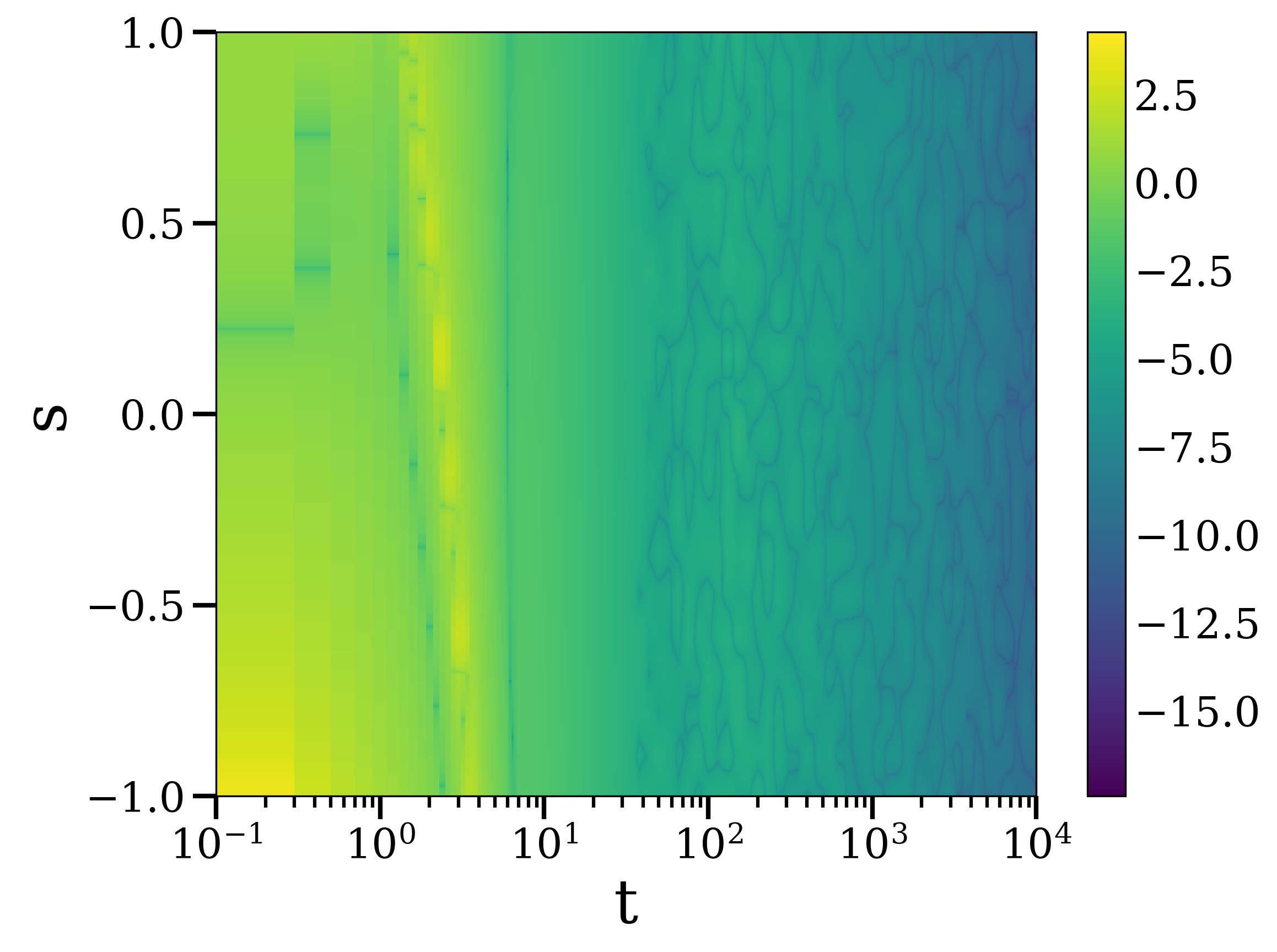}
    \end{subfigure}
    \begin{subfigure}[b]{0.48\textwidth}
        \includegraphics[width=\linewidth]{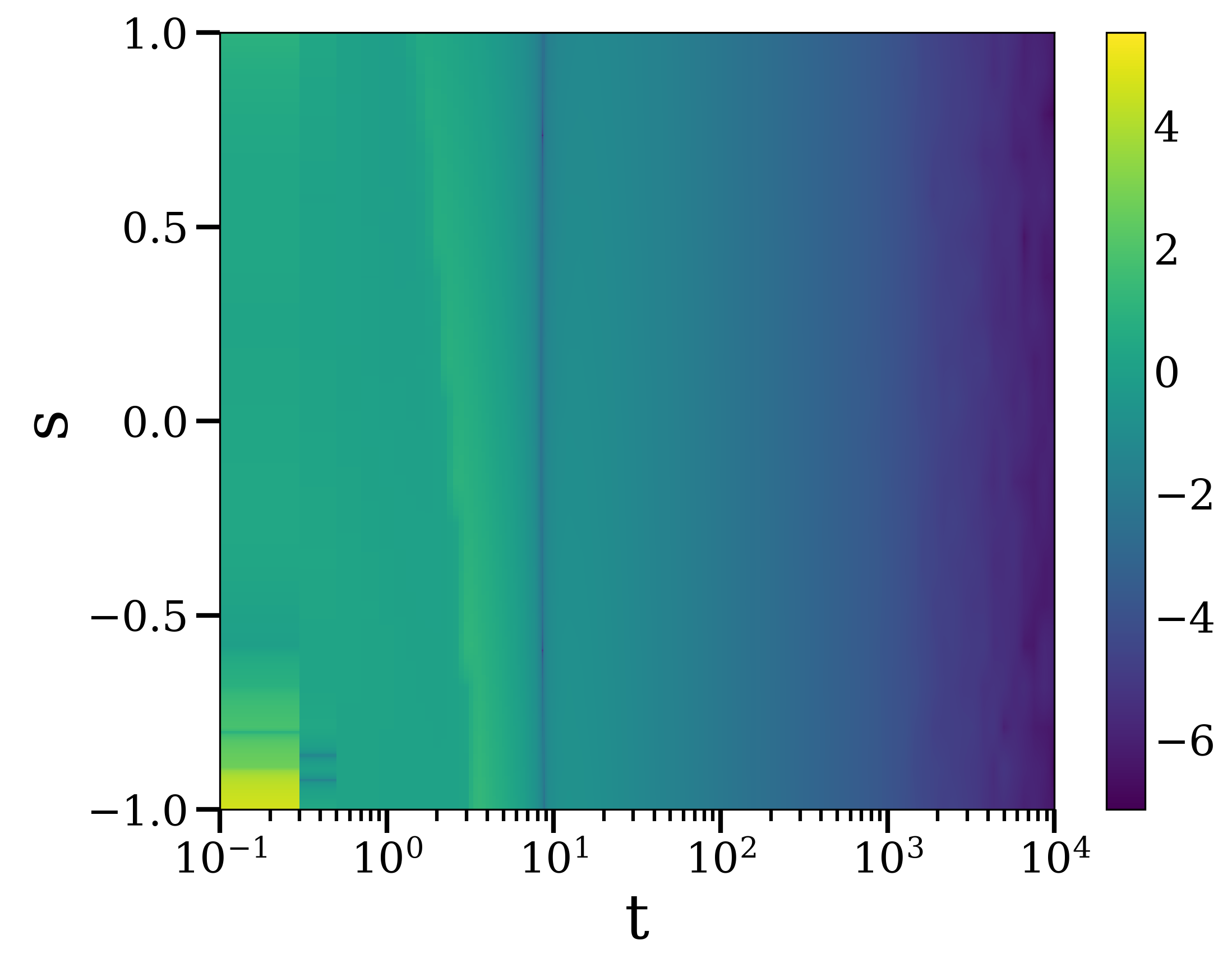}
    \end{subfigure}
    
    \begin{subfigure}[b]{0.48\textwidth}
        \includegraphics[width=\linewidth]{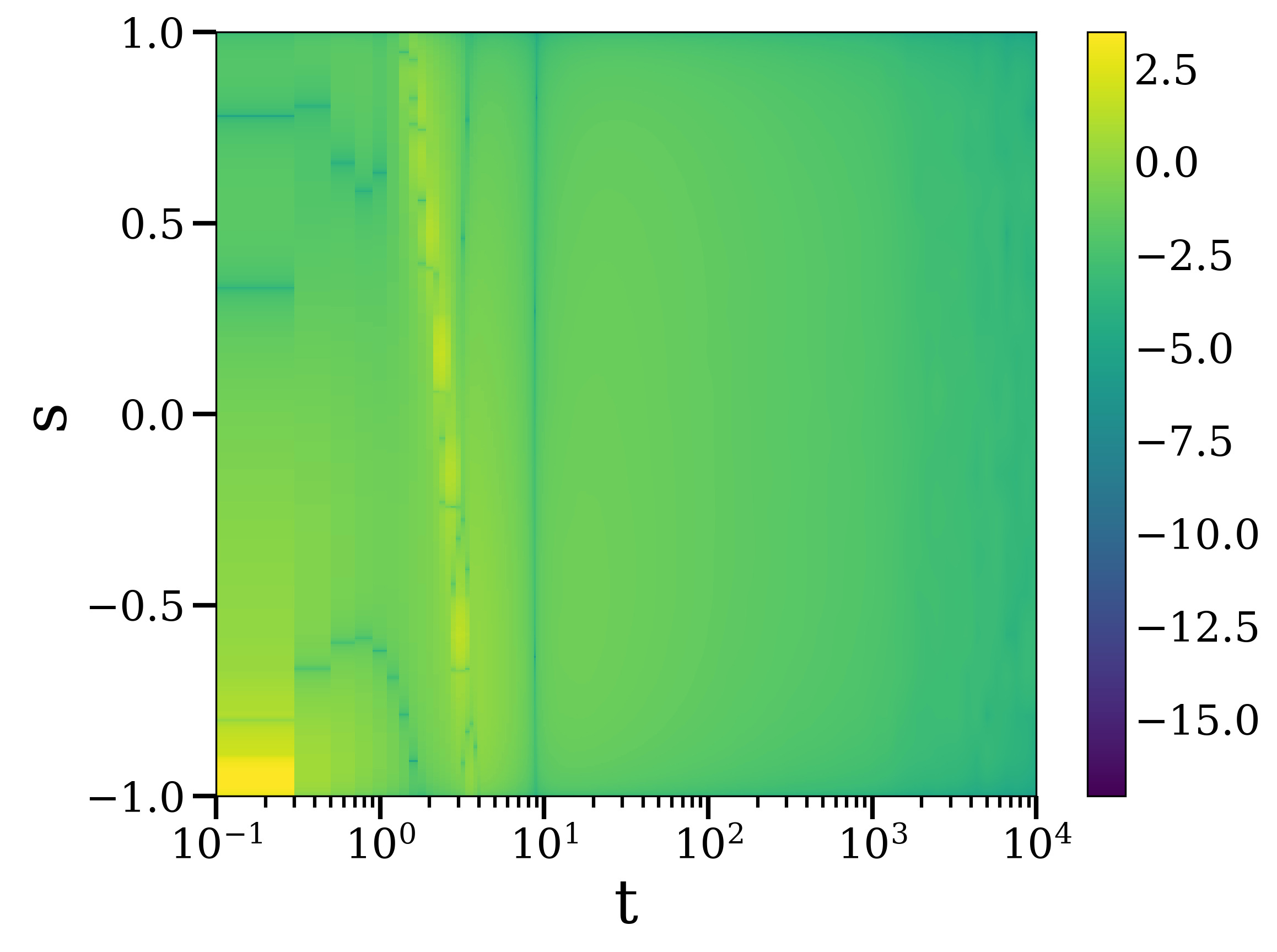}
    \end{subfigure}
    \begin{subfigure}[b]{0.48\textwidth}
        \includegraphics[width=\linewidth]{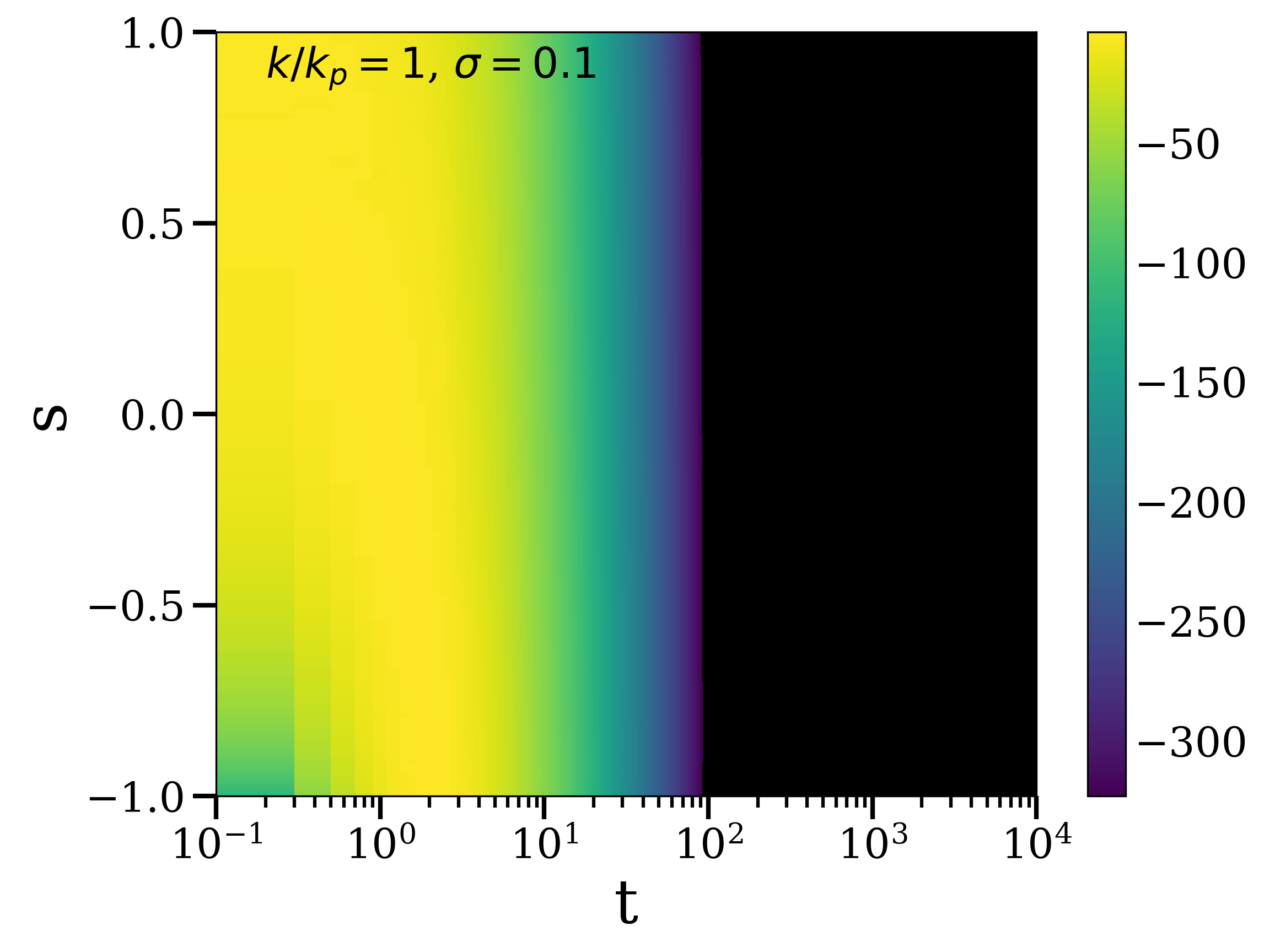}
    \end{subfigure}
    
    \begin{subfigure}[b]{0.48\textwidth}
        \includegraphics[width=\linewidth]{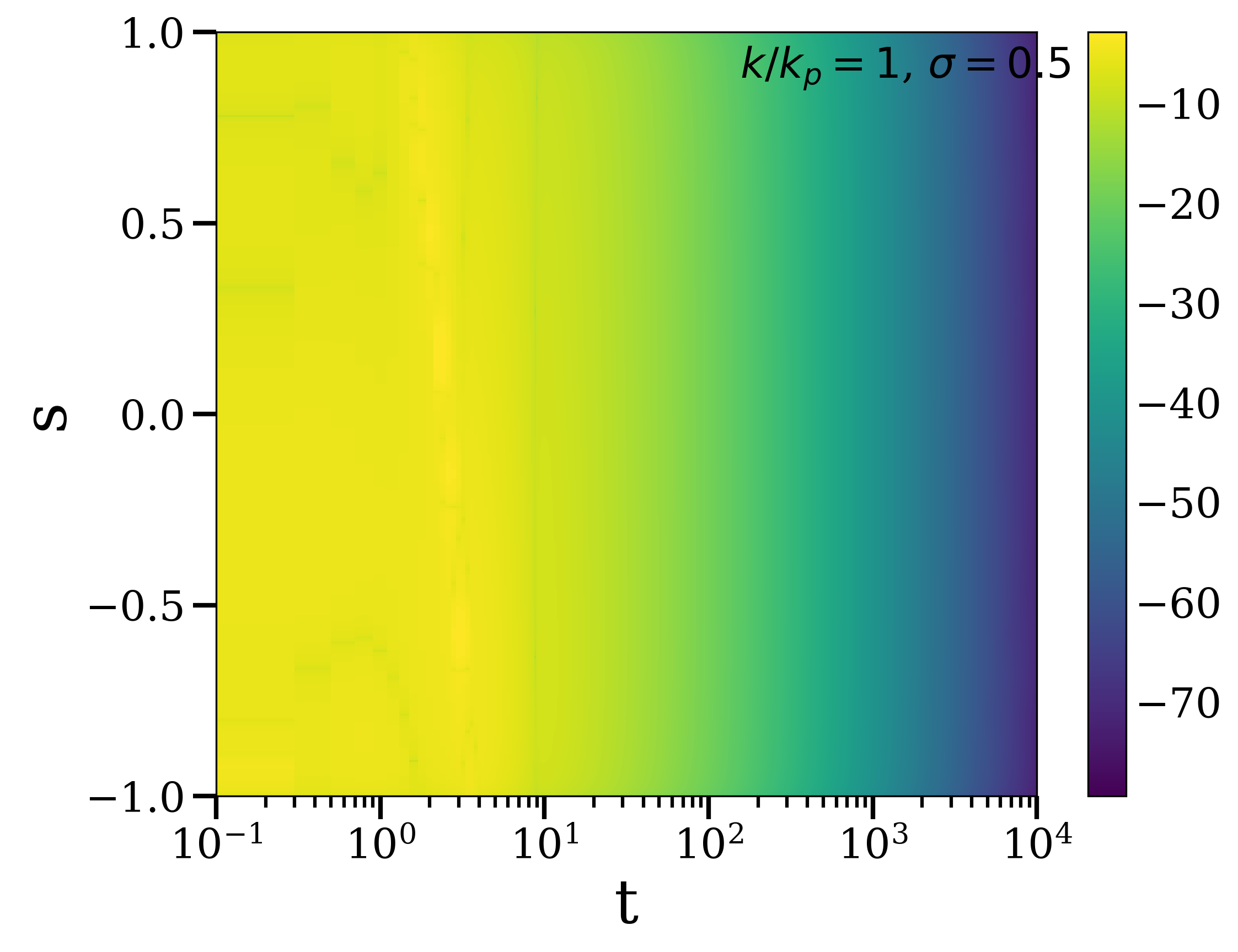}
    \end{subfigure}
    \begin{subfigure}[b]{0.48\textwidth}
        \includegraphics[width=\linewidth]{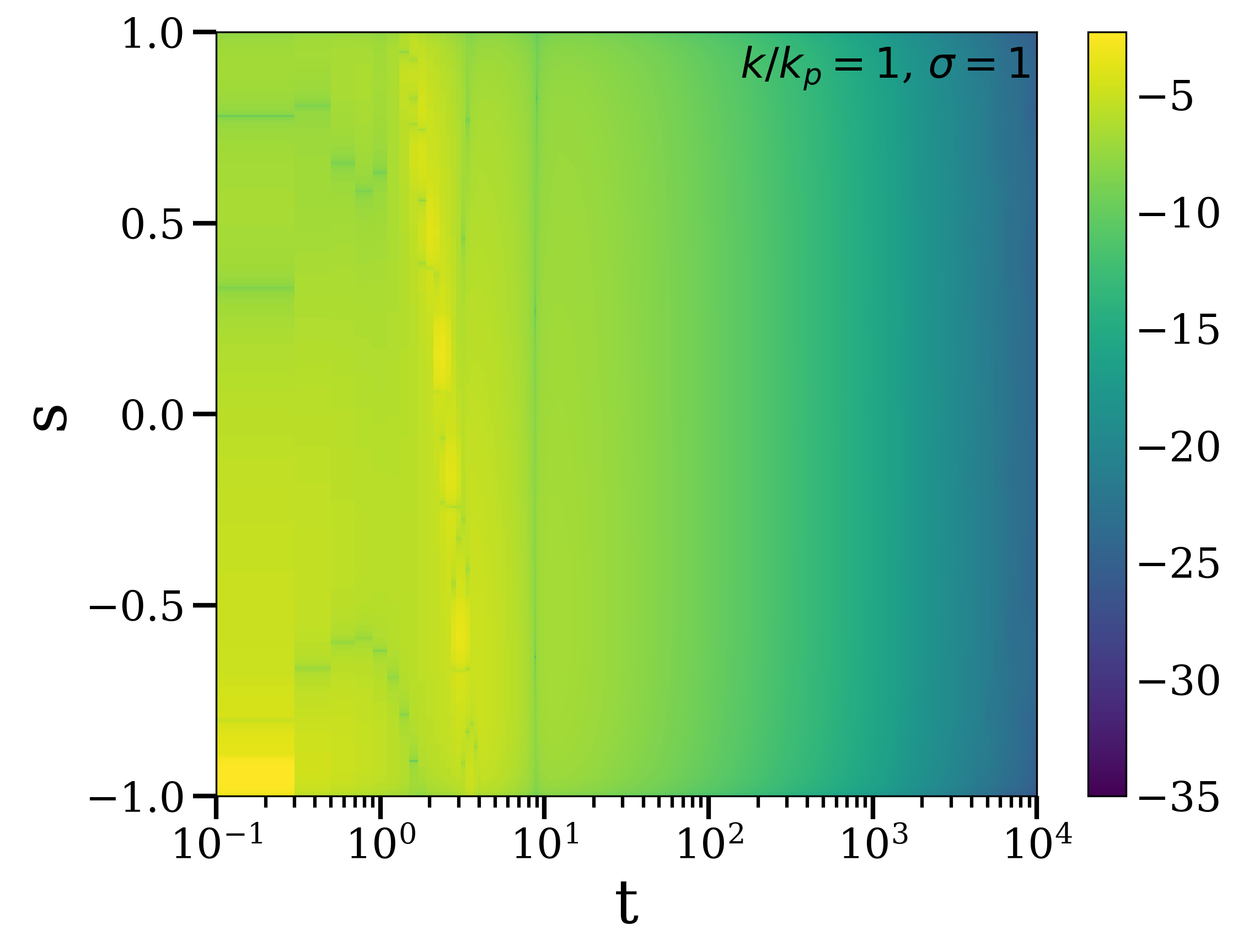}
    \end{subfigure}
    
    \caption{\footnotesize{Coloured mesh plots for the source term containing second-order vectors. (\textit{Top left}) First kernel corresponding to Eq.~\eqref{kernelb2rd}. (\textit{Top right}) Second kernel from Eq.~\eqref{kernelb2rd}. (\textit{Middle left}) Plot of the integrand in Eq.~\eqref{finalpsb2psi1} without the power spectra. (\textit{Middle right}) Plot of the integrand in Eq.~\eqref{finalpsb2psi1} with the power spectra included with $\sigma =0.1$. (\textit{Bottom left}) $\sigma = 0.5$ (\textit{Bottom right}) $\sigma = 1$. The two kernels exhibit an enhancement in the UV limit which couples large wavelength scalar perturbations to short wavelength tensor modes. This enhancement gets larger as the width of the primordial peak of scalars gets larger, we therefore conclude that this term diverges.} }
    \label{fig:b2psi1um_meshplots}
\end{figure}

\begin{figure}[H]
    \centering
    \begin{subfigure}[b]{0.48\textwidth}
        \includegraphics[width=\linewidth]{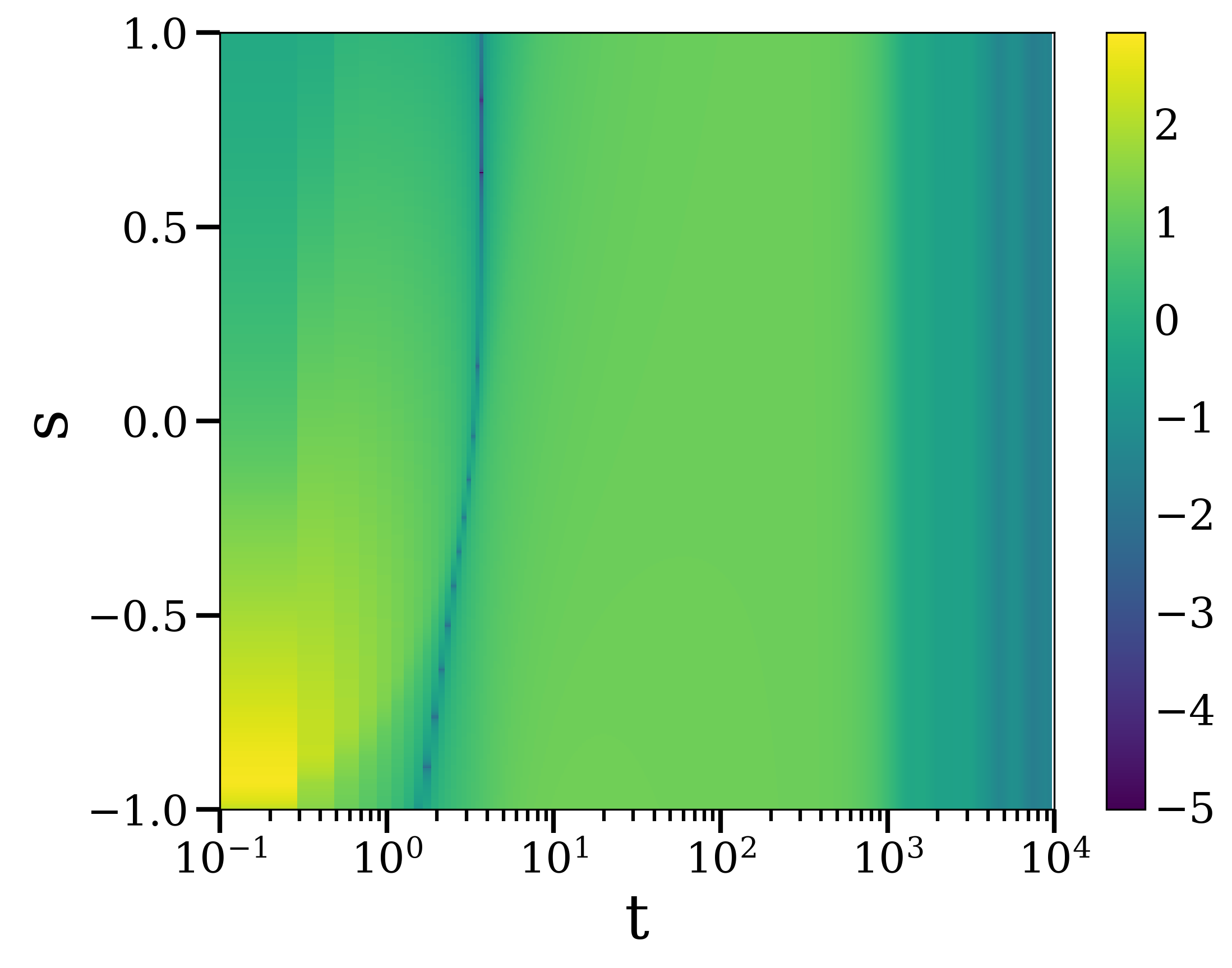}
    \end{subfigure}
    \begin{subfigure}[b]{0.48\textwidth}
        \includegraphics[width=\linewidth]{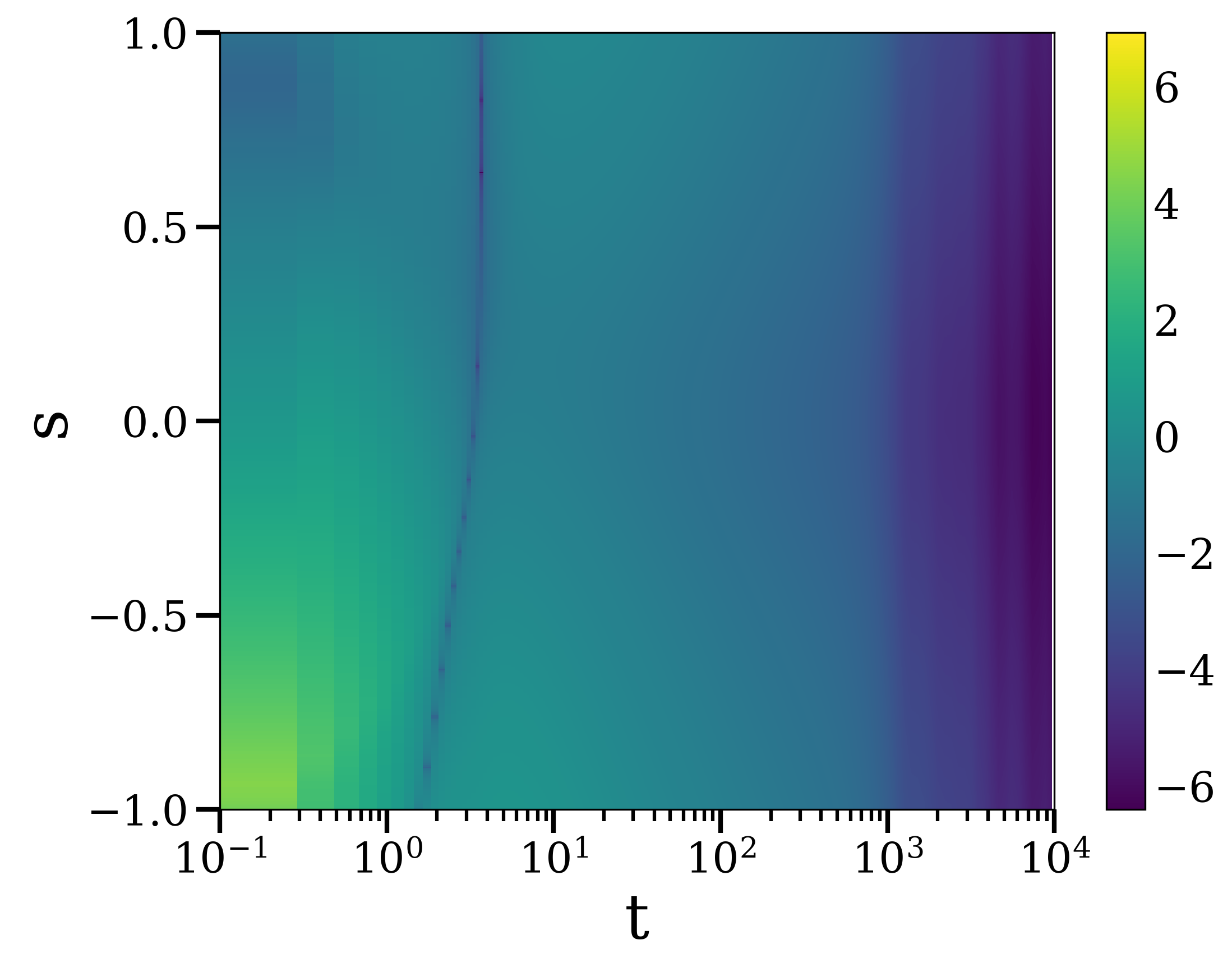}
    \end{subfigure}
    
    \begin{subfigure}[b]{0.48\textwidth}
        \includegraphics[width=\linewidth]{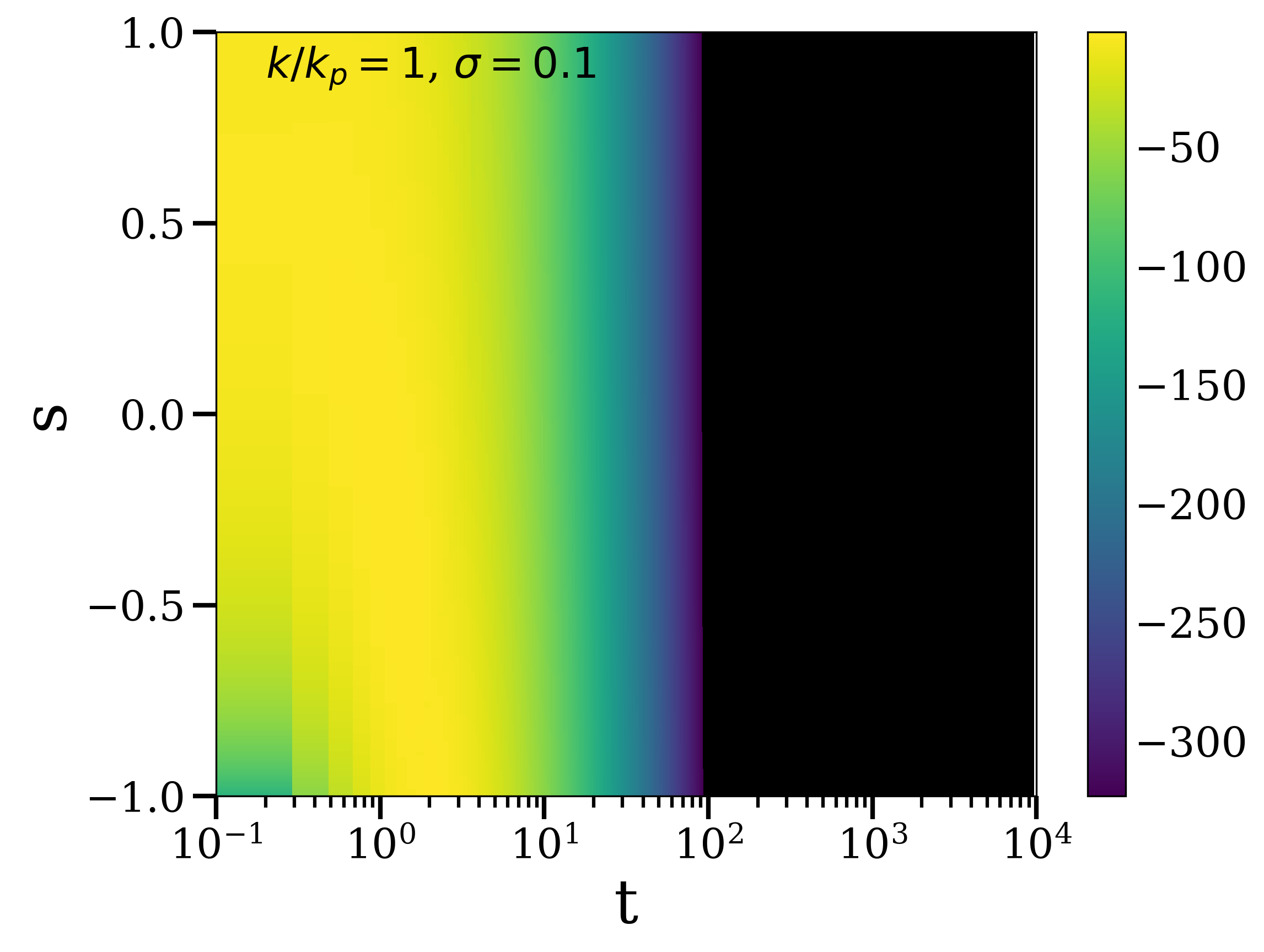}
    \end{subfigure}
    \begin{subfigure}[b]{0.48\textwidth}
        \includegraphics[width=\linewidth]{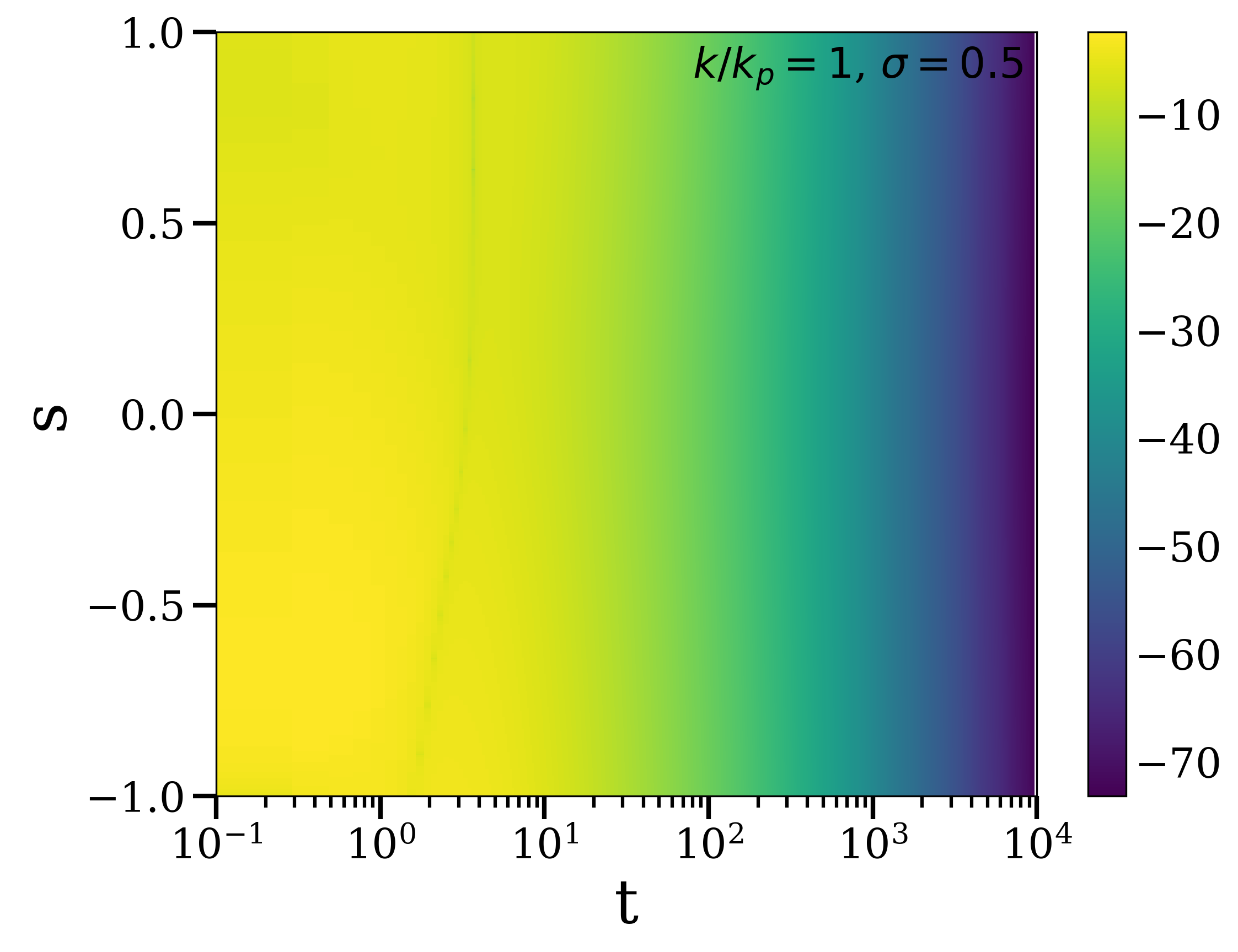}
    \end{subfigure}
    
    \begin{subfigure}[b]{0.48\textwidth}
        \includegraphics[width=\linewidth]{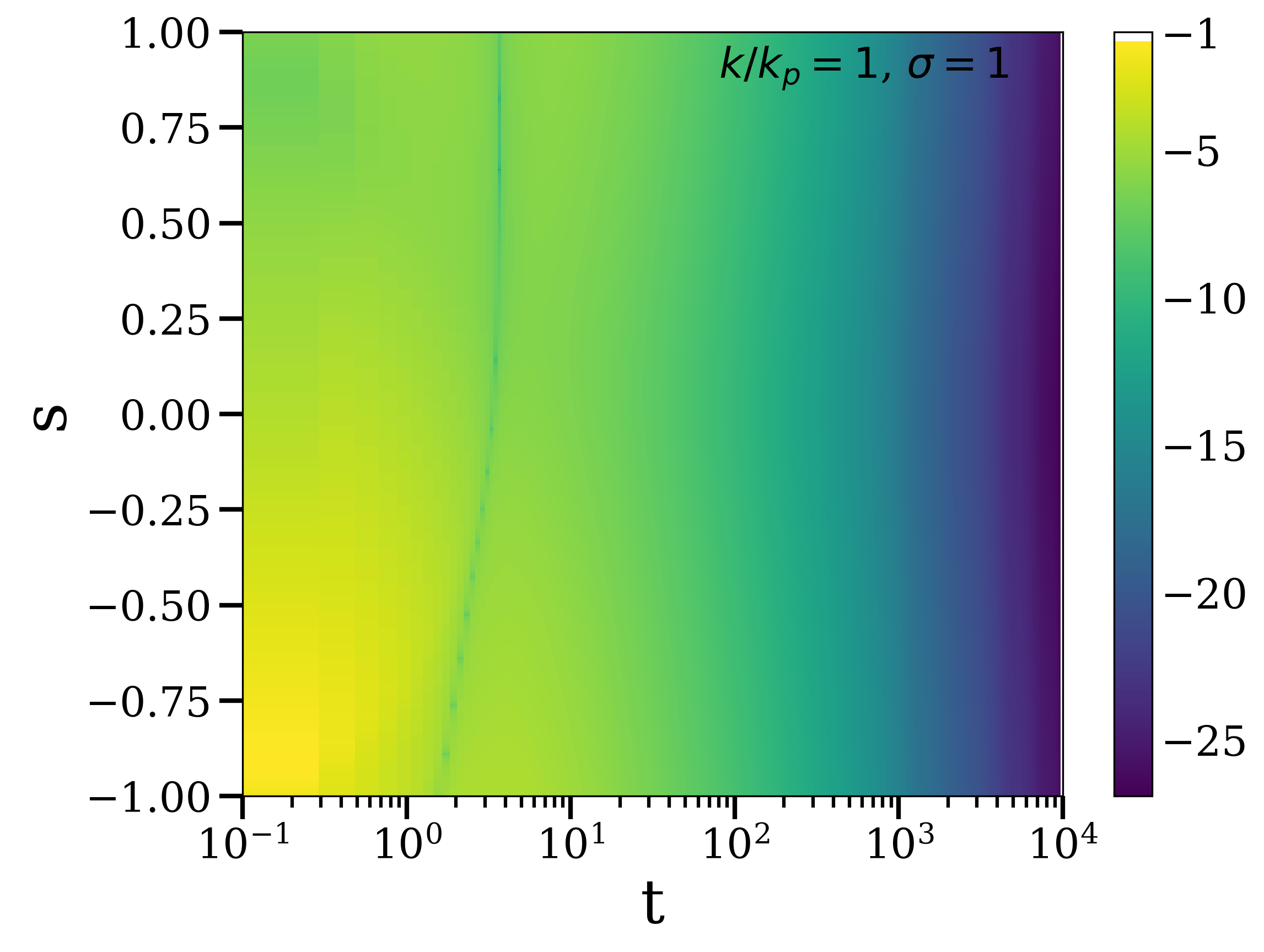}
    \end{subfigure}
    
    \caption{\footnotesize{Coloured mesh plots for the source term containing second-order tensors. (\textit{Top left}) Numerical kernel shown in Eq.~\eqref{finalkernelh2}.   (\textit{Top right}) Numerical part of the integrand shown in Eq.~\eqref{psh2psi1} without the primordial spectra. (\textit{Middle left}) Same as before but with inclusion of the first-order spectra, with $\sigma = 0.1$. (\textit{Middle right}) $\sigma =0.5$. (\textit{Bottom row}) $\sigma =1$. From these plots we can determine that the integrand diverges in the UV limit. The kernel exhibits an enhancement in the limit $v \rightarrow 1$ and $u \rightarrow 0$, which is also present when we include the polarisation functions. This enhancement disappears when the the scalar power spectrum is sufficiently peaked, however it reappears as $\sigma$ gets larger. In fact, it diverges at a similar rate than the second-order scalar-tensor iGWs (with opposite sign), however it is not possible to conclude whether this is also as $u^{-4}$.}}
    \label{fig:h2psi1um_meshplots}
\end{figure}

\newpage

\bibliography{refs} 

\providecommand{\href}[2]{#2}\begingroup\raggedright\begin{thebibliography}{10}

\bibitem{NANOGrav:2023gor}
{\scshape NANOGrav} collaboration, \emph{{The NANOGrav 15 yr Data Set: Evidence for a Gravitational-wave Background}}, \href{https://doi.org/10.3847/2041-8213/acdac6}{\emph{Astrophys. J. Lett.} {\bfseries 951} (2023) L8} [\href{https://arxiv.org/abs/2306.16213}{{\ttfamily 2306.16213}}].

\bibitem{EPTA:2023sfo}
{\scshape EPTA} collaboration, \emph{{The second data release from the European Pulsar Timing Array - I. The dataset and timing analysis}}, \href{https://doi.org/10.1051/0004-6361/202346841}{\emph{Astron. Astrophys.} {\bfseries 678} (2023) A48} [\href{https://arxiv.org/abs/2306.16224}{{\ttfamily 2306.16224}}].

\bibitem{EPTA:2023fyk}
{\scshape EPTA, InPTA:} collaboration, \emph{{The second data release from the European Pulsar Timing Array - III. Search for gravitational wave signals}}, \href{https://doi.org/10.1051/0004-6361/202346844}{\emph{Astron. Astrophys.} {\bfseries 678} (2023) A50} [\href{https://arxiv.org/abs/2306.16214}{{\ttfamily 2306.16214}}].

\bibitem{EPTA:2023xxk}
{\scshape EPTA, InPTA} collaboration, \emph{{The second data release from the European Pulsar Timing Array - IV. Implications for massive black holes, dark matter, and the early Universe}}, \href{https://doi.org/10.1051/0004-6361/202347433}{\emph{Astron. Astrophys.} {\bfseries 685} (2024) A94} [\href{https://arxiv.org/abs/2306.16227}{{\ttfamily 2306.16227}}].

\bibitem{Xu:2023wog}
H.~Xu et~al., \emph{{Searching for the Nano-Hertz Stochastic Gravitational Wave Background with the Chinese Pulsar Timing Array Data Release I}}, \href{https://doi.org/10.1088/1674-4527/acdfa5}{\emph{Res. Astron. Astrophys.} {\bfseries 23} (2023) 075024} [\href{https://arxiv.org/abs/2306.16216}{{\ttfamily 2306.16216}}].

\bibitem{Reardon:2023gzh}
D.~J. Reardon et~al., \emph{{Search for an Isotropic Gravitational-wave Background with the Parkes Pulsar Timing Array}}, \href{https://doi.org/10.3847/2041-8213/acdd02}{\emph{Astrophys. J. Lett.} {\bfseries 951} (2023) L6} [\href{https://arxiv.org/abs/2306.16215}{{\ttfamily 2306.16215}}].

\bibitem{Zic:2023gta}
A.~Zic et~al., \emph{{The Parkes Pulsar Timing Array third data release}}, \href{https://doi.org/10.1017/pasa.2023.36}{\emph{Publ. Astron. Soc. Austral.} {\bfseries 40} (2023) e049} [\href{https://arxiv.org/abs/2306.16230}{{\ttfamily 2306.16230}}].

\bibitem{NANOGrav:2023hvm}
{\scshape NANOGrav} collaboration, \emph{{The NANOGrav 15 yr Data Set: Search for Signals from New Physics}}, \href{https://doi.org/10.3847/2041-8213/acdc91}{\emph{Astrophys. J. Lett.} {\bfseries 951} (2023) L11} [\href{https://arxiv.org/abs/2306.16219}{{\ttfamily 2306.16219}}].

\bibitem{Bartolo:2016ami}
N.~Bartolo et~al., \emph{{Science with the space-based interferometer LISA. IV: Probing inflation with gravitational waves}}, \href{https://doi.org/10.1088/1475-7516/2016/12/026}{\emph{JCAP} {\bfseries 12} (2016) 026} [\href{https://arxiv.org/abs/1610.06481}{{\ttfamily 1610.06481}}].

\bibitem{Tomita:1967wkp}
K.~Tomita, \emph{{Non-Linear Theory of Gravitational Instability in the Expanding Universe}}, \href{https://doi.org/10.1143/PTP.37.831}{\emph{Prog. Theor. Phys.} {\bfseries 37} (1967) 831}.

\bibitem{Matarrese:1992rp}
S.~Matarrese, O.~Pantano and D.~Saez, \emph{{A General relativistic approach to the nonlinear evolution of collisionless matter}}, \href{https://doi.org/10.1103/PhysRevD.47.1311}{\emph{Phys. Rev. D} {\bfseries 47} (1993) 1311}.

\bibitem{Matarrese:1993zf}
S.~Matarrese, O.~Pantano and D.~Saez, \emph{{General relativistic dynamics of irrotational dust: Cosmological implications}}, \href{https://doi.org/10.1103/PhysRevLett.72.320}{\emph{Phys. Rev. Lett.} {\bfseries 72} (1994) 320} [\href{https://arxiv.org/abs/astro-ph/9310036}{{\ttfamily astro-ph/9310036}}].

\bibitem{Matarrese:1997ay}
S.~Matarrese, S.~Mollerach and M.~Bruni, \emph{{Second order perturbations of the Einstein-de Sitter universe}}, \href{https://doi.org/10.1103/PhysRevD.58.043504}{\emph{Phys. Rev. D} {\bfseries 58} (1998) 043504} [\href{https://arxiv.org/abs/astro-ph/9707278}{{\ttfamily astro-ph/9707278}}].

\bibitem{Ananda:2006af}
K.~N. Ananda, C.~Clarkson and D.~Wands, \emph{{The Cosmological gravitational wave background from primordial density perturbations}}, \href{https://doi.org/10.1103/PhysRevD.75.123518}{\emph{Phys. Rev. D} {\bfseries 75} (2007) 123518} [\href{https://arxiv.org/abs/gr-qc/0612013}{{\ttfamily gr-qc/0612013}}].

\bibitem{Baumann:2007zm}
D.~Baumann, P.~J. Steinhardt, K.~Takahashi and K.~Ichiki, \emph{{Gravitational Wave Spectrum Induced by Primordial Scalar Perturbations}}, \href{https://doi.org/10.1103/PhysRevD.76.084019}{\emph{Phys. Rev. D} {\bfseries 76} (2007) 084019} [\href{https://arxiv.org/abs/hep-th/0703290}{{\ttfamily hep-th/0703290}}].

\bibitem{Padilla:2024cbq}
L.~E. Padilla, J.~C. Hidalgo, K.~A. Malik and D.~Mulryne, \emph{{Detecting the stochastic gravitational wave background from primordial black holes in slow-reheating scenarios}}, \href{https://doi.org/10.1088/1475-7516/2024/12/011}{\emph{JCAP} {\bfseries 12} (2024) 011} [\href{https://arxiv.org/abs/2405.19271}{{\ttfamily 2405.19271}}].

\bibitem{Inomata:2023zup}
K.~Inomata, K.~Kohri and T.~Terada, \emph{{Detected stochastic gravitational waves and subsolar-mass primordial black holes}}, \href{https://doi.org/10.1103/PhysRevD.109.063506}{\emph{Phys. Rev. D} {\bfseries 109} (2024) 063506} [\href{https://arxiv.org/abs/2306.17834}{{\ttfamily 2306.17834}}].

\bibitem{Clesse:2018ogk}
S.~Clesse, J.~Garc{\'\i}a-Bellido and S.~Orani, \emph{{Detecting the Stochastic Gravitational Wave Background from Primordial Black Hole Formation}},  \href{https://arxiv.org/abs/1812.11011}{{\ttfamily 1812.11011}}.

\bibitem{PhysRevD.95.043511}
T.~Nakama, J.~Silk and M.~Kamionkowski, \emph{Stochastic gravitational waves associated with the formation of primordial black holes}, \href{https://doi.org/10.1103/PhysRevD.95.043511}{\emph{Phys. Rev. D} {\bfseries 95} (2017) 043511}.

\bibitem{Nakama:2020kdc}
T.~Nakama, \emph{{Stochastic gravitational waves associated with primordial black holes formed during an early matter era}}, \href{https://doi.org/10.1103/PhysRevD.101.063519}{\emph{Phys. Rev. D} {\bfseries 101} (2020) 063519}.

\bibitem{Vaskonen:2020lbd}
V.~Vaskonen and H.~Veerm{\"a}e, \emph{{Did NANOGrav see a signal from primordial black hole formation?}}, \href{https://doi.org/10.1103/PhysRevLett.126.051303}{\emph{Phys. Rev. Lett.} {\bfseries 126} (2021) 051303} [\href{https://arxiv.org/abs/2009.07832}{{\ttfamily 2009.07832}}].

\bibitem{DeLuca:2020agl}
V.~De~Luca, G.~Franciolini and A.~Riotto, \emph{{NANOGrav Data Hints at Primordial Black Holes as Dark Matter}}, \href{https://doi.org/10.1103/PhysRevLett.126.041303}{\emph{Phys. Rev. Lett.} {\bfseries 126} (2021) 041303} [\href{https://arxiv.org/abs/2009.08268}{{\ttfamily 2009.08268}}].

\bibitem{Yuan:2021qgz}
C.~Yuan and Q.-G. Huang, \emph{{A topic review on probing primordial black hole dark matter with scalar induced gravitational waves}}, \href{https://doi.org/10.1016/j.isci.2021.102860}{\emph{iScience} {\bfseries 24} (2021) 102860} [\href{https://arxiv.org/abs/2103.04739}{{\ttfamily 2103.04739}}].

\bibitem{Domenech:2021ztg}
G.~Dom\`enech, \emph{{Scalar Induced Gravitational Waves Review}}, \href{https://doi.org/10.3390/universe7110398}{\emph{Universe} {\bfseries 7} (2021) 398} [\href{https://arxiv.org/abs/2109.01398}{{\ttfamily 2109.01398}}].

\bibitem{Barnaby:2011qe}
N.~Barnaby, E.~Pajer and M.~Peloso, \emph{{Gauge Field Production in Axion Inflation: Consequences for Monodromy, non-Gaussianity in the CMB, and Gravitational Waves at Interferometers}}, \href{https://doi.org/10.1103/PhysRevD.85.023525}{\emph{Phys. Rev. D} {\bfseries 85} (2012) 023525} [\href{https://arxiv.org/abs/1110.3327}{{\ttfamily 1110.3327}}].

\bibitem{Thorne:2017jft}
B.~Thorne, T.~Fujita, M.~Hazumi, N.~Katayama, E.~Komatsu and M.~Shiraishi, \emph{{Finding the chiral gravitational wave background of an axion-SU(2) inflationary model using CMB observations and laser interferometers}}, \href{https://doi.org/10.1103/PhysRevD.97.043506}{\emph{Phys. Rev. D} {\bfseries 97} (2018) 043506} [\href{https://arxiv.org/abs/1707.03240}{{\ttfamily 1707.03240}}].

\bibitem{Dimastrogiovanni:2016fuu}
E.~Dimastrogiovanni, M.~Fasiello and T.~Fujita, \emph{{Primordial Gravitational Waves from Axion-Gauge Fields Dynamics}}, \href{https://doi.org/10.1088/1475-7516/2017/01/019}{\emph{JCAP} {\bfseries 01} (2017) 019} [\href{https://arxiv.org/abs/1608.04216}{{\ttfamily 1608.04216}}].

\bibitem{Chang:2022vlv}
Z.~Chang, X.~Zhang and J.~Z. Zhou, \emph{Gravitational waves from primordial scalar and tensor perturbations}, \href{https://doi.org/10.1103/PhysRevD.107.063510}{\emph{Phys. Rev. D} {\bfseries 107} (2023) 063510} [\href{https://arxiv.org/abs/2209.07693}{{\ttfamily 2209.07693}}].

\bibitem{Bari:2023rcw}
P.~Bari, N.~Bartolo, G.~Domènech and S.~Matarrese, \emph{Gravitational waves induced by scalar-tensor mixing},  \href{https://arxiv.org/abs/2307.05404}{{\ttfamily 2307.05404}}.

\bibitem{Picard:2023sbz}
R.~Picard and K.~A. Malik, \emph{{Induced gravitational waves: the effect of first order tensor perturbations}}, \href{https://doi.org/10.1088/1475-7516/2024/10/010}{\emph{JCAP} {\bfseries 10} (2024) 010} [\href{https://arxiv.org/abs/2311.14513}{{\ttfamily 2311.14513}}].

\bibitem{Yu:2023lmo}
Y.~H. Yu and S.~Wang, \emph{Primordial gravitational waves assisted by cosmological scalar perturbations},  \href{https://arxiv.org/abs/2303.03897}{{\ttfamily 2303.03897}}.

\bibitem{Chen:2022dah}
C.~Chen, A.~Ota, H.~Y. Zhu and Y.~Zhu, \emph{Missing one-loop contributions in secondary gravitational waves}, \href{https://doi.org/10.1103/PhysRevD.107.083518}{\emph{Phys. Rev. D} {\bfseries 107} (2023) 083518} [\href{https://arxiv.org/abs/2210.17176}{{\ttfamily 2210.17176}}].

\bibitem{Malik:2008im}
K.~A. Malik and D.~Wands, \emph{{Cosmological perturbations}}, \href{https://doi.org/10.1016/j.physrep.2009.03.001}{\emph{Phys. Rept.} {\bfseries 475} (2009) 1} [\href{https://arxiv.org/abs/0809.4944}{{\ttfamily 0809.4944}}].

\bibitem{Christopherson:2009fp}
A.~J. Christopherson and K.~A. Malik, \emph{{Practical tools for third order cosmological perturbations}}, \href{https://doi.org/10.1088/1475-7516/2009/11/012}{\emph{JCAP} {\bfseries 11} (2009) 012} [\href{https://arxiv.org/abs/0909.0942}{{\ttfamily 0909.0942}}].

\bibitem{Caprini:2018mtu}
C.~Caprini and D.~G. Figueroa, \emph{{Cosmological Backgrounds of Gravitational Waves}}, \href{https://doi.org/10.1088/1361-6382/aac608}{\emph{Class. Quant. Grav.} {\bfseries 35} (2018) 163001} [\href{https://arxiv.org/abs/1801.04268}{{\ttfamily 1801.04268}}].

\bibitem{Zhou:2021vcw}
J.~Z. Zhou, X.~Zhang, Q.~H. Zhu and Z.~Chang, \emph{The third order scalar induced gravitational waves}, \href{https://doi.org/10.1088/1475-7516/2022/05/013}{\emph{JCAP} {\bfseries 05} (2022) 013} [\href{https://arxiv.org/abs/2106.01641}{{\ttfamily 2106.01641}}].

\bibitem{Espinosa:2018eve}
J.~R. Espinosa, D.~Racco and A.~Riotto, \emph{{A Cosmological Signature of the SM Higgs Instability: Gravitational Waves}}, \href{https://doi.org/10.1088/1475-7516/2018/09/012}{\emph{JCAP} {\bfseries 09} (2018) 012} [\href{https://arxiv.org/abs/1804.07732}{{\ttfamily 1804.07732}}].

\bibitem{Domenech:2023fuz}
G.~Dom\`enech, \emph{{Lectures on Gravitational Wave Signatures of Primordial Black Holes}},  7, 2023, \href{https://arxiv.org/abs/2307.06964}{{\ttfamily 2307.06964}}.

\bibitem{Lepage:2020tgj}
G.~P. Lepage, \emph{{Adaptive multidimensional integration: VEGAS enhanced}}, \href{https://doi.org/10.1016/j.jcp.2021.110386}{\emph{J. Comput. Phys.} {\bfseries 439} (2021) 110386} [\href{https://arxiv.org/abs/2009.05112}{{\ttfamily 2009.05112}}].

\bibitem{Moore:2014lga}
C.~J. Moore, R.~H. Cole and C.~P.~L. Berry, \emph{{Gravitational-wave sensitivity curves}}, \href{https://doi.org/10.1088/0264-9381/32/1/015014}{\emph{Class. Quant. Grav.} {\bfseries 32} (2015) 015014} [\href{https://arxiv.org/abs/1408.0740}{{\ttfamily 1408.0740}}].

\bibitem{Sathyaprakash:2009xs}
B.~S. Sathyaprakash and B.~F. Schutz, \emph{{Physics, Astrophysics and Cosmology with Gravitational Waves}}, \href{https://doi.org/10.12942/lrr-2009-2}{\emph{Living Rev. Rel.} {\bfseries 12} (2009) 2} [\href{https://arxiv.org/abs/0903.0338}{{\ttfamily 0903.0338}}].

\bibitem{Obata:2016tmo}
I.~Obata and J.~Soda, \emph{{Chiral primordial Chiral primordial gravitational waves from dilaton induced delayed chromonatural inflation}}, \href{https://doi.org/10.1103/PhysRevD.93.123502}{\emph{Phys. Rev. D} {\bfseries 93} (2016) 123502} [\href{https://arxiv.org/abs/1602.06024}{{\ttfamily 1602.06024}}].

\bibitem{Bartolo:2017szm}
N.~Bartolo and G.~Orlando, \emph{{Parity breaking signatures from a Chern-Simons coupling during inflation: the case of non-Gaussian gravitational waves}}, \href{https://doi.org/10.1088/1475-7516/2017/07/034}{\emph{JCAP} {\bfseries 07} (2017) 034} [\href{https://arxiv.org/abs/1706.04627}{{\ttfamily 1706.04627}}].

\bibitem{Bartolo:2018elp}
N.~Bartolo, G.~Orlando and M.~Shiraishi, \emph{{Measuring chiral gravitational waves in Chern-Simons gravity with CMB bispectra}}, \href{https://doi.org/10.1088/1475-7516/2019/01/050}{\emph{JCAP} {\bfseries 01} (2019) 050} [\href{https://arxiv.org/abs/1809.11170}{{\ttfamily 1809.11170}}].

\bibitem{Picard:2024ekd}
R.~Picard and M.~W. Davies, \emph{{Effects of scalar non-Gaussianity on induced scalar-tensor gravitational waves}}, \href{https://doi.org/10.1088/1475-7516/2025/02/037}{\emph{JCAP} {\bfseries 02} (2025) 037} [\href{https://arxiv.org/abs/2410.17819}{{\ttfamily 2410.17819}}].

\bibitem{king_2017_438045}
T.~King, S.~Butcher and L.~Zalewski, \emph{{Apocrita - High Performance Computing Cluster for Queen Mary University of London}}, Mar., 2017.
\newblock 10.5281/zenodo.438045.

\bibitem{Pitrou:2013hga}
C.~Pitrou, X.~Roy and O.~Umeh, \emph{xpand: An algorithm for perturbing homogeneous cosmologies}, \href{https://doi.org/10.1088/0264-9381/30/16/165002}{\emph{Class. Quant. Grav.} {\bfseries 30} (2013) 165002} [\href{https://arxiv.org/abs/1302.6174}{{\ttfamily 1302.6174}}].

\bibitem{Carrilho:2015cma}
P.~Carrilho and K.~A. Malik, \emph{{Vector and tensor contributions to the curvature perturbation at second order}}, \href{https://doi.org/10.1088/1475-7516/2016/02/021}{\emph{JCAP} {\bfseries 02} (2016) 021} [\href{https://arxiv.org/abs/1507.06922}{{\ttfamily 1507.06922}}].

\bibitem{Lu:2007cj}
T.~H.~C. Lu, K.~Ananda and C.~Clarkson, \emph{Vector modes generated by primordial density fluctuations}, \href{https://doi.org/10.1103/PhysRevD.77.043523}{\emph{Phys. Rev. D} {\bfseries 77} (2008) 043523} [\href{https://arxiv.org/abs/0709.1619}{{\ttfamily 0709.1619}}].

\bibitem{Kohri:2018awv}
K.~Kohri and T.~Terada, \emph{{Semianalytic calculation of gravitational wave spectrum nonlinearly induced from primordial curvature perturbations}}, \href{https://doi.org/10.1103/PhysRevD.97.123532}{\emph{Phys. Rev. D} {\bfseries 97} (2018) 123532} [\href{https://arxiv.org/abs/1804.08577}{{\ttfamily 1804.08577}}].

\end{thebibliography}\endgroup
\bibliographystyle{JHEP}
\end{document}